# Using Deep Learning Techniques to Search for the MiniBooNE Low Energy Excess in MicroBooNE with $> 3\sigma$ Sensitivity

by

Jarrett S Moon

Submitted to the Department of Physics
in partial fulfillment of the requirements for the degree of

Doctor of Philosophy

at the

MASSACHUSETTS INSTITUTE OF TECHNOLOGY

September 2020

© Massachusetts Institute of Technology 2020. All rights reserved.

Author . . . . . . . . . . . . . . . . . . . . . . . . . . . . . . . . . . . . . . . . . . . . . . . . . . . . . . . . . . . . . . . .
Department of Physics
August 7, 2020

Certified by. . . . . . . . . . . . . . . . . . . . . . . . . . . . . . . . . . . . . . . . . . . . . . . . . . . . . . . . . . . .
Janet Conrad
Professor of Physics
Thesis Supervisor

Accepted by . . . . . . . . . . . . . . . . . . . . . . . . . . . . . . . . . . . . . . . . . . . . . . . . . . . . . . . . . . .
Nergis Mavalvala
Associate Department Head of Physics, MIT

# Using Deep Learning Techniques to Search for the MiniBooNE Low Energy Excess in MicroBooNE with $> 3\sigma$ Sensitivity

by

Jarrett S Moon

Submitted to the Department of Physics
on August 7, 2020, in partial fulfillment of the
requirements for the degree of
Doctor of Philosophy

## Abstract

This thesis describes an analysis developed for the MicroBooNE experiment to investigate an anomalous excess of electron-like events observed in the MiniBooNE detector. The hypothesis investigated here is that the MiniBooNE anomaly represents appearance of electron neutrinos. Using an amalgam of novel Deep Learning and standard algorithmic techniques this analysis reconstructs and identifies a highly pure sample of charged current quasi-elastic muon neutrino and electron neutrino interactions. This thesis describes the steps in the analysis chain and provides data-to-simulation comparisons for each step that establish confidence in the final prediction. When interpreted in the context of a $\nu_e$ appearance like model, this analysis predicts a $3.2\sigma$ sensitivity to exclude a standard model fluctuation which would appear as a MiniBooNE like anomaly using $7\times10^{20}$ protons on target of MicroBooNE Data.

Thesis Supervisor: Janet Conrad
Title: Professor of Physics

# Acknowledgments

I want to thank Janet Conrad for being an exceptional adviser these years. Thank you for your enthusiasm, support, guidance, and perpetual willingness to fight for your students.

I also thank my wife, Katie, for her continual patience and support and for following me around the country wherever this journey took us. Thanks for everything you've done to make this possible!

I also want to thank my friends and colleagues from the MicroBooNE deep learning group with whom I've had the pleasure of working with and learning from: Adrien Hourlier, Taritree Wongjirad, Lauren Yates, Nick Kamp, Rui An, Davio Cianci, Josh Mills, Katie Mason, Kazuhiro Terao, Ran Itay, Ralitsa Sharankova, Vic Genty, and Elizabeth Hall.

And finally thanks to my parents who set me down a path of insatiable curiosity and who have supported me ever since.

# Contents

# List of Figures

# List of Tables

# Chapter 1

# Introduction

The Standard Model of particle physics is arguably the most successful and comprehensive description of nature ever uncovered. Within its framework, all of nature, save gravity, is described by 12 elementary particles of matter, the fermions, and 4 force-carrying particles, the mediator bosons. [1] However, despite decades of highly accurate predictions, the first hints that the Standard Model required modification and expansion came from unexpected behaviors observed in 3 of those 12 matter particles, the neutrinos [2].

The neutrino was initially hypothesized by Wolfgang Pauli in 1930 to explain a peculiar feature observed in radioactive $\beta$-decay: $\beta$ emission resulted in an electron with a continuous energy spectrum. As it was then understood, $\beta$ decay yielded only two bodies in the final state, and thus should produce a monoenergetic electron if energy and momentum are conserved. Rather than abandon conservation laws, Pauli postulated the existence of a third unseen daughter particle. This invisible particle must be neutral, half spin-1/2, and have a very small mass [3].

The specifics of this process were explored further by Enrico Fermi who, in 1933, proposed that $\beta$ emission involved the decay of the recently discovered neutron into a proton, electron, and neutrino [4]. Fermi's work laid the theoretical foundation needed to put neutrino detection on firm experimental grounds. However, neutrinos interact so rarely with matter that direct confirmation would not come for over two decades until Frederick Reines and Clyde Cowan managed to experimentally measure

"inverse $\beta$ decay," $\nu + p \rightarrow n + e^+$, using a very high flux of antineutrinos from the Savannah River nuclear reactor [5].

In the following years, further experiments revealed that the neutrino was not actually just one particle. Much like there are three charged leptons, the electron, muon, and tau, there are three parallel flavors of neutrino. The neutrinos that had thus far been observed were what we today call the electron neutrino. The muon neutrino and tau neutrino were directly detected in, respectively, 1975 and 2000 [6] [7]. Further measurements of the Z-boson decay suggested that these three are the only flavors of neutrinos which actively participate in weak interactions. [8]

But a mystery remained: it was known that nuclear fusion within the sun would produce an enormous number of electron neutrinos–could we see them? The Homestake experiment was built to detect these solar neutrinos. And detect them it did, but nowhere near enough of them–depending on the exact solar model employed, between only a third to half as many as anticipated [9]. Attempts to find major problems in the standard solar model were unable to resolve the discrepancy. It was eventually proposed that the issue lay with our understanding of neutrinos themselves.

Neutrinos in the standard model were defined to have zero mass. A consequence of this is that a neutrino born as an electron neutrino is destined to forever remain an electron neutrino. But if the neutrinos had a small but finite mass, then electron neutrinos born in the sun could transmute into muon or tau neutrinos in flight to Earth. Experiments like Homestake, which had spotted the anomaly, were sensitive only to electron neutrinos. They were blind to the fact that the rest were not missing, but had simply transformed into a different type. This hypothesized solution came to be known as neutrino oscillation.

Direct evidence for oscillations came first from Super Kamiokande and the Sudbury Neutrino Observatory (SNO) in the late 1990s and early 2000s. Super Kamiokande observed evidence of oscillations in atmospheric neutrinos [11], and SNO, which was sensitive to all three types of neutrinos, verified that the expected solar flux was correct if all types of neutrinos were counted [10]. Together SNO and Super Kamiokande were awarded the 2015 Nobel Prize for their discovery of neutrino oscillation. The discovery requires that at least two of the three neutrinos cannot be massless. This was the first direct observation of physics beyond the Standard Model and it remains one of the most significant finds in neutrino physics and one of the most fertile areas of continuing research.

However, hot on the heels of this triumph was a a fresh set of anomalies. The first hint came from the LSND experiment at Los Alamos in the late 1990s. In contrast with the results from SNO and Super Kamiokande, LSND's results, if interpreted as oscillations, implied a significantly larger gap between the masses of the neutrinos [12]. This possibly implied the existence of a fourth neutrino with no weak interaction, a so called "sterile" neutrino. MiniBooNE was one of the first detectors to investigate the LSND signal, and it, too, found a low energy excess of electron neutrino events which is partially compatible with the sterile neutrino oscillation hypothesis [13]. Other experiments including reactor experiments and source experiments also noted deficits that can be interpreted as a sterile neutrino signal [15] [14].

However, it would be wrong to assume that all signs point one way. There is some tension between the experiments with anomalies within a sterile neutrino picture. Also, there have been null results from experiments searching for the anomalies. In particular, there is strong tension between experiments which search for $\nu_\mu$ disappearance, which have seen no signal, and those which search for $\nu_e$ appearance and disappearance, which has observed anomalies. If then sterile neutrino picture is correct, then at least some of the anomalies must be affected by unknown systematic effects. If the sterile neutrino picture is not correct, then all anomalies must be explainable by systematic effects.

This is where the MicroBooNE experiment enters the picture. MicroBooNE was constructed and run at Fermi National Accelerator Laboratory [16] to independently check the MiniBooNE anomaly. The MiniBooNE anomaly is a $4.7\sigma$ excess of $\nu_e$-like events in a $\nu_\mu$ beam–so it is quite significant [13]. However, the energy dependence of the excess is in tension with other experiments. The goal of MicroBooNE is to test the energy dependence of the MiniBooNE signal. The result may be

- Confirmation of the shape of the excess, in which case sterile neutrino models

will need to be maintained, but modified;

- Exclusion of the signal, in which case most sterile neutrino models will be ruled out; or

- Correction to the measured energy dependence in MiniBooNE, in which case we conclude that MiniBooNE did see a signal, but it was distorted by a systematic effect.

Any of these three outcomes will have major consequences for the future direction of short baseline oscillation studies.

MicroBooNE uses the same neutrino source as MiniBooNE, but different detection technology [21] that addresses MiniBooNE's limitations as a large Cherenkov detector. Specifically, MicroBooNE is the first large scale Liquid Argon Time Projection Chamber (LArTPC). This provides a new way to identify both electron and muon neutrino interactions. A measurement of the electron neutrino rate at various energies allows MicroBooNE to investigate the low energy MiniBooNE excess from a different perspective.

LArTPC's lend themselves to physics information encoded in the form of images, reminiscent of those from a bubble chamber. These high resolution images provide information that can be used for particle identification, topological reconstruction, and energy reconstruction. This thesis details an end-to-end reconstruction and selection framework which takes advantage of a novel mixture of traditional algorithms and Deep Learning, a machine learning technique particularly adept at image analysis. This analysis has currently been validated on a set of $\nu_\mu$ data and a limited set of $\nu_e$ samples. In the near future, pending collaboration agreement to unblind, it will be applied to $7\times 10^{20}$ protons on target (POT) with an anticipated exclusion sensitivity of $3.2\sigma$. In the following chapters, we describe how we obtained this predicted sensitivity [17].

# Chapter 2

# Physics of Neutrinos and Oscillations

This chapter provides an overview of neutrino physics as it is currently understood in a Standard Model + oscillation paradigm. A discussion of the status of 3+1 models (that is, 3 active + 1 sterile neutrino) is also provided.

## 2.1 Neutrinos in the Standard Model

The Standard Model contains three neutrinos: the electron neutrino, muon neutrino, and tau neutrino [2]. They are massless fermions that carry neither electromagnetic nor color charge. They do, however, carry weak isosopin. Consequently, neutrinos participate exclusively in weak interactions.

These weak interactions are mediated by three gauge bosons, $W^{\pm}$ and $Z^0$. Uniquely in the Standard Model, the force mediating particles for the weak force may be charged or neutral. This allows a useful categorization of interactions: charged current (CC) or neutral current (NC), corresponding respectively to the exchange of a W or Z boson [1]. CC interactions involve charge exchange, and, accordingly, explicitly reveal conservation of lepton number in that a CC interaction involving an electron neutrino will result in an electron in the final state. NC interactions, on the other hand, do not discriminate between lepton or quark flavor. Examples illustrating some simple

CC and NC interactions are given in Figs. 2-1 & 2-2, where time runs from top to bottom.

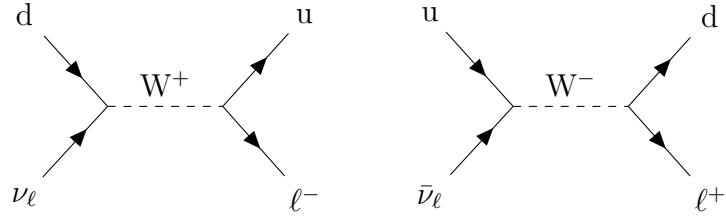

Figure 2-1: Feynman diagrams illustrating charged current interactions via the exchange of a $W^{\pm}$ boson between the neutrino and quark.

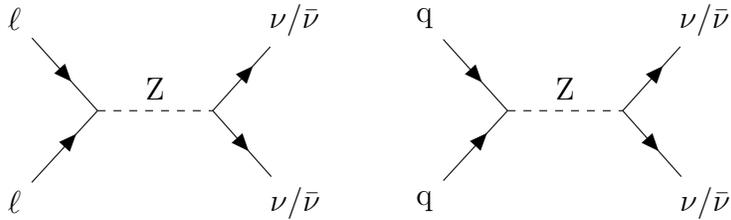

Figure 2-2: Feynman diagrams illustrating neutral current interactions via the exchange of a Z boson between any (anti)neutrino flavor and any lepton (left) or quark (right).

## 2.2 Neutrino Oscillations

By the late 20th century it was firmly and directly confirmed that there are three active flavors of neutrinos [8]. The electron, muon, and tau flavors had been detected explicitly, and results from the LEP collider which measured the decay of the $Z^0$ boson indicated that there were no more than these three active flavors to be found with light or zero masses.

The Standard Model states that these three flavor eigenstates all have zero mass. This allows for an explanation of parity violation observed in neutrino interactions, but it is a strange feature since all of the other leptons and quarks have masses. As a thought experiment, let us set aside the issue of explaining parity violation, and allow neutrinos to have mass. They must have light masses, less than $\sim 1$ eV, since no experiment has yet directly measured neutrino mass. Let us also assume that, as in the quark sector, the flavor eigenstates do not align with the flavor eigenstates–an effect called "mixing." With these assumptions, let's see where this takes us...

### 2.2.1 Neutrino Mixing

Neutrino mixing, in general, assumes that the flavor eigenstates are a non-trivial linear combination of mass eigenstates. That is, for three flavors, if we define $|\nu_\alpha\rangle$ as a flavor eigenstate where $\alpha \in [e, \mu, \tau]$ and $|\nu_i\rangle$ as a mass eigenstate where $i \in [1, 2, 3]$ then for some constants c, the following relationship is true.

$$|\nu_\alpha\rangle = c_{\alpha 1} |\nu_1\rangle + c_{\alpha 2} |\nu_2\rangle + c_{\alpha 3} |\nu_3\rangle \tag{2.1}$$

More generally, we can specify the link between the flavor and mass eigenbasis by adopting a unitary transformation matrix $\mathcal{U}$ which satisfies the following

$$|\nu_\alpha\rangle = \sum_i \mathcal{U}_{\alpha i} |\nu_i\rangle \tag{2.2}$$

### 2.2.2 Two Neutrino Oscillations: A Simple Picture

In an interaction or decay, a neutrino is produced in a flavor eigenstate. But the neutrino will propagate in its mass state. As it does, if we allow mass and mixing, the probability that the neutrino will be measured as the original flavor can be lower than unity. The simplest type of flavor change will occur in a vacuum and is called "neutrino oscillations." This thesis will focus on this type of flavor change. The presence of matter can also introduce a flavor potential that can distort the oscillatory behavior, but still lead to flavor conversion. Flavor change due to such "matter effects" has been observed, and we will briefly mention this below, but is not central to the story told in this thesis.

Neutrino oscillation involving three neutrinos is a complicated mathematical picture. Therefore, we often start by reducing the discussion to a simplified case in which only two neutrinos significantly mix. Call these flavor eigenstates $|\nu_\alpha\rangle$ and $|\nu_\beta\rangle$ and the mass eigenstates $|\nu_1\rangle$ and $|\nu_2\rangle$. In this case the mixing matrix becomes

$$\mathcal{U} = \begin{pmatrix} cos\theta & sin\theta \\ -sin\theta & cos\theta \end{pmatrix}. \tag{2.3}$$

Because the mass eigenstates are eigenstates of the free propagation hamiltonian, a flavor eigenstates evolves through space as in eq. 2.4

$$\begin{aligned}
|\Psi(t)\rangle &= e^{-i\mathcal{H}t} |\nu_\alpha\rangle \\
&= e^{-i\mathcal{H}t}(cos\theta |\nu_1\rangle + sin\theta |\nu_2\rangle) \\
&= e^{-it\sqrt{p_1^2+m_1^2}} cos\theta |\nu_1\rangle + e^{-it\sqrt{p_2^2+m_2^2}} sin\theta |\nu_2\rangle \\
& \quad p_i >> m_i \\
&\approx e^{-itE}\left(e^{\frac{-itm_1^2}{2E}} cos\theta |\nu_1\rangle + e^{\frac{-itm_2^2}{2E}} sin\theta |\nu_2\rangle\right).
\end{aligned} \tag{2.4}$$

In the ultra relativistic limit $t \approx L$, the distance traveled. The probability that a $|\nu_\alpha\rangle$ state is then observed as a $|\nu_\beta\rangle$ a distance $L$ away is obtained as

$$\begin{aligned}
P_{\alpha \to \beta}(L) &= |\langle\nu_\beta|\Psi t\rangle|^2 \\
&= sin^2(2\theta) sin^2\left(\frac{\Delta m^2 L}{4E}\right) \quad &\text{(Natural Units)} \\
&= sin^2(2\theta) sin^2\left(1.27\frac{\Delta m^2 L}{4E} \frac{eV^2 \cdot km}{GeV}\right) \quad &\text{(SI Units)}
\end{aligned} \tag{2.5}$$

This simplified model highlights several important general features about oscillations. The mixing angle shows up as a prefactor on the total probability. The behavior of the probability is oscillatory in $L/E$, making decisions about where to locate your experiment - commonly called the baseline - for a given energy very important. The periodicity of the probability depends on the difference of the square masses, $\Delta m^2 = |m_2^2 - m_1^2|$, explicitly demonstrating why the existence of oscillation mandates that there be some nonzero masses, else $P$ would be 0 everywhere.

### 2.2.3 Three Neutrino Oscillations: The Known Picture

Expanding the oscillation picture to more than two neutrinos is straightforward and proceeds in the same fashion as illustrated in the two neutrino case. The derivation is, however, significantly more cumbersome. While it is known that there are only three active flavors of neutrinos [8], we will shortly consider the possibility of sterile neutrinos which may oscillate alongside the active flavors, but do not participate in the weak interactions. It is therefore convenient to consider oscillations for an arbitrary number of neutrinos with mixing matrix $\mathcal{U}$. The probability that a $|\nu_\alpha\rangle$ state is observed as a $|\nu_\beta\rangle$ a distance L away is then found to be given by Eq. 2.6

$$\begin{aligned} P_{\alpha \to \beta} =&_{\alpha\beta} -4 \sum_{i>j} \mathcal{R}e\big(\mathcal{U}_{\alpha i}^* \mathcal{U}_{\beta i} \mathcal{U}_{\alpha j} \mathcal{U}_{\beta j}^*\big) sin^2\left(\frac{\Delta m_{ij}^2 L}{4E}\right) \\ &+ 2\sum_{i>j} \mathcal{I}m\big(\mathcal{U}_{\alpha i}^* \mathcal{U}_{\beta i} \mathcal{U}_{\alpha j} \mathcal{U}_{\beta j}^*\big) sin\left(\frac{\Delta m_{ij}^2 L}{2E}\right) \end{aligned} \quad (2.6)$$

As an example, Fig. 2-3 illustrates the oscillatory behavior in $L/E$ of the probability of observing each of the three varieties of active neutrino from an initially pure beam of $\nu_e$.

In a three neutrino picture, as in the two, $\Delta m^2$ controls the frequency of the oscillating probability. In the three neutrino picture, however, three different mass differences must be considered, $\Delta m_{21}^2 = m_2^2 - m_1^2$, $\Delta m_{32}^2 = m_3^2 - m_2^2$, and $\Delta m_{31}^2 = m_3^2 - m_1^2$. However, only two of these are actually independent because $\Delta m_{31}^2 = \Delta m_{21}^2 + \Delta m_{32}^2$. There are significant experimental challenges to directly measuring the neutrinos' minute mass, so the absolute scale of $m$ remains unknown, but oscillation measure-

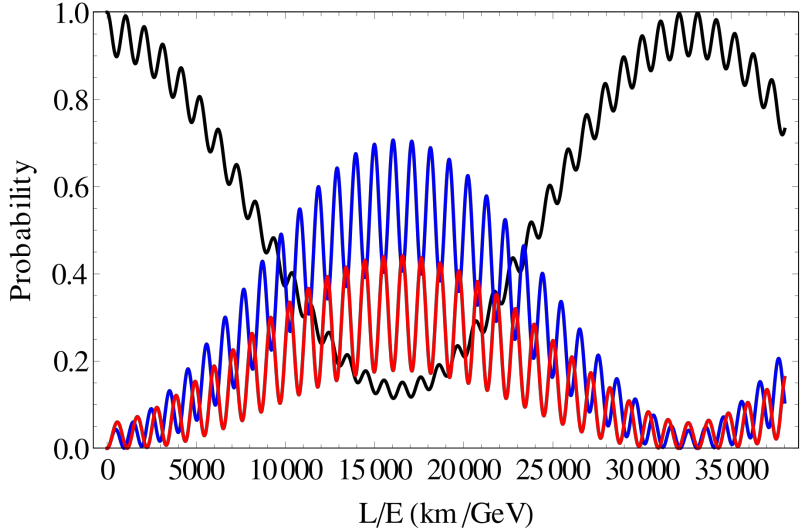

Figure 2-3: An example illustrating the probability of observing each flavor of neutrino as a function of L/E coming from an initial electron neutrino beam. Black is $\nu_e$, red is $\nu_\mu$, and blue is $\nu_\tau$.

ments have determined the mass splittings [22]:

$$\begin{aligned}\Delta_{21}^2 &= (7.53 \pm 0.18) \times 10^{-5} eV^2 \\ \Delta_{32}^2 &= (2.453 \pm 0.034) \times 10^{-3} eV^2.\end{aligned} \quad (2.7)$$

A combination of oscillation experiments and those that have exploited matter effects have determined the $3 \times 3$ matrix, known as the PNMS (after Pontecorvo, Maki, Nakagawa, and Sakata. This matrix can be thought of an analog to the CKM matrix for quark mixing. Notably different from the CKM matrix, however, the PMNS matrix has large off diagonal elements.

$$\mathcal{U} = \begin{pmatrix} \mathcal{U}_{e1} & \mathcal{U}_{e2} & \mathcal{U}_{e3} \\ \mathcal{U}_{\mu 1} & \mathcal{U}_{\mu 2} & \mathcal{U}_{\mu 3} \\ \mathcal{U}_{\tau 1} & \mathcal{U}_{\tau 2} & \mathcal{U}_{\tau 3} \end{pmatrix} \quad (2.8)$$

In principle a 3×3 unitary matrix such as this has nine free parameters, but in reality only four of these generate oscillation, while the others can be absorbed as phases. It is therefore common to conceptualize this matrix as comprising three unitary rotations which depend on three neutrino mixing angles, $\theta_{13}$, $\theta_{12}$, and $\theta_{23}$, as well as a CP violating phase, $\delta$. There is additionally a phase matrix which is the unit matrix if neutrinos are Dirac particles, but can contain two additional phases if neutrinos are Majorana. However, this phase matrix does not introduce oscillation in either the Dirac or Majorana case. For brevity, define $c_{ij} = cos\theta_{ij}$ and $s_{ij} = sin\theta_{ij}$.

$$\mathcal{U} = \begin{pmatrix} 1 & 0 & 0 \\ 0 & c_{23} & s_{23} \\ 0 & -s_{23} & c_{23} \end{pmatrix} \begin{pmatrix} c_{13} & 0 & e^{-i\delta}s_{13} \\ 0 & 1 & 0 \\ -e^{i\delta}s_{13} & 0 & c_{13} \end{pmatrix} \begin{pmatrix} c_{12} & s_{12} & 0 \\ -s_{12} & c_{12} & 0 \\ 0 & 0 & 1 \end{pmatrix} \begin{pmatrix} e^{\frac{i\alpha_1}{2}} & 0 & 0 \\ 0 & e^{\frac{i\alpha_2}{2}} & 0 \\ 0 & 0 & 1 \end{pmatrix} \quad (2.9)$$

These parameters are known from global fits with current best values [22]:

$$\begin{aligned} \sin^2(\theta_{12}) &= 0.307 \pm 0.013 \\ \sin^2(\theta_{23}) &= 0.545 \pm 0.021 \\ \sin^2(\theta_{13}) &= (2.18 \pm 0.07) \times 10^{-2} \\ \delta^o_{CP} &= 244 \pm 30. \end{aligned} \quad (2.10)$$

### 2.2.4 Four (or More) Neutrino Ocillation: New Physics?

While it is well established that only three flavors of neutrinos are active under the weak interaction, it remains possible that there exist additional neutrinos that can participate in oscillation but do not weakly interact. We term such a neutrino a *sterile neutrino*. The existence of one or more sterile neutrinos would mean that current oscillation experiments are sensitive to only a subset of all existing neutrinos. We have been here before. The earliest neutrino experiments were typically sensitive to only electron neutrinos, and thus observed deficits due to insensitivity to other flavors. Likewise, a sterile neutrino would be directly undetectable to any existing neutrino detection technology.

The presence of a sterile neutrino would, however, be indirectly detectable due to it's impact on oscillations in the active sector. This would arise through expansions of the PMNS matrix. If we posit the simplest model in which 1 sterile neutrino exists - a so called 3+1 model - then the PMNS matrix must be modified as shown in Eq. 2.11.

$$\mathcal{U} = \begin{pmatrix} U_{e1} & U_{e2} & U_{e3} \\ U_{\mu 1} & U_{\mu 2} & U_{\mu 3} \\ U_{\tau 1} & U_{\tau 2} & U_{\tau 3} \end{pmatrix} \rightarrow \mathcal{U}_{3+1} = \begin{pmatrix} U_{e1} & U_{e2} & U_{e3} & U_{e4} \\ U_{\mu 1} & U_{\mu 2} & U_{\mu 3} & U_{\mu 4} \\ U_{\tau 1} & U_{\tau 2} & U_{\tau 3} & U_{\tau 4} \\ U_{s1} & U_{s2} & U_{s3} & U_{s4} \end{pmatrix} \quad (2.11)$$

Motivation to include sterile neutrinos has arisen due to several anomalous observations that can be interpreted as oscillations with a similar $L/E \sim 1$ km/GeV, hence $\Delta m^2 \sim 1$ eV$^2$. Two examples of these experiments, which we will discuss in more detail in Chapter 3, are LSND and MiniBooNE [12] [13]. One cannot explain an additional large mass splitting with only three neutrinos, thus a fourth is introduced. Because of the LEP limits on the number of light active flavors, if a much larger mass state exists, it must be predominantly composed of the sterile flavor. We illustrate a model in Fig. 2-4 in which the anomalous results and the three neutrino results can be reconciled.

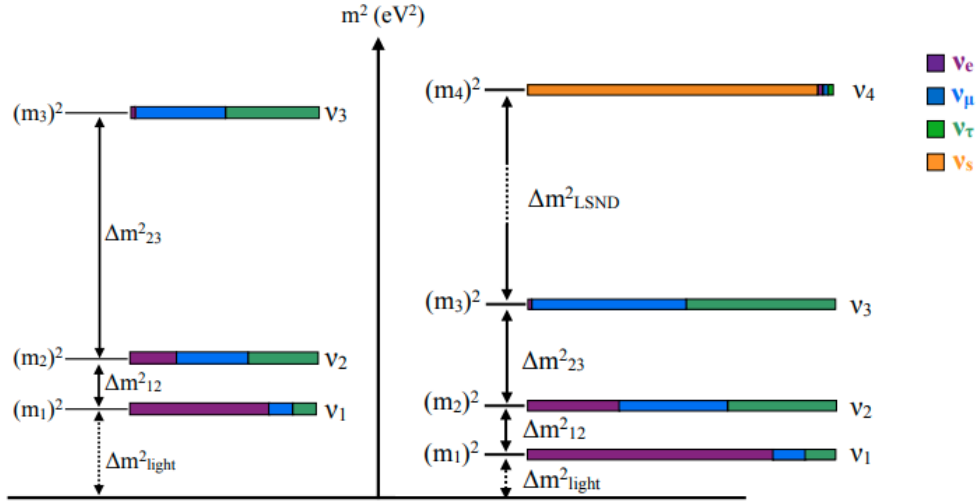

Figure 2-4: Graphical representation of the mass splittings between the active eigenstates and the relative flavor content of each mass state (left). The picture is modified if a fourth sterile neutrino is introduced with a much larger $\Delta m^2$ (right) [44]

An important point is that if the flavor splitting between the mostly sterile neutrino and the three light neutrinos is very large, then the mass difference between the three light neutrinos can be neglected. In that case, the four neutrino model reduces to an effective two-neutrino model.

## 2.3 This Thesis: An Agnostic Search for Flavor Change at Short Baseline

The above phenomenology is important to understanding the history that has led to proposal of the MicroBooNE experiment. The experiment, and the entire short-baseline program at Fermilab [18], was proposed based on the premise that the MiniBooNE anomaly is due to oscillations. Indeed, when physicists discuss MicroBooNE, they often refer to its primary goal as a "sterile neutrino oscillation search" that is described by by the concepts in Sec. 2.2.4.

Indeed, sterile neutrinos <u>could</u> be the explanation for the MiniBooNE anomaly. However, it is important to keep an open mind. The study presented here addresses sterile neutrino oscillation models, but does not depend upon them. Although flavor change through sterile oscillations is the most-discussed explanation for the MiniBooNE result, this model is not the only explanation, and, indeed, it does not fit the picture that well. As a result, beyond this point, we will leave behind the explicit discussion of sterile neutrino oscillations.

In this thesis, we will take an agnostic approach to the source of the signal, which can then be compared to oscillations or to any other model. We will assume that some phenomenon, which may or may not be oscillations, leads to a conversion of $\nu_\mu$ flavor to $\nu_e$ flavor, resulting in an observed excess in the MiniBooNE. We can then unfold the observed MiniBooNE result to get the a prediction for the flux arriving at an experiment located on the same beamline and almost the same location, MicroBooNE. We can use that flux to predict the rate of neutrino interactions in the detector. This allows us to predict the sensitivity to the signal with only one assumption: the signal is due to electron neutrinos appearing in the neutrino beam at a rate given by the MiniBooNE observation.

# Chapter 3

# The LSND and MiniBooNE Anomalies

As already introduced, a collection of short-baseline experiments observe flavor-change that can be explained with $\Delta m^2 \sim 1$ eV$^2$ [19]. Of these, the experiment with the highest significance is the MiniBooNE Experiment, where the excess is measured to be $4.7\sigma$ of electron neutrino interactions in a muon neutrino beam [13]. This experiment was motivated by an earlier experiment, LSND, which observed a similar $3.8\sigma$ excess [12]. The concept behind MiniBooNE was to assume that the LSND result was due to oscillations, hence the signal should have the same $\Delta m^2$ as the LSND result, as long as the same $L/E$ ratio was used in the design. The MiniBooNE design maintained that ratio, but with significantly different $L$ and $E$. The higher beam energy changed the signature of the neutrino interaction and the sources of systematic uncertainty. In other words: the MiniBooNE attack on LSND was to change everything except the assumption of oscillations.

An anomaly was observed in MiniBooNE, but the shape as a function of energy did not perfectly match the LSND-based oscillation prediction. This has led to confusion in interpreting both MiniBooNE and LSND results within an oscillation frame-work. Yet both have anomalies, and there are similarities between the results, so some new physics may be appearing in an unexpected way. Because of this, the MicroBooNE collaboration decided to mount another attack with a different strategy.

MicroBooNE kept many aspects of the MiniBooNE design the same, including utilizing the same beamline and sitting at almost the same $L$. But MicroBooNE makes use of a detector with a significantly different method of detecting neutrino interactions, thereby changing the backgrounds from MiniBooNE and also changing the sources of systematic uncertainty again.

## 3.1 The LSND Anomaly

The first observed anomaly came from a short baseline experiment at Los Alamos, the Liquid Scintillator Neutrino Detector (LSND). LNSD was an appearance experiment looking for $\bar{\nu}_e$ in a $\bar{\nu}_\mu$ beam. The beam was produced via an ∼800 MeV proton beam on a beam dump producing a $\pi^+$ dominated flux. The $\pi^-$ contribution is suppressed at the $10^{-4}$ level due to the capture in dump. Most of the $\pi^+$ come to rest before decaying, leading to the decay chain:

$$\begin{aligned}\pi^+ &\to \mu^+ + \nu_\mu \\ \mu^+ &\to e^+ + \nu_e + \bar{\nu}_\mu\end{aligned} \tag{3.1}$$

Thus this is an ideal source to search for $\bar{\nu}_\mu \to \bar{\nu}_e$ oscillations since there is only a very small component of "intrinsic" $\nu_e$ in the beam.

The detector itself was filled with 50,000 gallons of mineral oil and an organic scintillator as a target. The $\bar{\nu}_e$, which array up to 53 MeV of energy, could undergo a charged current interaction with a free proton from hydrogen in the oil, via

$$\bar{\nu}_e + p \to e^+ + n.$$

The free neutron could then capture on a proton via

$$n + p \to d + \gamma \text{ (2.2 MeV)}.$$

This produced a detection sigature comprising a fast positron signal followed by a delayed 2.2 MeV photon signal. The lightly-doped scintillator oil allowed observation of the Cherenkov ring from the positron as well as scintillation light from both the positron and photon as it Compton scattered.

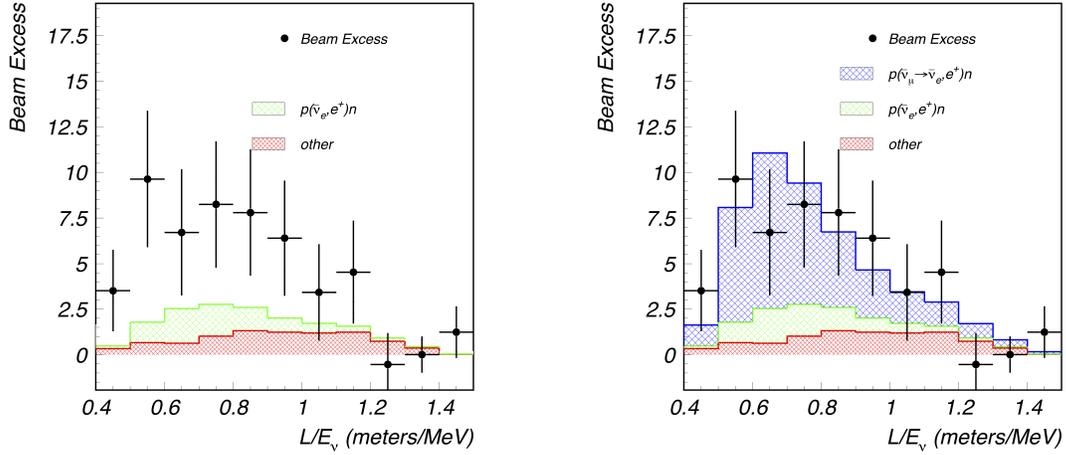

Figure 3-1: Observed $\bar{\nu}_e$ signal with beam backgrounds (Left). Observed $\bar{\nu}_e$ signal with 2 neutrino best fit. (Right)

The number of observed $\bar{\nu}_e$ interactions represents a $> 3\sigma$ excess over anticipated backgrounds. Given the baseline and typical beam energy, $\sim 30$ m and $\sim 40$ MeV, the excess $\bar{\nu}_e$ rate could be explained with a simple two-neutrino oscillation fit as shown in Fig. 3-1. This can occur within a four-neutrino oscillation model, if the $\Delta m^2$ of the mostly sterile state is more than an order of magnitude higher than currently favored values for $\Delta m^2_{12}$ and $\Delta m^2_{23}$, so that these can be effectively set to zero. This turns out to be the case for the LSND fits. This LSND signal was seen as potential evidence for oscillations involving a sterile neutrino.

## 3.2 MiniBooNE

### 3.2.1 The MiniBooNE Experiment

The Mini Booster Neutrino Experiment (MiniBooNE) was proposed to follow up on the observed LSND excess [20]. MiniBooNE is based at Fermilab and utilizes the Booster Neutrino Beam, which is discussed more fully in Chapter 4, as its source of neutrinos. It was designed to explore the same oscillation parameter space in $L/E$ as LSND. However both the baseline and neutrino energy were an order of magnitude higher than LSND with neutrino energies $\sim$ 700 MeV and baseline of $\sim$540m. MiniBooNE was also capable of running in either neutrino or antineutrino mode and was thus sensitive to both $\nu_e$ or $\bar{\nu}_e$ appearance in a relatively pure $\nu_\mu$ or $\bar{\nu}_\mu$ beam.

The detector itself - illustrated in Fig. 3-2 - comprises a spherical shell filled with 818 tons of mineral oil as the target. The mineral oil was not doped with scintillator. The shell is surrounded with 1280 inward facing photomultiplier tubes (PMTs) and 240 surrounding veto PMTs to reject cosmic ray muons [23].

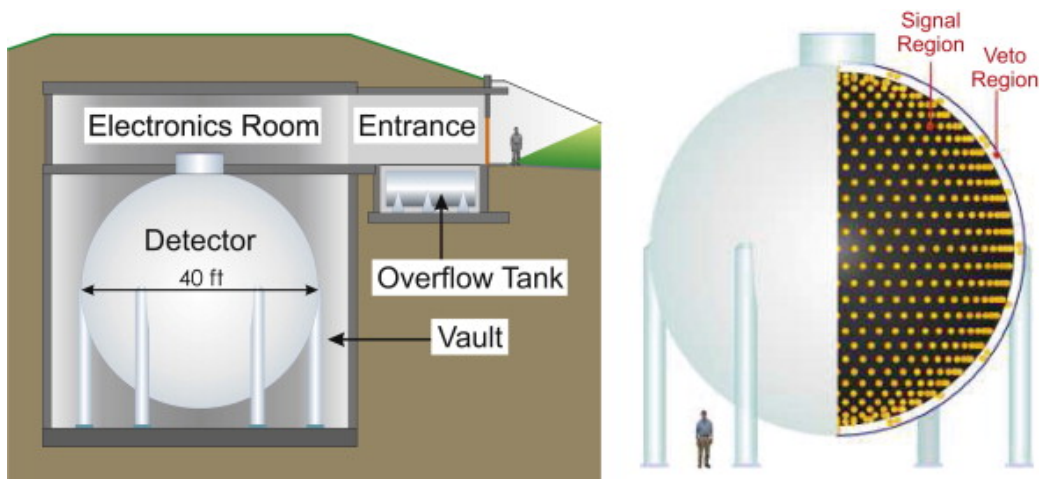

Figure 3-2: Illustration of the MiniBooNE detector hall underground (left). A cutaway illustrating the spherical configuration of the signal and veto PMTs (right).

A signal in MiniBooNE came from Cherenkov light produced by muons and electrons resulting from charged current interactions. Different final states can be differentiated by different patterns in the Cherenkov light. Muons produce a clean circular shape. Electrons, or a high energy photon that converts to an $e^+e^-$ pair, will produce less distinct rings. The rings from an electron and a photon are indistinguishable. Neutral pions will decay to two photons which will each produce rings. These are illustrated in Fig. 3-3. It is worth noting that with this type of detector an electron is indistinguishable from a photon. There is also a limitation imposed by the Cherenkov threshold that prevents protons from being visible except at very high energies. By reconstructing these rings, the energy of the interacting neutrino can be determined, up to energy that was lost below the Cherenkov threshold. Calibration of visible energy was performed using a sample of stopping muons which have a well known energy from their range. Michel electrons from muon decay and $\pi^0$ decay provided additional calibration points.

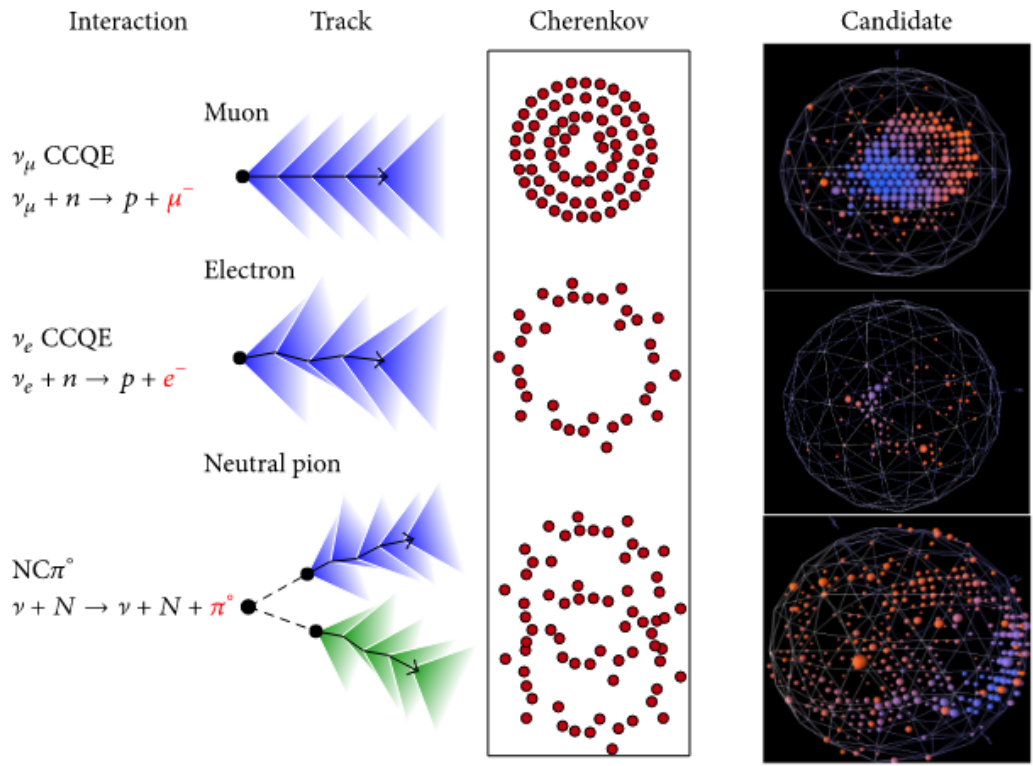

Figure 3-3: Illustrations of what the Cherenkov rings from different types of particles would appear like in the MiniBooNE experiment.

MiniBooNE's signal was charged-current quasi-elastic interactions (CCQE), which dominates in the energy range below $E_\nu \sim 1$ GeV. These interactions will be characterized by a single charged outgoing lepton. The most powerful technique to identify these events came from timing. Spills from the booster beam arrived in a well known pattern, or "spill" allowing backgrounds to be reduced by requiring beam-timing. A veto on optical activity in the outer volume of the shell further reduced backgrounds. Using features such as radiation topology, vertex position, and kinematic base cuts, a set of maximum likelihood algorithms was used to distinguish CCQE interactions from other backgrounds such as NC$\pi^0$ interactions. Assuming a CCQE interaction, the neutrino energy could then be reconstructed from the energy and angle relative to the neutrino beam of the outgoing lepton.

### 3.2.2 The MiniBooNE Excess

The results released by MiniBooNE over 17 years of running have consistently showed an excess at low energies [13]. This excess was apparent in both neutrino and antineutrino mode. The results are shown in Fig. 3-4. One immediately sees important differences between the MiniBooNE and LSND results: the MiniBooNE signal, which has very high statistics, sits on a very high background while the LSND signal, that had relatively low statistics, had minimal background. The backgrounds in Mini-BooNE fall into two categories: those due to the beam, which are shown in green colors, and those due to misidentification of particles in the detector, which are shown in the brown/red colors.

When interpreting this excess, it is important to note that a certain number of $\nu_e$ events were expected even in the absence of oscillation based appearance and these are the beam-related backgrounds. While the beam is dominated by $\nu_\mu$ arising from $\pi^+$ decay, a certain amount of $\nu_e$ are also produced via secondary decays in the beam, following the same chain as shown in Eq. 3.1, but, in this case, for decay-in-flight. Additional contamination via the direct decay $\pi^+ \to e^+ \nu_e$ occurs, but this is strongly suppressed relative to $\nu_\mu$ production. There are additional contributions to both $\nu_\mu$ and $\nu_e$ from Kaon decays, but these predominantly occupy the higher energies at several GeV. All together, these $\nu_e$ that are expected to be in the beam from Standard Model processes are termed *intrinsic* $\nu_e$ is predominantly at low energy and this leads to it being called the "low energy excess." As with LSND, the MiniBooNE excess can be interpreted within an oscillation framework. Examples of possible signals are overlaid on the data in this figure. One can see that the MiniBooNE signal is not perfectly described by oscillations. In particular, in the 200-300 MeV bin, there is an excess of events that is not well described by the example models. Using a two-neutrino model as a proxy for a four-neutrino model, the best fit parameter space is illustrated in Fig. 3-6 alongside results from LSND [24]. The $1\sigma$ best fit regions for the MiniBooNE signal are consistent with the 90% confidence regions from the LSND signal. LSND's and MiniBooNE's results are both in tension with $\nu_\mu$ disappearance data as seen in global fits. Thus, the oscillation explanation has important deficiencies.

Fig, 3-5 shows the MiniBooNE signals after background subtraction. The signal

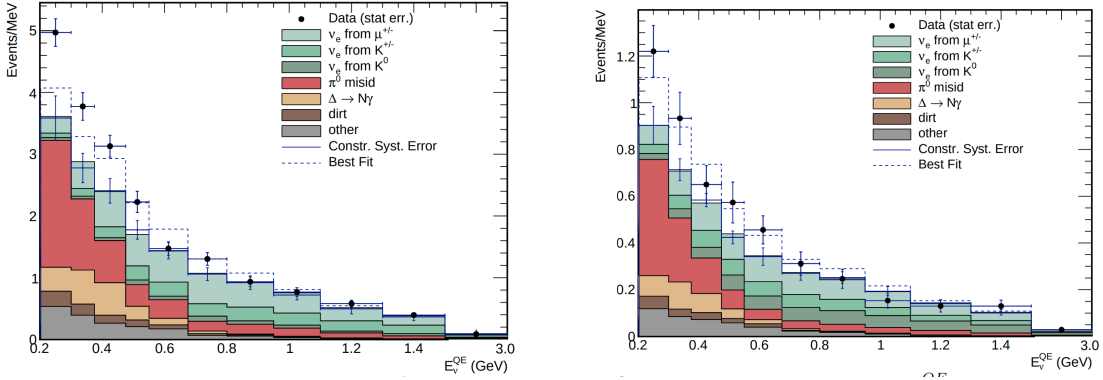

Figure 3-4: Observed events in neutrino mode (left) and antineutrino mode (right). The observed data is compared to a stacked prediction of all anticipated contributions [13].

Because MiniBooNE was unable to differentiate photons from electrons, the observed excess may be interpreted either as single electron or single photon events. This work will focus on the hypothesis that the observed excess arises from electron activity. Interpreting the MiniBooNE result as an excess appearance of electron neutrinos, then the signal will be visible in MicroBooNE. However, reconstruction, selection, and detector effects mean that the same signal will not appear identically in both detectors. Thus, when we say that MicroBooNE is examining a "MiniBooNE like excess" we actually mean that we are checking whether or not MicroBooNE sees results commensurate with MiniBooNE's once detector and reconstruction effects are deconvolved. To do this, we unfold the MiniBooNE signal, removing reconstruction and selection effects, to estimate the underlying event rates. This is ultimately described as an energy dependent reweighting of the intrinsic $CC\nu_e$ interaction rate.

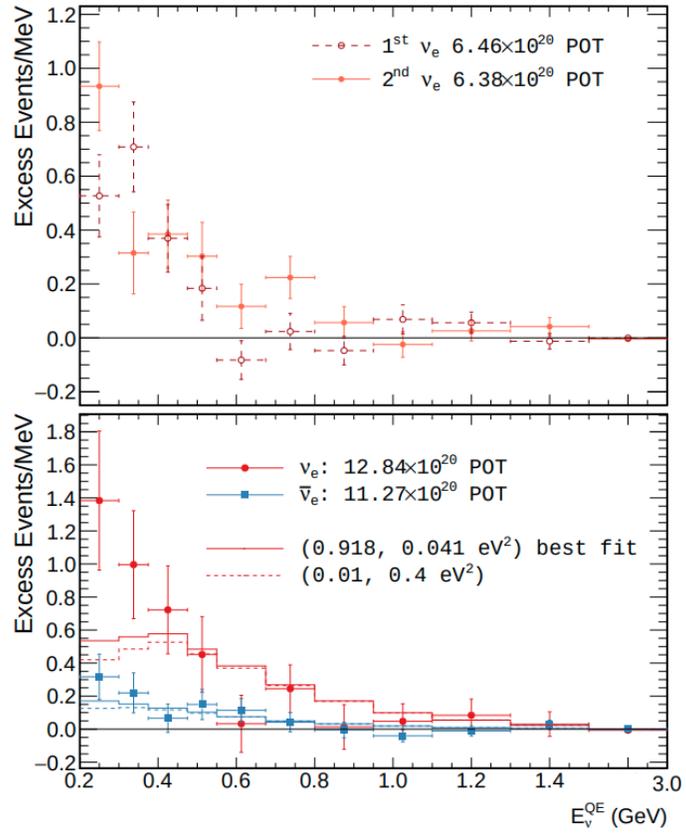

Figure 3-5: The MiniBooNE event excess in neutrino mode (top) vs $E_\nu$ with background subtracted. The bottom plot shows the background subtracted excess with antineutrino added in [13].

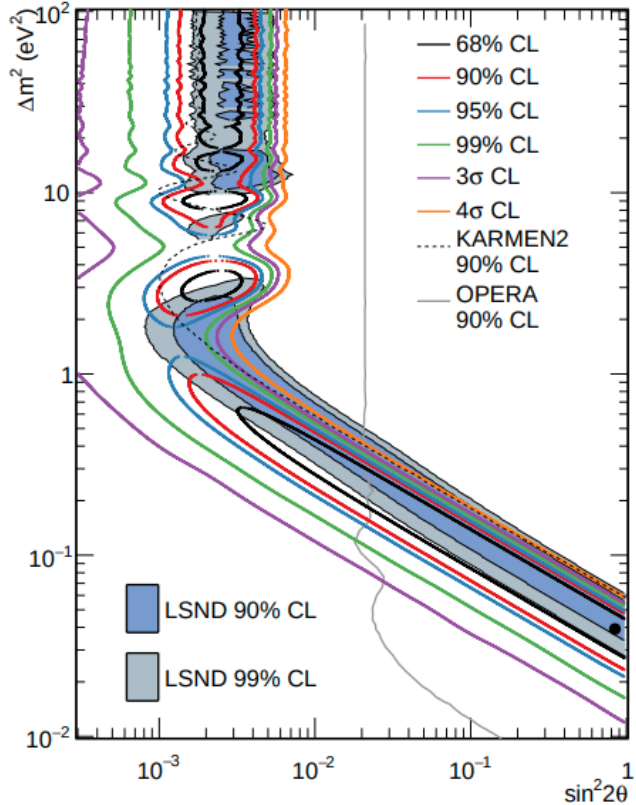

Figure 3-6: MiniBooNE allowed region in neutrino mode. Derived from $12.8 \times 10^{20}$ POT of data. The shaded regions illustrate allowed regions from the LSND $\bar{\nu}_\mu \to \bar{\nu}_e$ allowed regions. The black point at $(\Delta m^2, \sin^2 2\theta) = (0.039 \text{ eV}^2, 0.84)$ is the MiniBooNE best fit point [13].

## 3.3 Translating the MiniBooNE Excess for MicroBooNE

The goal of this analysis is an agnostic search for the MiniBooNE excess. In other words, we do not want to rely on an oscillation model. We want to use the MiniBooNE signal, itself, as the model. Specifically, the MiniBooNE signal derived from the 2018 analysis will be used [13]. Therefore, for this analysis, we need to translate the MiniBooNE Excess to MicroBooNE. Given a model for the physical origin of the signal seen by MiniBooNE, we can unfold the MiniBooNE data to produce a raw underlying event excess [26]. This is a necessarily model dependent enterprise as both detector effects and the assumption of it being electron like are folded in. The excess is parameterized in terms of reconstructed neutrino energy. Call the MiniBooNE observed spectrum $\mathcal{S}^{MB}(E_\nu)$, the excess $\mathcal{E}^{MB}(E_\nu)$, and the background $\mathcal{B}^{MB}(E_\nu)$. We then have

$$\mathcal{S}^{MB}(E_\nu^{reco}) = \mathcal{E}^{MB}(E_\nu^{reco}) + \mathcal{B}^{MB}(E_\nu^{reco}) \qquad (3.2)$$

We want to link this to a raw, underlying excess. This is the true excess which exists independent of MiniBooNE specific detection efficiencies. It is this raw excess which MicroBooNE could measure. Call this raw excess $\mathcal{R}^{MB}(E_\nu^{true})$, where we parameterize the raw excess in terms of the true neutrino energy. The observed excess will be linked to the raw excess via a detector response function which can be represented by the following matrix equation.

$$\mathcal{E}^{MB}(E_\nu^{reco}) = \mathcal{M}^{MB}(E_\nu^{reco}, E_\nu^{true})\mathcal{R}^{MB}(E_\nu^{true}) \qquad (3.3)$$

Here, $\mathcal{M}$ corresponds to the detector, reconstruction, and selection response which causes there to be some probability of a given $E_\nu^{true}$ yielding a particular $E_\nu^{reco}$. The desired fractional excess, $\mathcal{F}^{MB}$ parameterized in terms of $E_\nu^{true}$, is given by comparing the unfolded $\mathcal{R}^{MB}(E_\nu^{true})$ to the null spectrum, $\mathcal{H}_0^{MB}(E_\nu^{true})$, as in Eq. 3.4

Now that we have cast the raw excess as a fractional deviation from a central value prediction, the expected excess for MicroBooNE is given relative to it's central value prediction.

$$\mathcal{F}^{MB}(E_\nu^{true}) = \frac{\mathcal{R}^{MB}(E_\nu^{true})}{\mathcal{H}_0^{MB}(E_\nu^{true})} = \frac{\left(\mathcal{M}^{MB}\right)^{-1}(E_\nu^{reco}, E_\nu^{true})\mathcal{E}^{MB}(E_\nu^{reco})}{\mathcal{H}_0^{MB}(E_\nu^{true})} \qquad (3.4)$$

$$\mathcal{R}^{\mu B}(E_\nu^{true}) = \mathcal{F}^{MB}(E_\nu^{true})\mathcal{H}_0^{\mu B}(E_\nu^{true}) \qquad (3.5)$$

We will thus ultimately compare the two models, the null hypothesis will correspond to a $\nu_e$ spectrum derived from $\mathcal{H}_0^{\mu B}$ whereas the alternative hypothesis will contain a MiniBooNE like signal. Once the different detector, reconstruction, and selection effects of MicroBooNE are accounted for this amounts to a comparison of the reconstructed spectra corresponding to the two models in Eq. 3.7.

$$\mathcal{S}_{\text{null}}^{\mu B}(E_\nu^{reco}) = \mathcal{B}^{\mu B}(E_\nu^{reco}) \tag{3.6}$$

$$\mathcal{S}_{\text{MB signal}}^{\mu B}(E_\nu^{reco}) = \mathcal{B}^{\mu B}(E_\nu^{reco}) + \mathcal{M}^{\mu B}(E_\nu^{reco}, E_\nu^{true})\mathcal{F}^{MB}(E_\nu^{true})\mathcal{H}_0^{\mu B}(E_\nu^{true}) \tag{3.7}$$

Unfolding the MiniBooNE excess amounts to finding the inverse of $\mathcal{M}^{MB}$ in Eq. 3.4 which is done using singular value decomposition. The result is a set of weights that must be applied to the MicroBooNE intrinsic $\nu_e$ spectrum to produce the excess. These weights are illustrated in Fig. 3-7

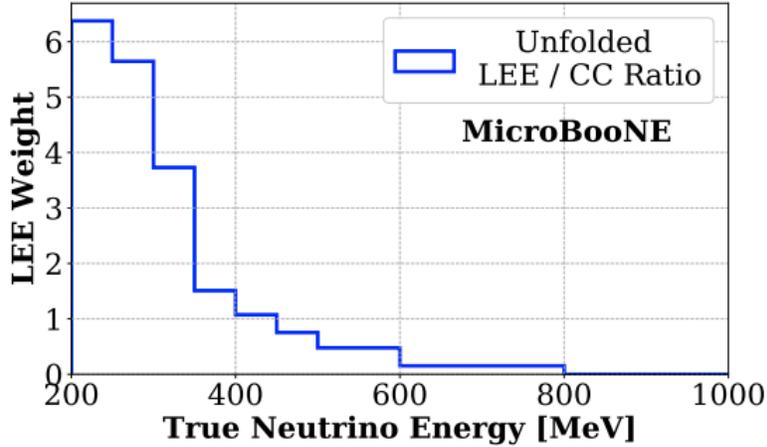

Figure 3-7: The ratio of unfolded MiniBooNE excess, under a $\nu_e$-like hypothesis, to the intrinsic $\nu_e$ rate in MicroBooNE as a function of true neutrino energy [44].

# Chapter 4

# The Shared Beamline: The Fermilab Booster Neutrino Beam

The concept behind the MicroBooNE design is that the detector is changed compared to MiniBooNE, but the $L$ and $E$, along with the $L/E$ ratio, are not. This chapter provides an overview of the Booster Neutrino Beam (BNB) located at Fermi National Laboratory (FNAL) that is the source of neutrinos for both MiniBooNE and MicroBooNE. The overall layout for the experiments with respect to this beamline is shown in Fig. 4-1.

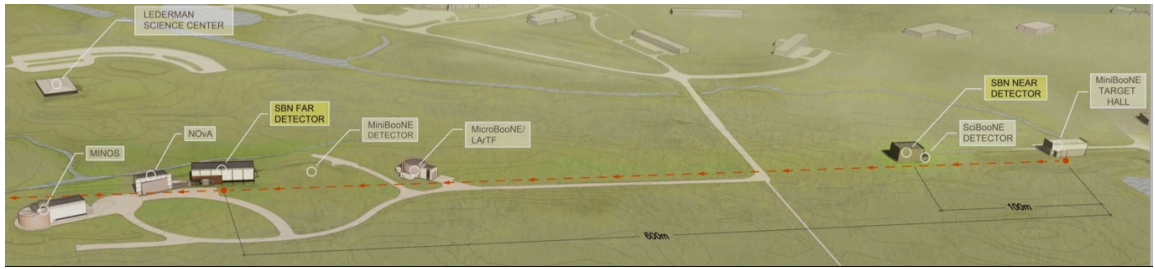

Figure 4-1: Overview of the experiments that lie on the Booster Neutrino beam at Fermilab as part of the Short Baseline Program. MicroBooNE lies just slightly upstream of MiniBooNE. To date, only MiniBooNE and MicroBooNE have taken data, and the other experiments are now finalizing construction and commissioning.

Protons of 8 GeV kinenetic energy impinge on a beryllium target and the resulting charged hadrons are focused down a decay pipe via a magnetic horn. The secondary hadron decays give rise to the neutrino beam which is primarily, though not exclusively, muon neutrinos.

## 4.1 The Proton Beam

Free protons are generated from $H_2$ gas which is ionized to $H^-$. These are then fed into a linear accelerator (LINAC) which initially raises the protons to 400 MeV via a set of radio frequency cavities. At 400 MeV, the protons are extracted from the LINAC and injected into a synchrotron which further accelerates them to 8 GeV. The protons are bunched into units 2 ns wide and spaced 1.6 $\mu$s apart. The number of protons is approximately 4-5 $\times$ $10^{12}$ per pulse. The number of protons directed onto the target (POT) is measured by a pair of toroids upstream of the target. The number of POT will ultimately serve as a readily measured proxy to normalize the expected number of neutrino interactions in the detector.

Once at 8 GeV, the protons are directed at a beryllium target. Beryllium is used because of it's high pion yield and low energy deposition per unit length which minimizes cooling concerns. The residual radioactivity is also comparatively low making it a safer long term choice.

When the proton beam hits the target, it primarily produces charged pions and kaons. The beryllium target is located inside of a large focusing magnet, called the horn. The horn is a pulsed electromagnet made from an aluminum alloy. The horn carries a maximum current of 170 kA with a pulse width of 143 $\mu$s coincident with the arrival of the proton beam on the target. This produces a field strength peaking at 1.5 T. This field focuses the charged mesons. The direction of the current permits a selection of positive or negative charge, ultimately producing either a primarily neutrino or antineutrino beam.

Beyond the horn lies a 50 meter long steel decay pipe which is filled with air and terminating in a beam stop made of steel and concrete. Beyond this stop the beam is essentially pure neutrino. The neutrinos then pass 470 m to the detector. An illustration of this process is given in Fig. 4-2.

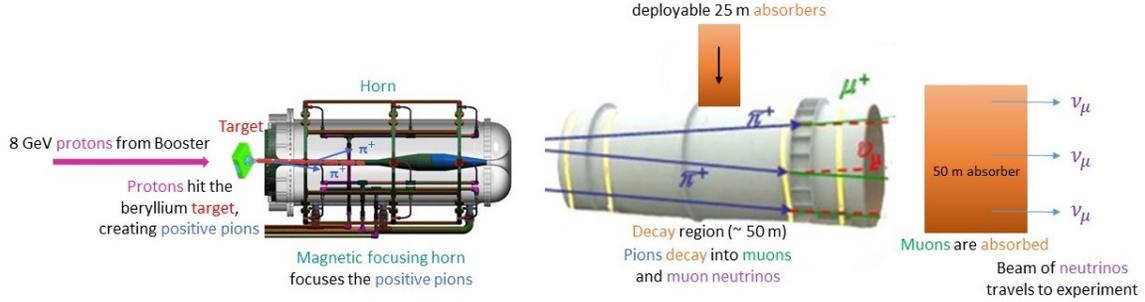

Figure 4-2: Illustration of the production of the neutrino beam used in this analysis. The 8 GeV protons impact a beryllium target producing charged mesons. The magnetic horn focuses the positive mesons down a decay pipe yielding a neutrino beam.

## 4.2 Beam Composition

The resulting neutrino beam is dominated by $\nu_\mu$ arising from the decay of $\pi^+ \to \mu^+ + \nu_\mu$. There is some contribution from $K^+ \to \mu^+ + \nu_\mu$ that populate higher energies. There is contamination of $\nu_e$ arising primarily from the secondary decay $\mu^+ \to e^+ + \nu_e + \bar{\nu}_\mu$. At this point it is important to distinguish between two types of $\nu_e$. Those that arise from decays in the beam are termed *intrinsic* $\nu_e$ and are an inherent part of the beam. These are different from $\nu_e$ which can arise from flavor transformation that my cause the MiniBooNE excess. These intrinsic $\nu_e$ are an irreducible background to a $\nu_\mu \to \nu_e$ oscillation measurement. The sources of the $\nu_\mu$ and $\nu_e$ fluxes at MiniBooNE are shown in Fig. 4-4

Simulation of MicroBooNE's neutrino flux uses the Booster Beam Monte Carlo developed by MiniBooNE. [28] This simulation includes a full simulation of the beam line, horn, and decay region. The parameters of the flux model are reused from MiniBooNE except where detector specific parameters must be adjusted. The resulting simulated flux, incident at MicroBooNE, is illustrated in Fig. 4-3. [29]

## 4.3 Uncertainties and Constraints on Intrinsic $\nu_e$

The *ab initio* uncertainties on the neutrino flux used in this analysis come from MiniBooNE studies. Uncertainties on the kaon contribution are constrained by mea-surements made by SciBooNE. Other uncertainties related to pion production are estimated by propagating errors via splines through HARP data and MiniBooNE data.

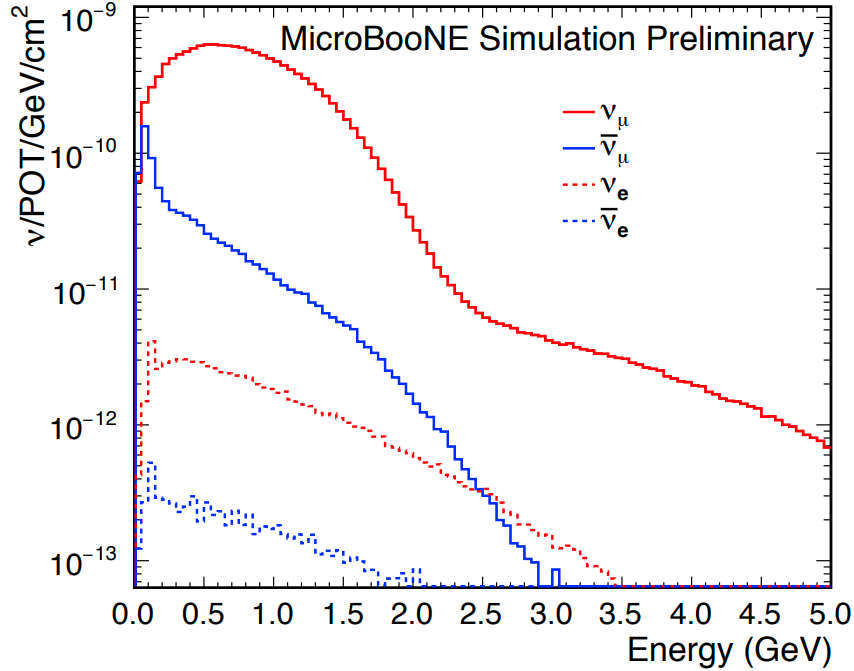

Figure 4-3: Booster Neutrino Beam simulated neutrino flux. Adapted from MiniBooNE monte carlo for the MicroBooNE detector [29].

An important aspect of the MiniBooNE and MicroBooNE analyses is that the *ab initio* uncertainties on the intrinsic $\nu_e$ flux can be constrained through measurement of the $\nu_\mu$ rate. This is because, at low energies, the $\nu_\mu$ and $\nu_e$ flux are connected through a common pion spectrum. The pion is the parent of the $\nu_\mu$: $\pi^+ \to \mu^+ + \nu_\mu$ and the grandparent of the $\nu_e$ that come from muon decay: $\mu^+ \to \bar{\nu}_\mu e^+ \nu_e$. The two-body nature of the pion decay to the $\nu_\mu$ leads to a strong correlation between the $\nu_\mu$ energy spectrum and the pion energy spectrum. Given the pion spectrum, it is straightforward to simulate the chain that leads to intrinsic $\nu_e$ . Because the beam produces a high rate of $\nu_\mu$ interactions, the rate of these can provide a strong constraint on the relative rate of intrinsic $\nu_e$. [13] [27]

The implementation of this constraint is discussed further in Chapter 8.

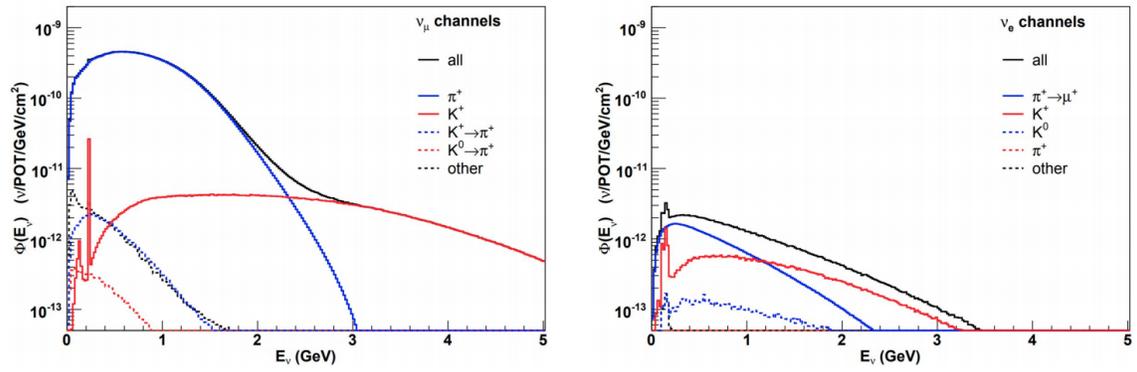

Figure 4-4: Breakdown of contributions to the $\nu_\mu$ and $\nu_e$ flux in the Booster Neutrino Beam by parent meson [29].

# Chapter 5

# The MicroBooNE Experiment

This chapter introduces the MicroBooNE experiment in more detail. MicroBooNE technology is completely different from MiniBooNE. MicroBooNE is a liquid argon time projection chamber (LArTPC). A time projection chamber uses wire planes to determine position in two views and drift time to determine the third component of the spatial position. MicroBooNE consists of a time projection chamber with three readout wire planes, along with an optical system of 32 PMTs, and an external cosmic ray tagging system. LArTPCs are capable of providing high resolution interaction "images" as well as calorimetry for particle identification and energy reconstruction. MicroBooNE has been collecting data since 2015 and has collected over $13 \times 10^{20}$ POT of data, of which $7 \times 10^{20}$ POT will be used in the analysis described in this thesis.

## 5.1 Introduction to the MicroBooNE Detector

Using liquid argon to build a TPC offers several uniquely powerful benefits. As a relatively massive nucleus, argon is well suited as a neutrino target. As a noble element, argon is an ideal medium to build large scale detectors because ionization electrons will not readily recombine with the bulk material. Argon is also a bright scintillator, discussed more in Sec 5.1.1, which is utilized to determine the interaction timing. Argon is also relatively cheap compared to heavier noble elements such as Xenon.

Detection and reconstruction of neutrinos from the Booster Beam, detailed in Section 4, is performed jointly by the TPC and the optical system. The optical system detects the prompt flash from a neutrino interaction and provides event timing. Final state particles produce tracks of ionization that are drifted by an electic field past two wire planes and collected on the third. The 3D reconstruction and calorimetry are then done by merging the optical timing information with the wire plane images. The result is a high resolution 3D image of charged particle tracks and showers.

### 5.1.1 The Optical System

The optical system in MicroBooNE collects scintillation light resulting from the excitation of argon. Argon scintillates brightly, emitting ∼50,000 photons per MeV deposited. Light is produced when an argon atom is ionized or excited, forms an $Ar_2^+$ dimer with a neighboring argon atom, and then breaks apart. $Ar_2^+$ can exist in either a singlet or triplet state, both of which decay to produce scintillation light at 128 nm. The two states do, however, have differing time constants with the singlet decaying promptly witih $\tau \approx 6$ ns (so-called "early light") and the triplet decaying with $\tau \approx 1600$ ns ("late light") [30]. These mechanisms are illustrated in Fig. 5-1. The fact that the scintillation comes from argon molecules, not individual atoms, is crucial because it means that bulk argon itself is transparent to this scintillation light wavelength. This is one of the key features that makes large LArTPCs feasible.

Collection of this light is crucial for two reasons. Firstly, the early scintillation light provides a much more accurate interaction time than simply using beam spill timing. Secondly, collecting scintillation light helps remove cosmic induced background in interaction images. This is necessary because the drift time is long enough, due to the 114 cm / ms drift speed, that a typical image will contain a non negligible amount of cosmic contamination. Using matching between optical patterns and wire image features enables the removal of many of these cosmics. This is detailed more in Section 6.4.

The collection of scintillation light is accomplished by a system of 32 optical units situated behind the anode wire plane. (In fact, during data-taking, one optical unit failed, and so some of the data sets makes use of 31 optical units). Each optical unit consists of an 8 inch Hamamatsu cryogenic PMT and an acrylic plate with a tetraphenyl butadiene (TPB) based coating. TPB is an organic wavelength shifter which absorbs the 128 nm scintillation light and re-emits at 425 nm. This is necessary because the PMTs are insensitive to 128 nm, but 425 is near their peak sensitivity as illustrated in Fig. 5-3 [32]. A diagram of an optical unit and an image of them installed inside the cryostat prior to filling is shown in Fig. 5-2 [21].

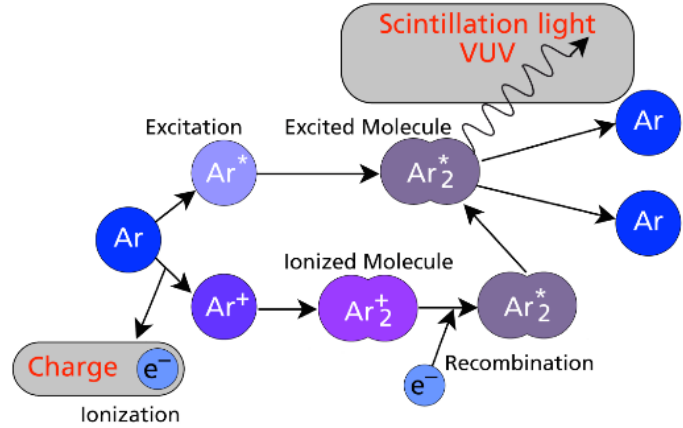

Figure 5-1: An excited or ionized argon atom can form a short lived argon dimer. The decay of this dimer yields scintillation light at 128 nm.

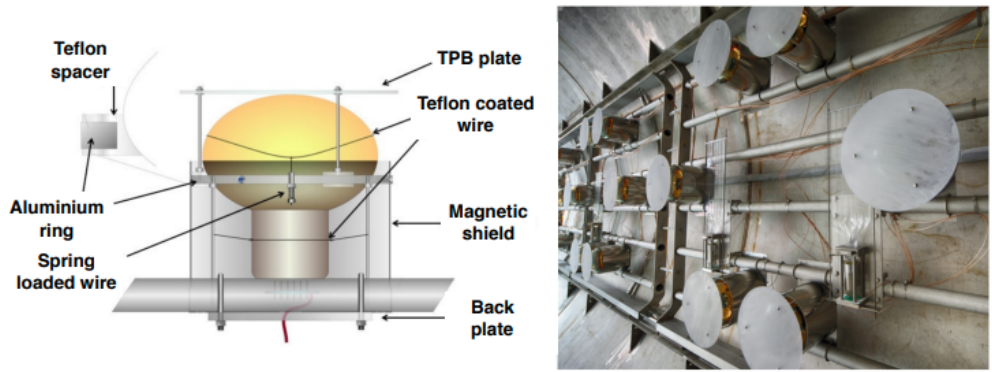

Figure 5-2: Schematic of a MicroBooNE optical unit (left). Photo from inside the cryostat of a set of optical units installed (right) [44].

We apply a data driven correction to optical data that allows a well calibrated energy scale from the PMTs. This produces better uniformity between PMTs and reduces differences between data and simulation. Gain tuning of the PMT system was a role played by the author of this thesis. The methodology is discussed in detail in Appendix B [32].

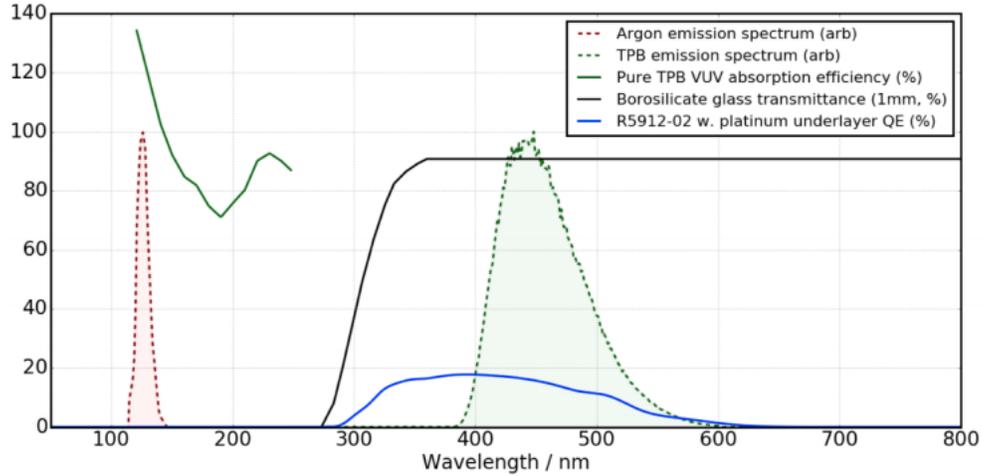

Figure 5-3: Liquid argon scintillation spectrum, TPB emission and absorbance spectrum, and PMT quantum efficiency vs wavelength [32].

### 5.1.2  The Time Projection Chamber

The MicroBooNE TPC is located within a cylindrical cryostat filled with 170 t total mass of liquid argon. The TPC active volume is formed by a rectangular solid enclosed in the tank with size 2.3m × 2.6m × 10.4m in $x,y,z$, corresponding to an active target mass of $\sim 85$ t. The coordinate system is defined with positive $x$ pointing against the drift direction, and with zero located at the anode plane. Positive $y$ is up with zero located at center. Positive $z$ along the beam axis with zero at the upstream face of the detector.

The TPC itself comprises three key elements: the wires, cathode plane, and field cage. These are illustrated in Fig. 5-5. The cathode is made of a steel sheet and is held at $-70$ kV. There are three sets of detection wires opposite the cathode plane, termed the U,V, and Y planes. These are made of gold plated copper. The Y plane is made from 3456 vertically oriented wires spaced at 3 mm. The U & V are made of 2400 wires oriented $60^o$ clockwise and counterclockwise respectively of vertical also with 3 mm spacing. The U & V planes are referred to as induction planes as they register the inductive signal caused by drifting electrons, but do not typically capture them. The Y plane is referred to as the collection plane, or alternatively as the anode plane, and is held at ground. This generates a uniform electric field of 273 V/cm within the active volume. The field cage is made of steel tubes which connect the anode and cathode via resistor chain, slowly stepping the voltage down.

The result of this setup is three different views of an interaction. The $x$ distance from the wires is known from the prompt optical flash combined with the known drift velocity of the electrons. From these three 2D views and the known initial position, full 3D reconstruction of the event is obtained. A schematic of this process is illustrated in Fig. 5-4.

As we discuss event reconstruction and analysis we will variously use cartesian or spherical coordinates where most natural. The (x,y,z) coordinates form a right handed system with the +z direction corresponding to the initial neutrino beam. This makes +y vertical and +x point from anode to cathode (opposite the electron drift). Where spherical coordinates are more natural, the polar angle ($\theta$) is measured relative to the neutrino beam, i.e. $\theta = 0$ aligns with the cartesian +z. The azimuthal angle ($\phi$) is conventionally defined from $-\pi$ to $\pi$ where $\phi = 0$ corresponds to the cartesian +x. These are illustrated in Fig. 5-6.

### 5.1.3 Readout and Triggering

It is not practical, either from a data acquisition or storage perspective, to read out data continuously, so once a neutrino interaction occurs, the detector must be triggered to readout the information. The signals from both PMTs and wires are initially analog, and so this readout must also be digitized and properly processed for analysis.

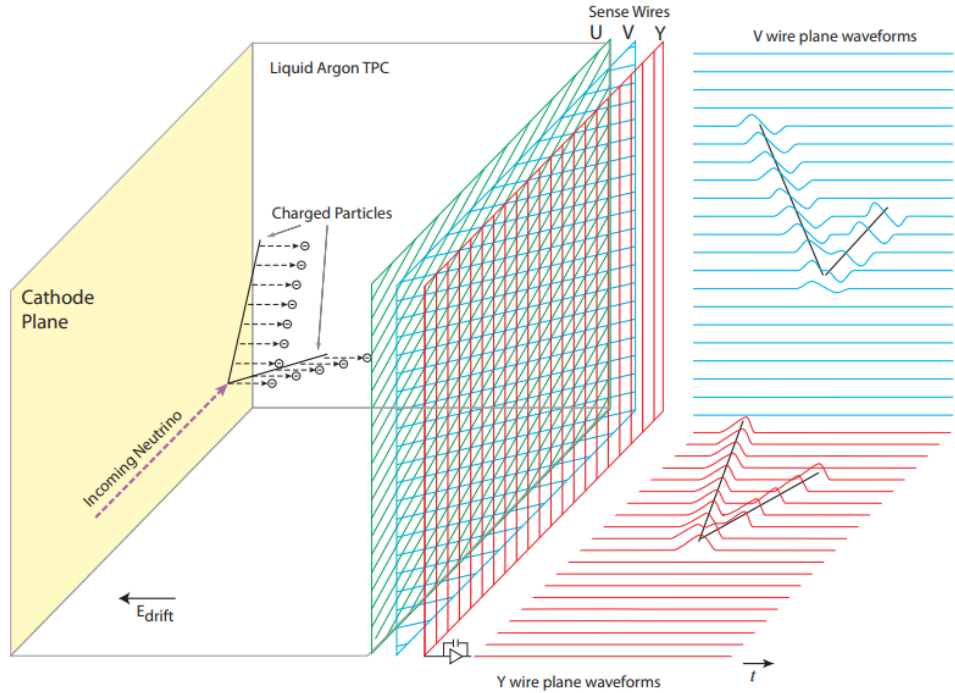

Figure 5-4: An illustration of how a LArTPC produces interaction images from optical and wire plane information. An incoming neutrino interacts with an argon nucleus producing final state charged particles. In this schematic, perhaps a proton and a muon. These charged particles traverse the argon producing ionization electrons and scintillation light. The optical system captures this light in $\mathcal{O}$(ns) and provides t=0 for the event. The electric field causes the ionization tracks to drift past the two induction wire planes and collect on the anode wire plane. The resulting 2D wire images are combined to produce a 3D interaction.

**Triggering**

Both TPC and PMT signals are read out once a trigger, or combination or triggers, is sent to the readout crates. There are several possible trigger signals that MicroBooNE uses. The booster beam provides a signal corresponding to a beam dump. This is known as the BNB trigger. The BNB trigger causes a 1.6 $\mu$s readout window called the beam gate window to be opened. Any neutrino interaction that does occur will be within this window, however most beam gate windows will still be empty. To prevent gathering empty windows, an additional trigger using the optical system is used. This software level trigger requires that 6.5 photoelectrons, summed across several PMTs, be observed within the 1.6 $\mu$s beam spill window. This additional requirement rejects

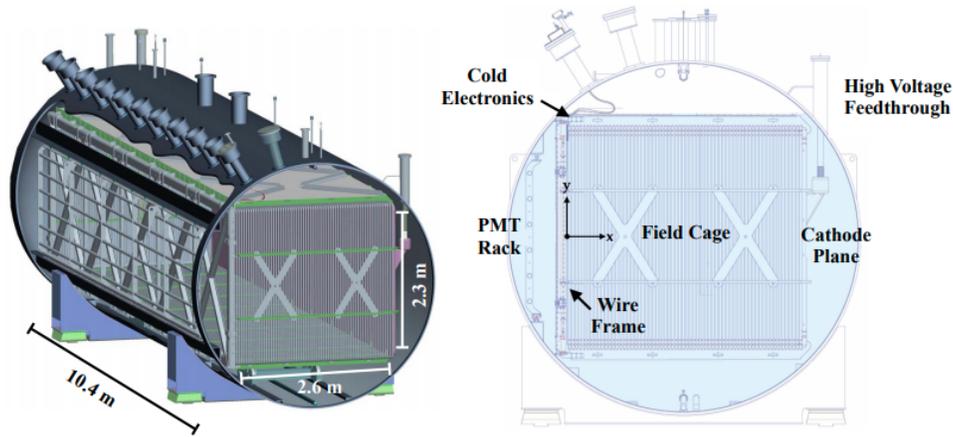

Figure 5-5: Schematic of the MicroBooNE cryostat and field cage. Ports at the top provide feedthrough for connection to the electronics (left). A cross section of the detector highlighting the key components. Viewed as if aligned with the neutrino beam and looking upstream. (right)

97% of empty beam spills.

Other triggers are used as well to help study cosmic activity. External BNB events (EXTBNB) are readout when there is sufficient optical activity to pass the optical software trigger, but outside of the time window when a beam event is present. This provides a set events representing cosmic activity which would trigger a readout and thus represents potential background. Unbiased triggers may also be taken which are manually triggered readouts at a certain rate.

Once a readout trigger has been issued, the full set of data is transferred to data acquisition (DAQ) machines for final packaging into specialty format. The data is then transferred to tape for storage.

**PMT Readout**

PMT signals first reshaped into a positive unipolar signal with 60 ns shaping time. The analog signal is digitized at 64 MHz. The beam readout is always produced in time with the beam spill and is 23.4 $\mu$s in duration. The information is transmitted via optical cabling from a transmit board to the DAQ. In order to avoid saturation and retain sensitivity both large and small optical signals, two copies of all PMT activity is read out. One has higher gain and one lower, with the high gain channel 10× the gain of the low. In situations where saturation in the high gain occurs the low is instead referenced.

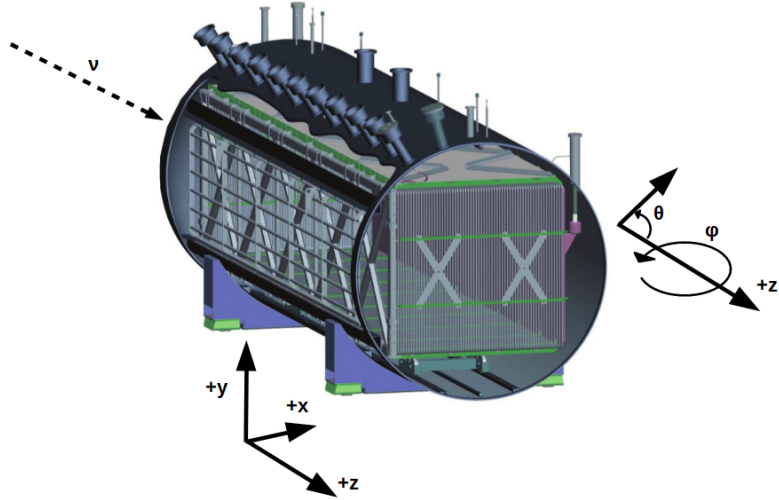

Figure 5-6: Illustration of how the cartesian and spherical coordinates used in this analysis are defined relative to the detector and neutrino beam.

**TPC Readout**

Readout and digitization of the wire information is done using a mixture of cold and warm electronics which, as the names suggest, refer to components which are placed inside or outside the cryostat. The first stage of the wire readout uses cold ASICs which attached to a dedicated motherboard. These ASICs perform signal amplification, applying a gain of 14 mV / fC, and signal shaping with a characteristic time of 2 $\mu$s. Cold cables then transmit the amplified signal through a set of feedthroughs on the top of the cryostat where they connect to warm electronics. The digitization of the signal is handled by the warm electronics. The TPC signals are sampled at 16 MHz sampling with 12 bit resolution. These digitized waveforms are then recorded in 1.6 ms long windows. 1.6 ms is chosen so as to be long enough to let electrons drift from anywhere in the detector past the wires. [33] [44].

| Property or Characteristic | MiniBooNE | MicroBooNE |
|---|---|---|
| Average $L$ | 540 m | 470 m |
| Average $E$ | 700 MeV | 700 MeV |
| Fiducial tonnage | 450 tonnes | 70 tonnes |
| Target | $CH_2$ | Argon |
| Readout method | Light observed by PMTs | Charge on wire planes |
| Detector stability | PMT response stable with time | Suffers from noise and dead wires |
| Cosmic Veto | Full coverage | Partial, added late in run |
| Proton reconstruction | None | Good capability for $> 25$ MeV |
| Cosmic Background | Negligible | Dominant initial background |
| Intrinsic $\nu_e$ Background | Major component | Primary final background |
| $\pi^0$ Mis-ID Background | Major component | Small |
| Other photon background | Minor component | Negligible |

Table 5.1: Similarities and differences for MiniBooNE vs. MicroBooNE, at a glance.

## 5.2 Pros and Cons of a LArTPC-based Design

There are pros and cons to switching from a Cherenkov-detector-based Design in MiniBooNE to the LAr-TPC-based design in MicroBooNE. We summarize some of these points in Table 5.1. In the following subsections, we highlight one "pro" and one "con" that are especially relevant to this thesis.

### 5.2.1 Pro: Image-like quality to ask: $e$ or $\gamma$?

As evident in Fig. 3-4, one of the major backgrounds that MiniBooNE suffered from was $\pi^0$ misidentification. Under certain circumstances, a final state $\pi^0$ can lead to a signal that looks like a CC$\nu_e$ interaction in a cherenkov detector. These troublesome cases can happen in several ways. All of these begin with an NC interaction which leads to a $\pi^0$. The $\pi^0$ will immediately decay via $\pi^0 \to 2\gamma$. The gammas themselves are not visible but once they pair produce they will yield two collinear electrons which produce a single ring, making the photon ultimately indistinguishable from an electron. Most of these $\pi^0$ will produce two rings, as illustrated in Fig. 3-3.

However, the second ring does not necessarily always appear. These 1-ring $\pi^0$s are the backgrounds which may be misidentified as an electron. These come from several sources. The conversion of the photon to an $e^+e^-$ pair is stochastic and may not occur before the photon leaves the detector. This situation accounts for approximately half of the $\pi^0$-misid background. The next largest background occurs when the $\pi^0$ decay is highly boosted with the decay axis along the boost. This leads to only one photon of sufficient energy to produce a distinct ring. Coherent $\pi^0$ production, which is typically very forward, produce 2 photons which merge and reconstruct as a single ring. Because the $\pi^0$ decay is virtually instantaneous, it is also possible for nuclear photoabsorption to prevent one of the photons from exiting the nucleus.

All of these scenarios will produce a signal which is topologically identical in a Cherenkov detector to a CC$\nu_e$ interaction. All except nuclear photoabsoption have non-background signatures that can be measured *in situ*, allowing very accurate simulation to constrain the background rate. Photoabsorption, however, relies on cross section models which contain more uncertainty and could not be directly measured.

Differentiating between $\pi^0$ and genuine CC$\nu_e$ interactions fundamentally requires a different technology which can reconstruct the event more fully and in more detail. LArTPCs offer such capabilities. A LArTPC offers photographic like event reconstruction which is capable of resolving details of the interaction. The LArTPC will reconstruct all of the particles in the interaction, unlike the Cherenkov detector, which is blind to protons. In a CC$\nu_e$ ineraction, the proton and the scattered electron will connnect at the vertex. In the case of the $\pi^0$ there is only a very small probability that there will be a connection between the photon and any other particle in the event. Ability to observe the proton, and resolve what is happening at the interaction vertex, is crucial to reducing the $\pi^0$ background to a minimal level.

In MicroBooNE, the wire spacing is 3 mm, and though the time information is finer-grained, we can group this information such that, when combined with the drift velocity, we are looking at the collected charge in 3 mm units. Thus for each plane of the TPC we can form images with 3 mm $\times$ 3 mm bins, where the bin-entry is the ADC charge. Each bin can be thought of as a pixel in an image, where the color of the pixel relates to the ADC charge in that pixel. Fig. 5-7 illustrates the power of these images, showing what neutrino interactions may look like in MicroBooNE with an electron and with a $\pi^0$. In images such as these, the difference between an electron and photon is evident because of the gap between the interaction vertex and the start of the electromagnetic shower. Because the pixel size is 3mm on a side, a very early conversion is required for a photon to appear connected to a proton with no pixel-gap, which has a $< 2\%$ probability of occuring.

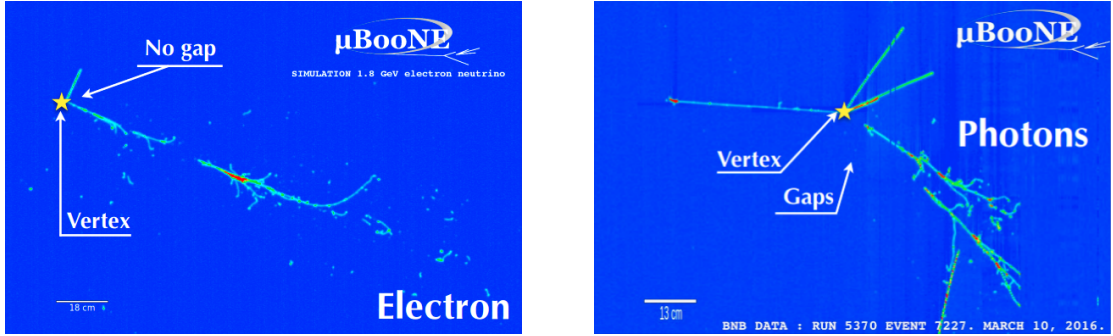

Figure 5-7: Simulated CC$\nu_e$ interaction (left) producing a proton and electron in the final state. There is no visible gap at the interaction vertex because the electron initiates an immediate electromagnetic shower. A CC$\nu_\mu$ interaction (right) identified in data which contains several proton tracks and a $\pi^0 \to 2\gamma$ in the final state. Both photons are visible and converted after some distance, producing a visible gap between the interaction vertex and the electromagnetic shower.

It is possible for photons from the $\pi^0$ to convert quickly enough that there is no visible gap, however in these cases the ionization density, as represented by the charge in the pixel, can still be used to distinguish a photon from an electron. An electron will be nearly minimum ionizing losing about 2.1 MeV/cm. A photon, however, pair produces leading to a higher ionization density. Identifying gaps and differences in $dE/dx$ will therefore give MicroBooNE enormously larger e/$\gamma$ separation power relative to MiniBooNE. Other analyses make direct use of the $dE/dx$ predicted versus observed in events in their cuts. As will be seen later, in this analysis, we make more general use of ionization topologies, relying more strongly on the no-gap requirement and ability to identify separated pixels that represent the second photon from a $\pi^0$ decay. Nevertheless, the technique of detailed $dE/dx$ comparison to prediction should be noted as a "pro" for LArTPCs.

## 5.2.2 Con: TPC Noise and Dead Wires

LArTPCs are state-of-the-art detectors. MicroBooNE was the first large-scale LArTPC constructed in the U.S. This is also a single-phase detectors that operates without signal-gain. This is unlike gas-based TPCs or two-phase LArTPCs that use a high voltage to accelerate electrons in a gasseous region proeducing an avalanche. This is also different from PMT based systems, like Cherenkov detectors, where a single photon produces signals of 10,000 electrons or more, depending on the gain. Thus, the MicroBooNE TPC must be sensitive to very low signals. Not surprisingly, because of this, the MicroBooNE detector has suffered from sources of noise. It has also suffered from dead wires, which will be apparent as blank regions in the images for the analysis. These effects needed to be addressed in this analysis, and were addressed successfully. The fact that we can address these problems shows that it will be possible to scale up a MicroBooNE-like detector to a very large size, such as is proposed for the DUNE Experiment.

**TPC Noise**

Initial noise filtering is needed before the waveforms can be used to produce a clean image. Several sources of noise were observed in MicroBooNE beyond what had been initially expected [33]. In order of significance, these are:

1. Noise from low voltage regulators. This is coherent noise that shows up on all wires which share a given regulator. This is mitigated by an offline filter which produces a correction waveform on a sample by sample basis and applied to all wires impacted by the regulator. This waveform is then subtracted.

2. Noise from the cathode high voltage power supply. This occurs due to a ripple in the HV power supply. It appears at odd harmonics of 36 kHz with the majority of the power near 36 and 108. An offline filter removes this noise by filtering out these harmonics in the frequency domain.

3. Burst Noise: This noise occurs in 900 kHz bursts. The exact origin of this has

not been certainly determined, but no specific mitigation is needed as the noise is typically attenuated to acceptable levels within the cold electronics.

The impact of this noise, and the success in removing it, are illustrated in Fig. 5-8. Updates to the electronsics signficantly improved the level of noise in the detector after Run 1. As a result, this analysis will be divided into 2 periods: Run 1 and post-Run 1, where the noise issues lead us to treat the data samples differently.

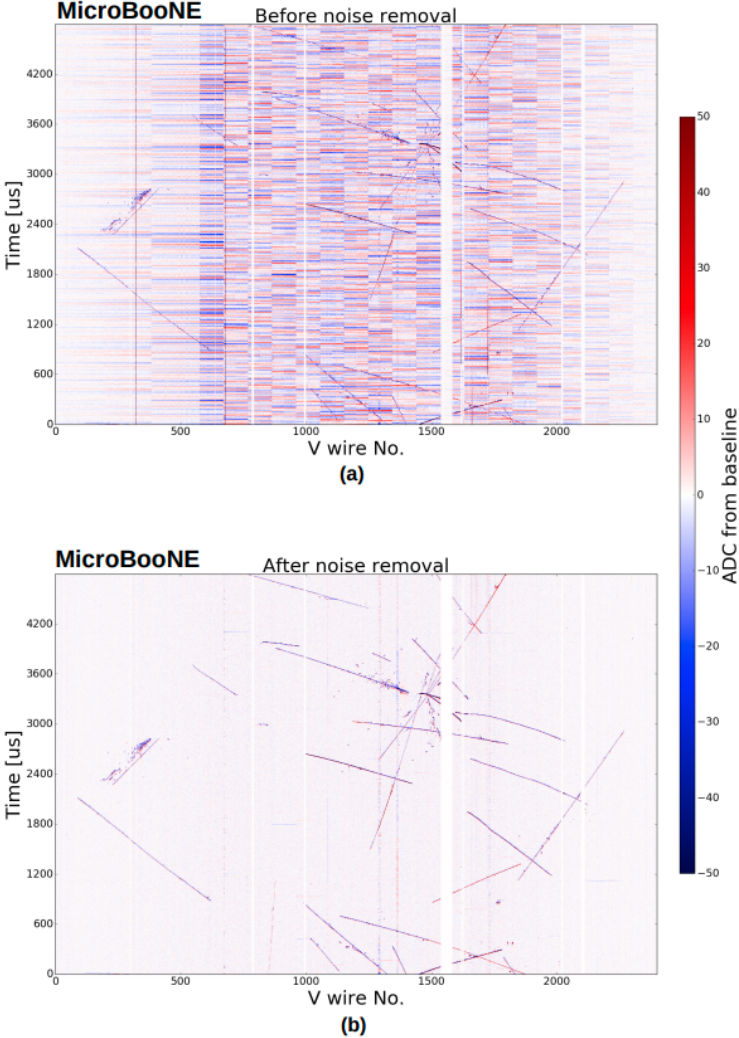

Figure 5-8: A 2D event display showing a raw signal before (a) and after (b) the offline noise filtering [33].

Information on noise is incorporated into the simulation through the use of "overlay" discussed below.

**Dead Wires**

During commissioning of the detector, there was some evidence that several V-plane wires had loosened [33]. Visual inspections were performed and no issue was identified, however later data suggested that one of the V wires was connected to certain U and Y wires. These connected wires affected several of the ASICs and produced channels that either produce no signal, the so called *dead wires*, or which have very high noise. A complete study was subsequently done with a signal pulser to identify live, dead, and noisy channels. Noisy channels were flagged via a threshold on the RMS of raw ADC signals as compared to the wire plane as a whole.

In total, it was determined that 862 channels are fully unusable: 433 on the U plane, 98 on the V plane, and 331 on the Y plane. While the positions and behaviors of these dead wires are known, they nonetheless have the potential to introduce reconstruction and selection difficulties. e.g. an event which crosses dead wires will appear to have ionization charge missing which can lead to underreconstruction of energy or failure to reconstruct ionization tracks if there are large gaps. An illustration of the regions which contain at least one dead wire is given in Fig. 5-9.

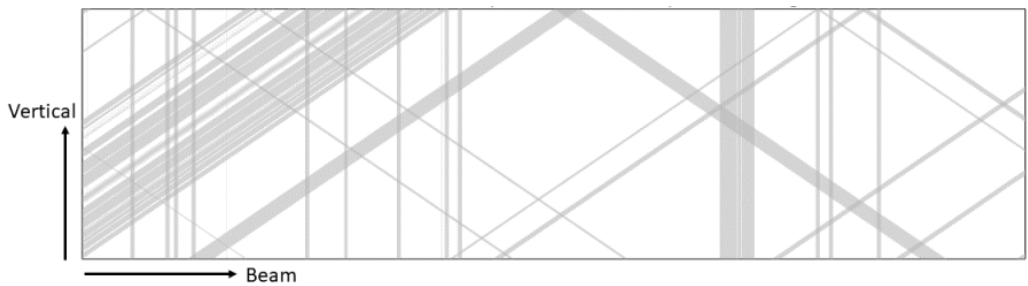

Figure 5-9: A side view of the detector (+z direction to the right). Grey regions indicate that at least one dead wire passes through that point [46].

Information on dead wires is incorporated into the simulation through the use of "overlay" discussed below.

## 5.3 Data Streams

MicroBooNE acquires data using several parallel readout streams. These are intended to acquire data under different triggering conditions for different purposes. In this section we introduce some explicit terminology for the data streams used in this analysis.

1. **BNB:** This is the primary neutrino data collection stream. In this analysis, a BNB sample or BNB event refers to an event which was taken in coincidence with a Booster Beam dump and which produced sufficient coincident optical activity to pass the software trigger. For this analysis, samples from the BNB stream are the only ones which have any significant probability of containing a neutrino interaction. It is not, however, guaranteed that a neutrino is present. BNB data has been split across five runs. Three are available for this analysis and correspond to a total POT of $6.9 \times 10^{20}$. We provide some key points about each run that is used in this analysis.

    (a) **Run 1** Feb-Sept 2016. POT = $1.7 \times 10^{20}$. Light yield was stable. PMTs suffered random exposure to pulses of UV light from an unshielded purity monitor. This produced ringing and PMT images on wires. Wires suffered noise from service board problems.

    (b) **Run 2** Oct 2016 - Oct 2017. POT = $2.6 \times 10^{20}$. Prior to this run, new service boards were installed significantly reducing wire noise. Light yield began to decline, although there is no evidence that this analysis is impacted.

    (c) **Run 3** Oct 2017 - Sept 2018. POT = $2.4 \times 10^{20}$. Light yield steadied at about half the level of Run 1. Otherwise no known differences from Run 2.

2. **EXTBNB:** This sample provides a study of cosmic ray behavior which is crucial for a data driven assessment of cosmic background rejection. EXTBNB interactions are external to the BNB readout window, but still require optical activity which would pass the software trigger. Being outside the beam spill guarantees that no neutrino is present.

3. **Unbiased BNB:** This sample provides a set of cosmic images which may or may not have produced an optical software trigger. These are, as the name suggests, read out at a predetermined rate without regard to the presence of a beam spill or optical signal. These images are critical the production of *overlay simulations*. As we discuss in the following section, MicroBooNE uses monte-carlo simulation to study and develop analysis on neutrino interactions. But real neutrino interactions will contain cosmic background activity. So after simulating the neutrino interaction, Unbiased BNB images are overlaid to produce realistic cosmic background, as discussed in the next section [40].

## 5.4 Simulations

This analysis makes extensive use of interaction and detector simulations to develop and, via data comparisons, validate all tools. The interactions on argon nuclei are modeled using GENIE [34] with tuned parameters intended to produce the most reliable low energy comparison possible. This latter point is particularly important. The low energy excess MicroBooNE seeks to explain must be understood relative to our current best understanding of low energy neutrino interactions. MicroBooNE relies on a good estimated central value prediction with which to spot an excess. As a result, the simulations adopted by MicroBooNE and discussed here have made particular effort to focus on low energy modeling [36], and on the modeling of $\nu_e$ vs $\nu_\mu$ interactions at low energies so that the systematics can be well constrained. In particular new models for nuclear effects, quasi elastic scattering, and meson-exchange interactions have been adopted which are more suitable for the low energy regime.

Below we describe the neutrino simulations. But before embarking on this, we describe what aspects we do not simulate, but instead extract from data and add directly to the simulations.

### 5.4.1 Overlay: Data Added to Simulated Events

There are three aspects of simulation that are particularly difficult to accurately represent: 1) Cosmic Rays, 2) Noisy Wires and 3) Dead Wires. While there are codes to simulate Cosmic Ray fluxes, including CORSIKA [35], we have found that the details depends on ability to simulate details of the surrounding hall to the level that we could not achieve satisfactory data-to-MC agreement. The experiment has found that rather than model all three of these effects, it is best to actually use data. We can use Unbiased BNB data to collect information on cosmic rays, noise and dead wires, and overlay this onto the simulated neutrino events.

A very large number of Unbiased BNB events are needed for the overlay. Due to limitations from livetime, the experiment collected most of the overlay events during long-shutdowns. This will result in slight differences between dead wire and noisy wire locations in the data and the simulation, that will lead to the data-to-Monte Carlo comparisons of the vertex distributions to have slightly lower $p$-values than might be expected, as will be shown below. However, there is no evidence that this has translated into an issue within the analysis.

### 5.4.2 GENIE Model and Parameter Tuning

The first step in the neutrino simulation is to describe the interaction of the neutrino with the argon. MicroBooNE makes use of the GENIE event generator package. Discussion of neutrino interactions within GENIE uses the following shorthand tha will appear later in this thesis:

- CCQE – Charge current elastic scattering events, as have already been introduced;

- MEC – Meson Exchange current events;

- NC$\pi$ and CC$\pi$ – within GENIE, this refers to resonance events producing the lowest mass $\Delta$ baryon resonance;

- DIS – this refers to all events from pions that are not coherent pion production nor resonant pion production from the first $\Delta$ resonance.

While not important to this thesis, GENIE contains many other interaction channels, including neutral current elastic scattering, coherent pion production, kaon-production channels and others. MicroBooNE is the first neutrino experiment to move to GENIE v3 for its "central value" simulation, specifically v3.0.6 G18 10a 02 11a. This version contains models more appropriate to lower energy regions including the Valencia model [37] [38] for the local Fermi gas momentum distribution as well as CCQE and CCMEC interactions. It also contains improved descriptions of nuclear final state interactions and treatment of pions. Fits to MiniBooNE data indicate a significantly better agreement with GENIE v3 vs v2 in the CC0$\pi$ channel.

The most up-to-date GENIE version adopted by MicroBooNE contains physics which is believed to be more correct than that in prior versions. However while the models employed fundamentally treat the low energy regime more theoretically correctly, there remains uncertainty in some of the parameters. Initial observations with MicroBooNE data, as well as data from T2K and MiniBooNE, indicated that the central value simulation was under predicting the CC interaction channels with no pions. To reconcile this, it was decided to retune the parameters for the CCQE & CCMEC models. These were chosen because there is significant theoretical uncertainty about their parameters to begin with. Other channels are not tuned at this time either because there is no evidence of a discrepancy or because the specific modeling of that channel is not crucial to the low energy excess. Tuning was done using T2K CC0$\pi$ cross section data. Four parameters were chosen for the tuning:

1. CCQE $M_a$: The axial dipole form factor in CCQE interactions

2. CCQE RPA: The random phase approximation. A correction to the CCQE cross section to account for nuclear screening cause by long range nucleon correlations

3. CCMEC Normalization: A parameter not inherently available in GENIE v3, but introduced by MicroBooNE to control the overall normalization of MEC interactions.

4. CCMEC Cross Section Shape: A parameter that morphs the shape of the predicted CCMEC cross section from the Nieves prediction (parameter = 0) into the empirical MEC model (parameter = 1). This parameter is also not by default in GENIE and was added by MicroBooNE

The optimal parameters found from the T2K fits are:

1. CCQE $M_a = 1.18 \pm 0.12$

2. CCQE RPA $= 0.4 \pm 0.4$

3. CCMEC Normalization $= 1.26 \pm 0.7$

4. CCMEC Cross Section Shape $= 0.22^{0.78}_{-0.22}$

For all simulations used in this analysis, GENIE v3.0.6 G18 10a 02 11a with these retuned values will be used. Hereafter it will be referred to generically as *simulation* or as the *MicroBooNE Tune*.

### 5.4.3 Specific Simulations Important to This Analysis

There are several simulations that are useful in developing and validating this analysis. There were hardware changes between early and later data collection runs, so these samples are all simulated independently with the appropriate detector simulation changes for the corresponding data time frames. These are enumerated here:

1. **BNB Overlay Cocktail:** This sample is a full beam simulation. It incorporates the proper $\nu_\mu$ and $\nu_e$ flux (without oscillation) and simulates all types of interactions. The simulated neutrino interaction image is overlaid with data cosmics drawn from Unbiased BNB

2. $\nu_e$ **Intrinsic:** This sample simulates only $\nu_e$ interactions from the intrinsic decay spectrum. In principle an analogous sample forms a part of the BNB cocktail sample. However this is a much higher statistics sample and is crucial for studying selections.

3. **Detector Variation Unisims:** To assess uncertainties associated with detector systematics, both the BNB cocktail and $\nu_e$ intrinsic samples above are also generated with varied detector simulations. These are detailed more in Chapter 8

# Chapter 6

# Reconstructing LArTPC Images

This chapter discusses how interaction information is represented as images in MicroBooNE and how physics information is extracted from these images. A fully automated procedure begins with the 2D wire plane images and the PMT information. A mixture of traditional algorithms and deep learning techniques is then used to label and reconstruct features within the image. This procedure identifies neutrino interactions, daughter particles, 4-momenta of the daughter particle, and ultimately the reconstructed neutrino energy. This wealth of reconstructed information is crucial because most of the features that are reconstructed will be backgrounds of some form. We will discuss how the output of this analysis is used to select a pure signal sample further in Section 7.

## 6.1 A Focused Hunt: 1 Lepton + 1 Proton Events

As discussed in Section 3.3 this analysis investigates the hypothesis that the MiniBooNE excess will appear as an increase in $\nu_e$-like interactions vs. the central value prediction. Background rejection will be critical, so this analysis restricts its focus to a specific topology containing 1 final state lepton and 1 proton ($1\ell1p$) arising from a CCQE interaction. Hence, we are searching for $1\mu1p$ topology from $\nu_\mu$ events and $1e1p$ topology frrom $\nu_e$ events. We discuss the motivation and advantages of this choice here.

| Final Particles | Kinetic Energy | Signal Topology |
|---|---|---|
| p,p,$\mu$ | 10,120,50 | yes |
| p,$\mu$ | 20, 60 | no |
| p,p,p,e | 200,5,12,80 | yes |
| p,p,e,$\pi^+$ | 80,15,110,80 | no |

Table 6.1: Example of various possible final states particles and energies and which do or do not qualify, topologically, as signal.

### 6.1.1 CCQE $1\ell1p$ Interactions

The subset of interactions which form the signal for this analysis are dually defined by the final state topology and interaction type. Topologically we require one visible proton and one visible lepton in the final state. In this context, visible means that the particle had sufficient kinetic energy to produce a reconstructible ionization pattern. This will produce a characteristic "V"-shape from the two visible ionizing particles. We use a threshold of 60 MeV for a proton and 35 MeV for a lepton corresponding to a length in liquid argon of of ∼3 cm for a proton and ∼6 cm for a muon [41] [42]. At higher energies, electrons will shower rather than behave like a track, but near this energy threshold they behave similarly to muons and thus will also be visible as a ∼6cm track. Additional protons may be present, so long as they are below this energy threshold. Thus, when we refer to $n\ell mp$ final states, $n$ and $m$ are referring to the number of *visible* leptons and protons. Neutrons ejected during final state interactions will not be reconstructed and are ignored in the analysis. No final state mesons are permitted in principle (but will form backgrounds).

The two topologies of interest to this analysis will therefore be $1e1p$ events arising from $\nu_e$ interactions and $1\mu1p$ arising from $\nu_\mu$ interactions. The $1e1p$ events are the signal, providing the explicit check for the MiniBooNE excess, while the $1\mu1p$ events will help to constrain systematics. Examples of what $1e1p$ and $1\mu1p$ interactions will look like are shown in Fig. 6-1.

In addition to topological requirements, we also focus on quasi-elastic interactions. As shown in Fig. 3-7, the MiniBooNE excess occurs between 200 and 800 MeV, with the bulk occuring in the lowest energy bins. This energy region is dominated by CCQE interactions in which the incoming neutrino interacts with a neutron producing a lepton and proton in the final state as in Eq. 6.1:

$$\nu_\ell + n \to \ell^- + p \tag{6.1}$$

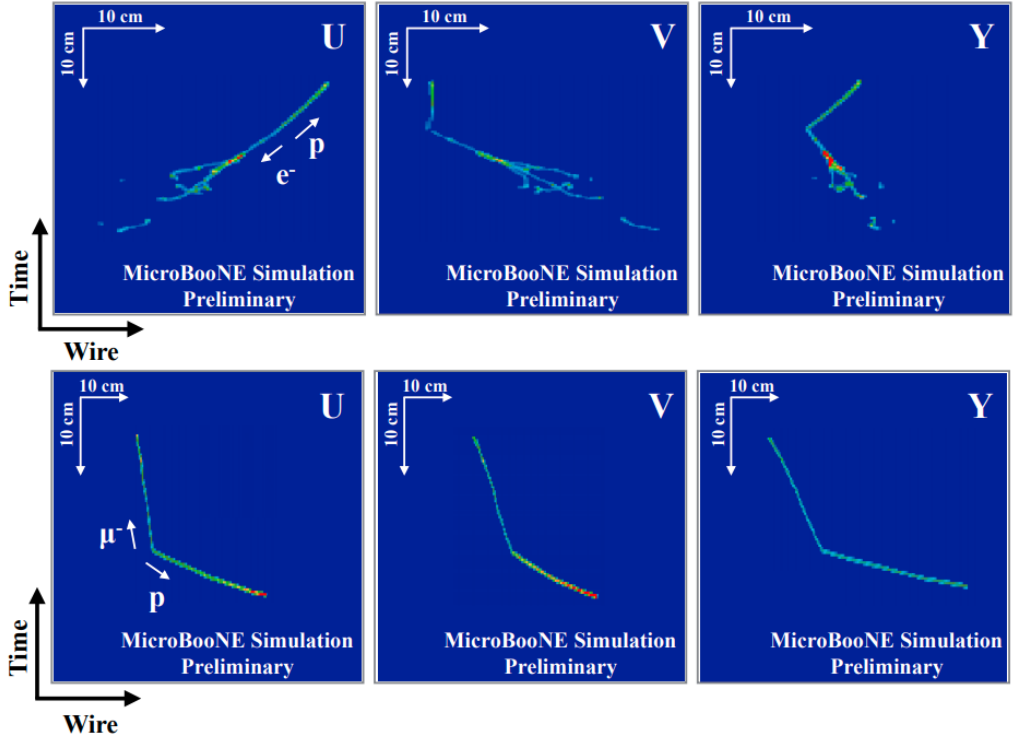

Figure 6-1: Simulated examples of a 1e1p interaction from a $\nu_e$ interaction (top) and a 1$\mu$1p interaction from a $\nu_\mu$ interaction (bottom). In both cases three images are shown, one for each 2D projection on the U,V,&Y wire planes.

The CCQE kinematics assume the neutrino interacts with a free neutron target at rest. Because the targets are within argon nuclei, and not free neutrons, thus will have Fermi momentum, CCQE kinematics will be approximate. Also, the CCQE kinematics disregard scattering and energy loss as the proton exits the nucleon.

### 6.1.2 Motivation

The primary driver for this signal choice is to minimize background. MiniBooNE was significantly systematics limited due to the inability to remove certain backgrounds.

Both this choice of topology and focusing on quasi-elastic interactions provide multiple handles by which to ultimately select a pure sample. The characteristic V shape of these events is difficult for cosmic backgrounds to mimic. Further, requiring the presence of a proton minimizes the risk of misidentification with cosmics and with other neutrino interactions which contain no proton. The simplicity of the topology increases the ease and accuracy with which the particles can be reconstructed, minimizing smearing in energy reconstruction. As we will discuss in more detail in Section 7, quasi-elastic scattering also offers characteristic kinematic behavior which helps distinguish it from backgrounds.

## 6.2 Physics in Pictures: Reconstruction Overview

The event reconstruction described here centers predominantly around the TPC data expressed as digital images. An event will yield three 2D images corresponding to the U, V, and Y wire planes. Each images is represented with the horizontal axis corresponding to wire number and the vertical axis corresponding to time. Each column is a single wire and the bins are filled with the digitized waveform value (ADC charge) from that particular wire. In raw form, the data is digitized as 3mm horizontally (wire spacing) and 0.55 mm along the vertical axis (time ticks). This analysis will further compress the image vertically by a factor of 6, summing the ADC charge, so as to produce pixels with similar vertical and horizontal scale, where the intensity represents the total ADC charge.

This analysis takes us from these pictures to a fully reconstructed and selected histogram of $\nu$ energies that is used to evaluate the potential MiniBooNE-like excess. The first couple steps, detailed in sections 6.3 & 6.4, make use of the PMT optical information. A set of optical precuts restrict the images we choose to analyze to those which likely contain neutrino activity. A set of tomographic optical matching algorithms, detailed in section 6.4, further helps to identify which features in an image are likely to be out-of-time from the beam and remove them prior to further analysis. A convolutional neural network is then used, which analyzes remaining charge in the image and determines which pixels are part of an ionization track or an electromagnetic shower, this network is detailed in section 6.5. The next step identifies the neutrino interaction vertex. This corresponds to the point in space at which the neutrino interacted with an argon nucleon and first began producing visible activity. The procedure for vertex finding is detailed in section 6.6. Once the vertex is identified, the event is reconstructed in 3D. This comprises two parallel steps in which electromagnetic showers and ionization tracks are reconstructed. Using cross plane matching, calorimetry, and reconstructed track ranges, this step provides the daughter particle information. The algorithms to perform this are discussed in sections 6.7& 6.8.1. The final reconstruction stage involves another neural network, the Multi-Particle Identification Network (MPID). This neural net is trained to identify the presence of different particles attached to the vertex.

The final steps beyond reconstruction, which are discussed in Chapter 7, use all this reconstructed information to perform both $1e1p$ and $1\mu1p$ event selection. A joint fit to both enables the $1\mu1p$ to constrain systematics on the $1e1p$ spectrum. The $1e1p$ spectrum is then used to test consistency with Standard Model central value prediction.

## 6.3 Optical Precuts

The first step does not explicitly involve image analysis. Rather, we want to use optical activity to restrict our focus to images that are likely to contain real neutrino activity. This helps prevent backgrounds from being reconstructed and reduces the processing load. The fundamental piece of information from which the precuts are built is the optical hit, or ophit. An ophit corresponds to a pulse observed in the digitized PMT waveforms. An ophit has both a known time and integrated peak area which is proportional to the number of incident photoelectrons. There are a total of 32 PMTs (although one PMT was lost after Run 1), producing potentially up to 32 ophits in any given time window. An example of both uncorrelated and correlated ophits are illustrated in Fig. 6-2.

A true neutrino interaction will produce a flash of light that will be larger than single photoelectron background noise and will be temporally correlated. Further, the activity will correlate with the known time of the accelerator beam spill. The goal of these optical precuts, then, is to find substantial activity that is correlated across PMTs and occurs at the right temporal location.

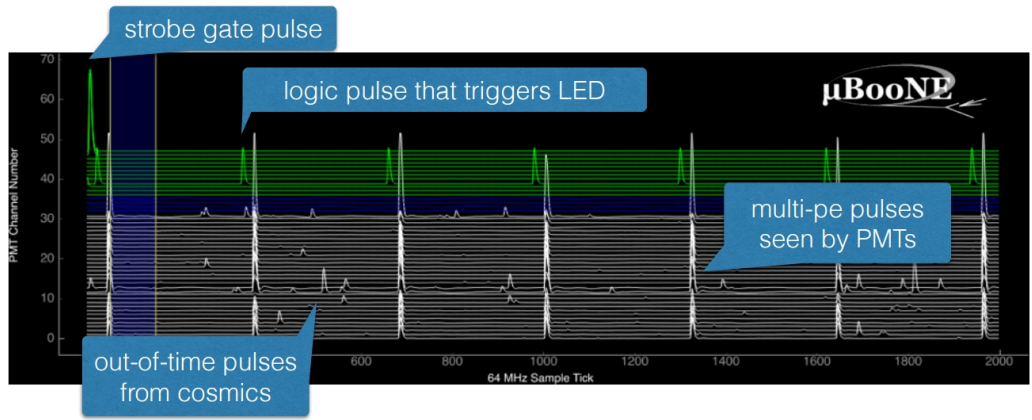

Figure 6-2: The 32 PMT's (white waveforms) are being driven by two different sources of light in this example. Artificial injection of light at a known time is caused by LEDs inside the cryostat and occurs shortly after the logic pulse (green). Shortly after this we see all 32 PMTs pulse, i.e. all 32 would form an ophit in the time window shortly after the LED. There are also smaller, uncorrelated pulses seen only on some waveforms between the LED spikes which occur at unpredictable times. These are caused by noise and cosmic activity.

The first step is to look at $\sim 2\mu$s time window which encloses the beam spill. This region is then broken into bins of 6 time-ticks, where a tick is defined by the digitization sampling frequency. 6 time-ticks corresponds to 93.75 ns. The ophits from each PMT are then assigned to the appropriate time bin and summed. This captures the total coincident light the PMTs observed and will be significantly larger when a neutrino is present. Note that because this uses $< 100$ ns time bins, this test is sensitive to the early component of the scintillation light. In order for an event to pass, 20 photoelectrons or more must be observed in the largest bin. An illustration of these summed ophits vs. time for an event containing neutrino activity is illustrated in Fig. 6-3.

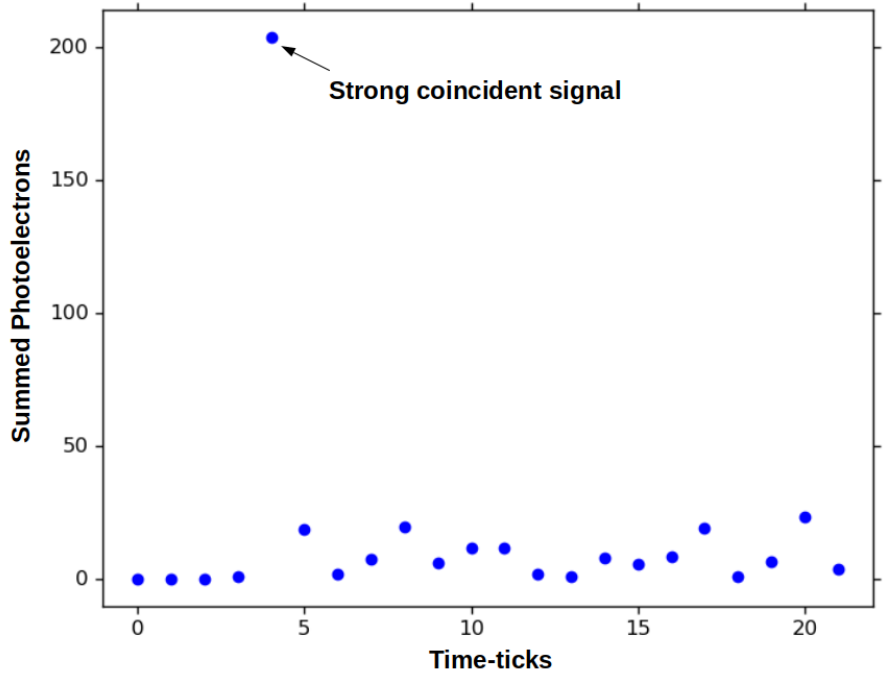

Figure 6-3: Pattern of optical hits, summed across all PMTs, vs time. A low level of random background is present at most times except when a large combined pulse corresponding to a neutrino interaction is seen.

Typically only an interaction within the beam spill window will yield significant optical activity within that time frame. However, one significant source of false positives comes from cosmic rays which enter the detector prior to the beam spill window, stop within the detector, and then produce a michel electron via $\mu^- \to e^- \nu_\mu \bar{\nu}_e$. If we only look within the beam window these michels can masquerade as a neutrino signal. To minimize these false positives, we look at a second window prior to the beam window and check if a second optical signal exists that could correspond to the entering cosmic muon which produced the michel. Similarly to our beam window requirement, if a bin exists in this prior window that exceeds 20 photoelectrons then the event is vetoed. An example of an a cosmic interaction which produces a michel signal is illustrated in Fig. 6-4.

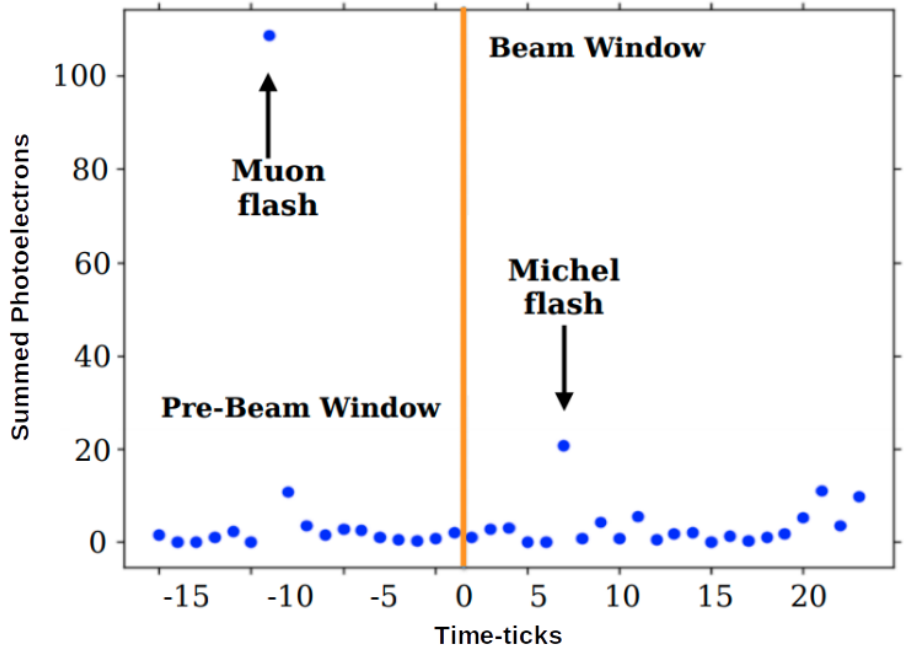

Figure 6-4: Example of a cosmic muon interaction which produces a flash both before and within the beam spill window. Tagging the existence of the first flash yields a veto which prevents the michel decay from triggering a false positive.

The efficiency of these optical cuts on the CCQE $1\ell1p$ signal events is very high with $> 98\%$ of signal interactions passing. Purity is significantly improved with 68% of empty events being rejected.

## 6.4 Cosmic Tagging With Wirecell

The second step merges information from the wire plane images and the PMTs to aid in cosmic tagging. MicroBooNE is on the surface and as such is subject to significant cosmic flux. In order to ensure that interactions can drift the length of the detector the readout is long enough that a typical image will contain a large amount of cosmic background activity. At an incident rate of 5.5 kHz, a typical event will contain 26 cosmic muons. As shown in Fig. 6-5, even when a neutrino interaction is relatively clear there is still cosmic interference. The presence of these cosmics poses a major problem. Cosmics that directly pierce neutrino activity can lead to reconstruction difficulties if the cosmic charge is confused with neutrino daughter charge. Cosmics themselves can also explicitly form a significant background category. A cosmic ray may be reconstructed as if it were a neutrino interaction even in the absence of any real neutrino. While a large majority of cosmics are very different topologically and kinematically from a neutrino, the relatively enormous number of cosmics vs signal means that extremely high discrimination power is needed. To this end, pixels that are likely cosmic in origin are tagged in this stage prior to event reconstruction and selection.

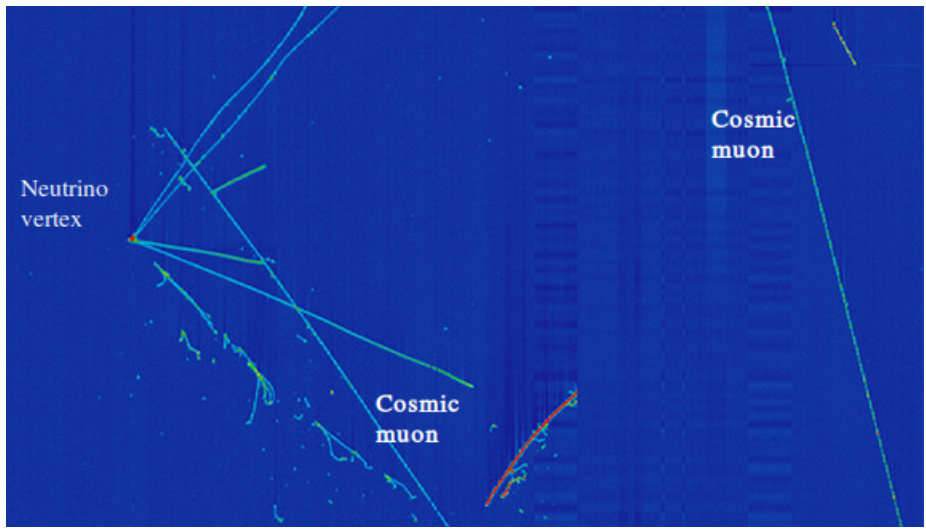

Figure 6-5: A typical neutrino interaction image will contain cosmic background activity in the image. In this interaction the neutrino vertex and daughter particles are clearly visible, but the event is also pierced by an unrelated cosmic track and contains other detached cosmic activity in the image. These can pose both a reconstruction problem and a background if reconstructed as neutrino activity.

Cosmic tagging is done with a set of algorithms collectively called WireCell [46], which proceeds as follows:

The first step is called geometric tiling. Along the x axis, 2D cross sectional image slices with width corresponding to $2\mu$s of drift are taken on each wire plane image. In each slide, any wires that fired continuously within it are merged to make a wire "bundle". If a bundle appears consistently on all three planes, this is then called a wire "blob".

In the next step, assuming that a blob actually represents three views of a single 3D event then the amount of ionization charge seen on each wire plane should be the same. This means that a linear equation linking the observed charge to an unknown true charge should have a solution. If the only viable solution is zero, then the blob is deemed spurious and removed from the image. The opposite situation, that of an underdetermined system, can also happen because of information loss when $n^2$ pixels is collapsed to $n$ wirie measurements in each 2D image. In these cases, constraints from sparsity, non-negativity, and connectivity are used to constrain the solution using the Compressed Sensing technique [47].

The next step is 3D imaging. In this stage a 3D image is constructed by directly combining the 2D cross sectional images in the time dimension.

At this stage, the reconstructed 3D image likely contains many thousands of blobs. The next step is to group them into clusters which will correspond to activity from a cosmic or neutrino daughter. Groups of blobs which are simply connected are now grouped.

We now have clusters of charge in 3D which may correspond to a neutrino daughters or to cosmic activity. The final step to tag the cosmic clusters takes advantage of the optical information and performs a many-to-many matching of light to charge cluster. For each cluster, an optical hypothesis can be made by assuming that scintillation light is proportional to the reconstructed ionization charge in a 3D voxel. This hypothesis is compared with all PMT flashes that were actually observed. If the prediction matches the observed light pattern then the hypothesis is accepted. All possible pairs of hypotheses between cluster and flash are checked. This again leads to a system of linear equations which can be solved to identify the optimal many-to-many match. However, the system may be underdetermined because of inefficiencies in the light detection system, the fact that light can be picked up from sources other than particle interactions, or because of clustering failures which lead to multiple good matches. To resolve the underdetermination, the compressed sensing technique is again used. Finally, clusters that do not match a PMT flash in the beam window are identified as cosmic clusters and are removed from the image.

The average accuracy of the charge-light matching is approximately 95% in both simulation (as determined by truth information) as well as in data (as determined by manual hand scanning) [46].

## 6.5 Pixel Labeling With Neural Nets

At this point, we have the data represented as images. The optical precuts have ensured that a large fraction of the neutrinoless images have been eliminated, and the wirecell cosmic tagging has cleaned cosmic background charge from the images. The next step is to perform additional pixel labeling using a convolutional neural network called *SparseSSNet* [48] [49] .

SparseSSNet is a semantic segmentation network whose purpose is to take an image and perform labeling on the individual pixels and assign them to a set of previously learned labels. A classic example of such a network is in the AI of self driving cars [43]. These cars image the surroundings and identify portions of the image which are road, other cars, pedestrians, etc.

### 6.5.1 Network Architecture & Training

SparseSSNet builds on a previous network, SSNet, which was first pioneered in MicroBooNE for image labeling [48]. It is sparse in the sense that it incorporates algorithms in which the image data is processed using a sparse rather than dense matrix. This offers a substantial increase in processing efficiency.

The architecture of the network itself (U-ResNet) is a hybrid of U-Net and ResNet. The network comprises 32 filters in the initial layer and has a total of 5 layers. A softmax classifier and cross entropy loss function which is summed over all non-zero pixels is used.

Training of the network is achieved via supervised learning using Monte Carlo simulated interactions with full detector simulation. The training sample consists of ∼120,000 images, and an additional ∼23,000 images are produced as a validation sample. The images make use of the "Particle Bomb" simulation. In this, A random 3D point within the detector volume is chosen to mimic an interaction vertex. A random number of particles is then drawn from [1,6] uniformly. The direction of each particle is chosen from a spherically isotropic distribution. The kinetic energy of each particle is drawn from a particle-type-dependent uniform distribution to reflect typical energies that would be observed in real interactions. For 85% of the sample, a large but realistic energy range is used. The remaining 15% of the sample is biased toward low energies commensurate with the eventual focus on a low energy signal. Finally, a random number of muons is generated and overlaid to mimic cosmic activity. The number, type, and kinetic energy distribution of simulated particles is shown in Table. 6.2.

| Particle | e | $\gamma$ | $\mu$ | $\pi^{\pm}$ | p | Cosmic $\mu$ |
|---|---|---|---|---|---|---|
| Allowed Multiplicity | 0-2 | 0-2 | 0-2 | 0-2 | 0-3 | 5-10 |
| KE [MeV] | 50-1,000 | 50-1,000 | 50-3,000 | 50-2,000 | 50-4,000 | 5,000-20,000 |
| P (Low E) [MeV/c] | 30-100 | 30-100 | -175 | 95-195 | 300-450 | 5,000-20,000 |

Table 6.2: The permitted particle contents, multiplicities, and kinetic energy ranges from which the particle bomb training is drawn in training SparseSSNet [49].

Prior to training, a simple masking based on pixel intensity is performed. Pixels with intensity $< 10$ ADC or $> 300$ are masked. This eliminates the majority of background pixels, reducing the total number of pixels in an image by $\sim$99.5%. The masked pixels are fully disregarded and do not get stored nor used in any way by the network.

There are five labels which the network is then trained to identify.

1. Highly Ionizing Tracks: Protons, typically shows up as a shorter, fatter, more ionizing tracks.

2. Minimum Ionizing Tracks: Muons and charged pions, typically shows up as thinner, longer, lower ionizing tracks.

3. Electromagnetic Shower: Cascades induced by electrons, positrons, and photons above $\sim$ 33 MeV.

4. Delta Rays: Electrons resulting from the hard scatter of charged particles, typically from a muon.

5. Michel Electron: Electron produced from the decay of a muon.

Future versions of this analysis will fully leverage all of these labels. However, for the analysis discussed here these labels are collapsed, post training, into two categories:

1. Track: Highly ionizing tracks or minimum ionizing tracks.

2. Shower: Shower, delta, or michel electron.

To prevent class imbalance - the situation where one class dominates the loss function - while training, a pixel weighting scheme is applied in the loss function

$$\mathcal{L} = \sum_i w_i (\vec{l_i} \cdot log(\vec{p_i}))  \quad (6.2)$$

where $w_i$ is the weight defined for each pixel, $\vec{l_i}$ is the label vector for pixel i, and $\vec{p_i}$ is the softmax probability vector for pixel i. Two types of weights are applied, and $w_i$ is the sum of them. The two weights used are Cluster Weighting and Vertex Weighting.

1. **Cluster Weighting**: This weight accounts for the fact that big clusters contain many pixels, typically of the same type. Thus labeling a large cluster correctly reduces the loss function more than correctly labeling a smaller cluster. In principle though, as both likely arise from a single particle it is equally important to label both properly. Numerically, the cluster weight is inversely proportional to the size of the cluster.

2. **Vertex Weighting**: Pixels which are in the middle of a cluster of similar pixels are easier to identify. Pixels that are proximate to a different type of pixel are more difficult. This is a crucial category to get correct because a neutrino vertex will commonly have two different particles attached. Pixels which are within three pixels of a different kind of pixel are weighted more heavily.

## 6.5.2 Accuracy

The accuracy with respect to non-zero pixels and the loss on training is illustrated in Fig. 6-6. The confusion matrix both before and after collapsing definitions to track & shower are provided in Figs. 6-7 & 6-8. Ultimately, based on results from the validation sample the neural network is capable of correctly identifying if a feature is track or shower like in >99% of cases. An example of the labeling in action for a simulated $1e1p$ and $1\mu1p$ event is shown in Fig. 6-9.

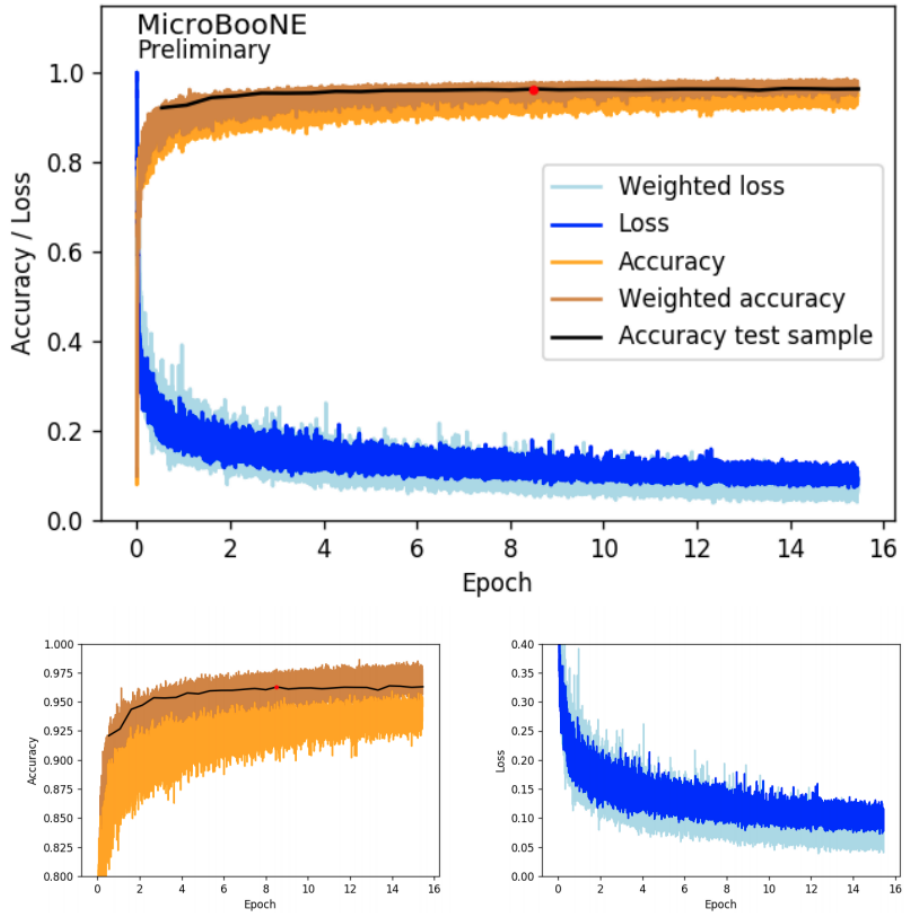

Figure 6-6: The accuracy and loss of the network on both training and inference for the collection wire plane images. The accuracy both before (orange) and after (brown) using the weighting scheme is illustrated. The ultimately selected network weights are taken from the point at the red dot [49].

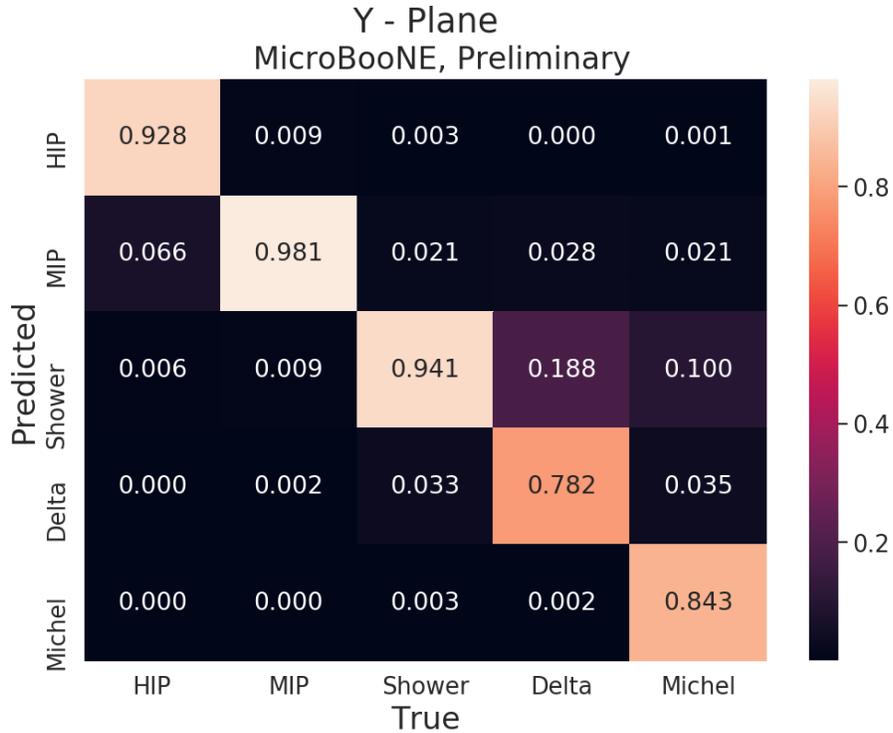

Figure 6-7: Full five class confusion matrix derived from the y-plane validation sample. The $i^{th}j^{th}$ box represents the fraction of pixels from truth class j which are labeled by the network as class i [49].

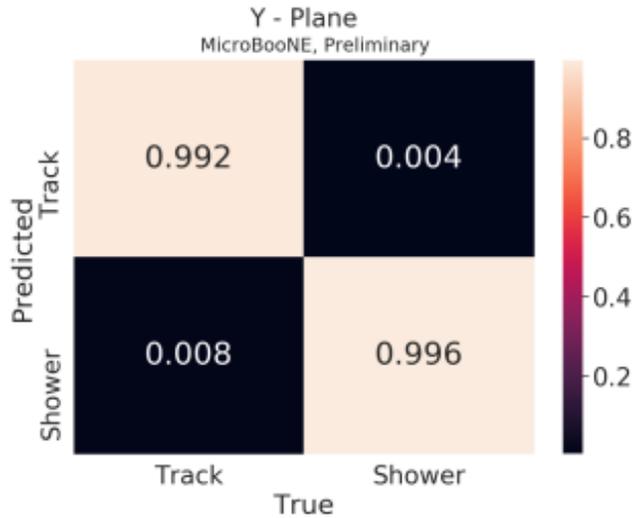

Figure 6-8: Collapsed two class confusion matrix derived from the y-plane validation sample. These results correspond to those in Fig. 6-7 but with [HIP,MIP] mapped to track and [michel,delta,shower] mapped to shower prior to computing the confusion. The $i^{th}j^{th}$ box represents the fraction of pixels from truth class j which are labeled by the network as class i [49].

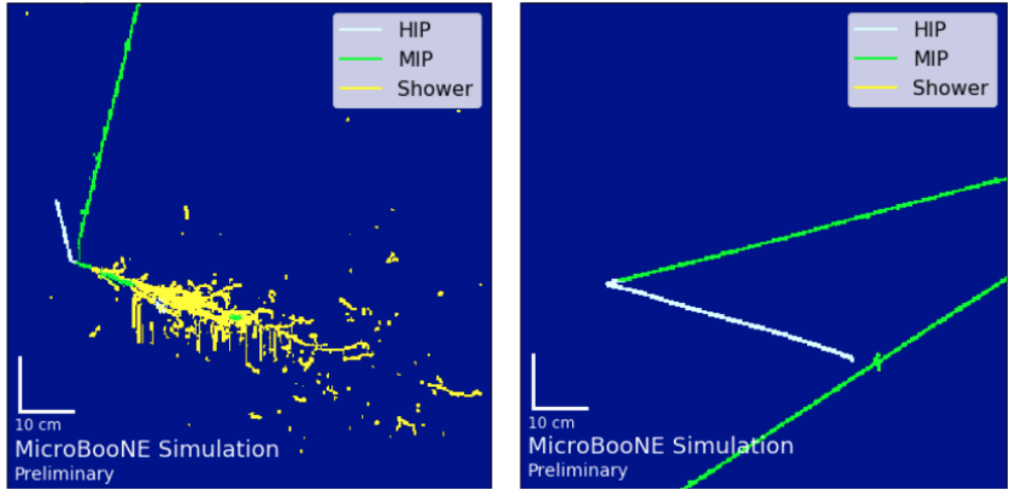

Figure 6-9: SparseSSNet pixel labeling applied to simulated validation events. We see a 1e1p event with a cosmic muon background (left) and a $1\mu 1p$ event with cosmic muon background (right). In both cases the minimum ionizing muons are correctly identified as MIP class, the protons are identified as highly ionizing particles, HIP class, and the electromagnetic shower is overwhelmingly classified as shower like, with a small amount of MIP contamination toward the beginning of the shower [49].

## 6.6 Interaction Vertex Identification

At this stage we now have images with pixel level labeling and much of the cosmic activity removed. The vertexing step [45] [44] will identify the specific location in the image at which the neutrino interacted with an argon nucleus. To begin, the wire images have an intensity threshold applied to retain only the major topological features. Pixels below 10 ADC are zeroed out. The three wire input images are then split into three subcategories. The ADC image will refer to the entire image, the track image has shower-like pixels set to zero, the shower image has track-like pixels set to zero. An example of these for a $1\mu 1p$ and $1e1p$ event is show in Fig. 6-10.

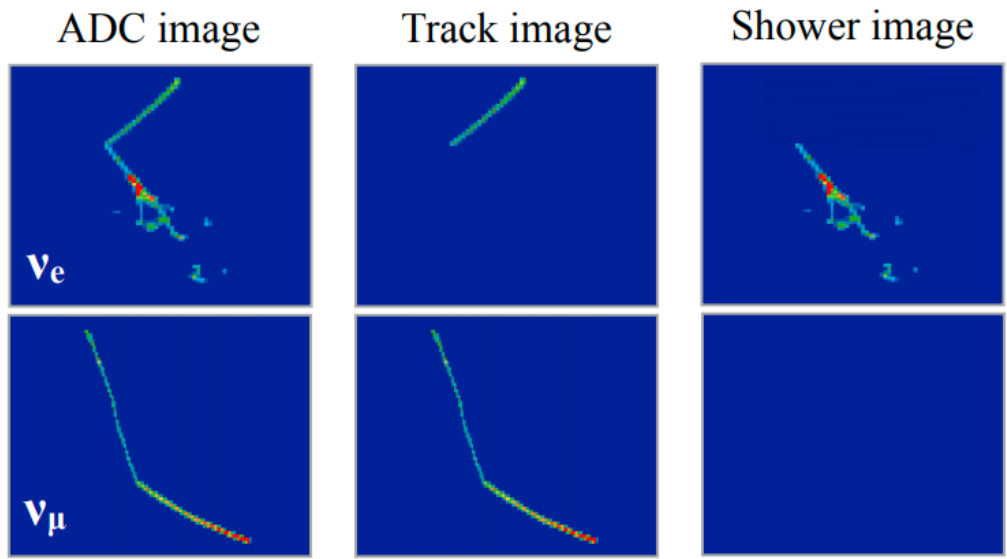

Figure 6-10: For either a 1e1p (top) or $1\mu$1p (bottom) interaction, a given wire image can be filtered to only include track or shower activity as informed by the SparseSSNet pixel labeling [44].

Two parallel algorithms are then run to identify vertices. The track-track algorithm is purely topological and seeks V shaped features. The shower-track algorithm takes advantage of the fact that the intersection of a cluster of shower-like pixels and track-like pixels is likely to be a vertex.

### 6.6.1 Track-Track Vertex Finding

The track-track algorithm searches for a vertex at the kink of a V shape in an image. This algorithm is run on the track only image. The first step is to identify vertex seeds. These seeds are points which show 2D features that plausibly correspond to a 3D consistent vertex. Frequently, this amounts to finding a 2D kinked shape which corresponds to the 2D projection of a 3D V shape. Vertex seeds are thus found as described subsequently on all three planes independently.

The first step is to use the pixel intensity to identify highly ionizing (HI) and minimally ionizing (MI) regions. This is motivated by the fact that we know a priori that the tracks we are interested in identifying are a highly ionizing proton and lower ionizing lepton. Accordingly the pixel intensity threshold which separates HI and MI regions is informed by studies on simulated protons. The response of each wire plane is not the same. A threshold for the U,V,& Y plane is set to 80, 120, and 140 ADC respectively.

Once pixels are separated by ADC value, contours that enclose these regions are identified using an image analysis software package called OpenCV [50]. This provided explicit groupings of pixels within the 2D image. To avoid the potential for hollow clusters, clusters identified as HI are grouped into clusters found as MI. So at this stage HI clusters are a strict subset of MI clusters.

The next step computes what is known as the convex hull. The convex hull is the smallest convex polygon which bounds the interaction. This is also performed using the OpenCV package. Given the convex hull, we then identify defect points. A defect point is the position on the polygon which maximizes the length of line segment which passes through the defect point and also bisects the opposing side. These points correspond to locations at which the cluster is potentially bending. If the bisector line segment is longer than five pixels, then the contour is broken into two along this bisector. New convex hulls and defect points are then found with the new broken shapes, iterating until no bisectors larger than 5 pixels can be found.

The result of this procedure is a set of straight line contours and a set of known defect points. These defect points are now saved as vertex seeds. The straight line contours are now fit using a Principle Component Analysis (PCA). The PCA is a linear estimator which minimizes the perpendicular deviation between pixels and the estimator. A PCA is performed for each of the straight clusters found. Any PCA intersection which aligns with a non zero pixel is also saved as a vertex seed.

For example, consider a $1\mu 1p$ interaction such as is illustrated in Fig. 6-11. In Step 1, the image is thresholded. In Step 2 the MI and HI regions are identified and enclosed within a contour by OpenCV. The HI cluster is then grouped back with the MI cluster. In Step 3, the convex hull is identified by OpenCV. In step 4, the defect point and line is found. The shape is then separated along the defect line. After this is done, Step 4 is repeated since there is a kink along the bottom track that will produce a second usable defect line. This ultimately leads to three segemented contours in Step 5. The contours from Step 5 yield the defect point vertex seeds shown in Step 6, while the intersection of linear PCA to the contours produces the additional seed shown in Step 7.

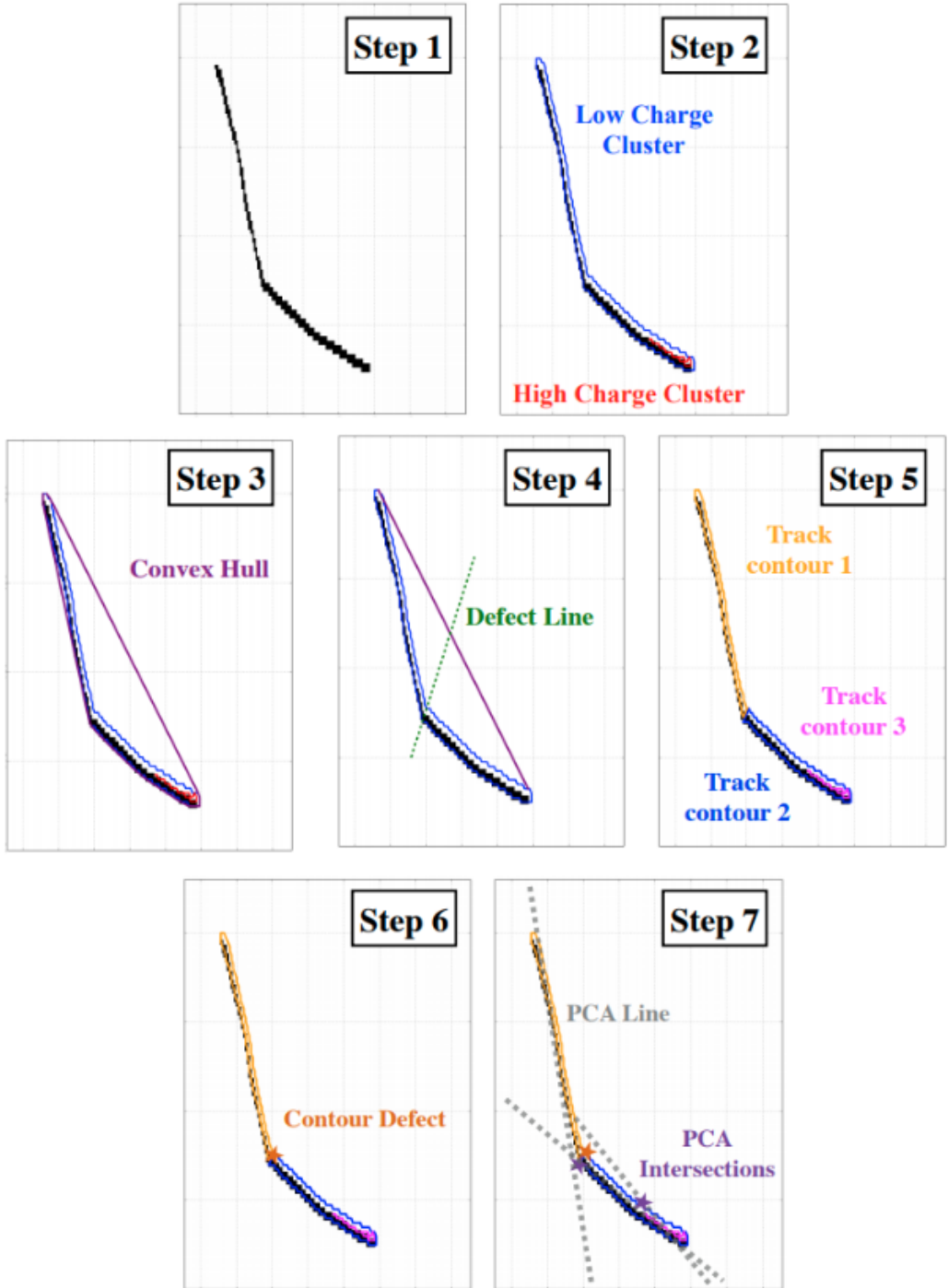

Figure 6-11: Illustration of vertex seed finding using a simulated $1\mu 1p$ event [44].

We now have a set of 2D vertex seeds on each of the U,V, & Y images. The next step seeks to refine the vertex seeks by finding the optimal kink point as informed by the local ionization track behavior. For each a 2D vertex seed, define a circle of radius 12 pixels centered at the seed. This circle will be pierced on the edges by ionization tracks. Using the points where the circle is pierced, define the following two quantities. The procedure to define these angles is illustrated in Fig. 6-12.

1. $\theta$: This is the minimum opening angle between points on the circle pierced by ionization tracks. (Fig. 6-12e)

2. $\phi$: At each point on the circle pierced by an ionization track, crop a 7×7 pixel region centered at that point. Perform a linear PCA fit to the charge within that image. This yields a good estimate of the direction of the track. $\phi$ is then the smallest opening angle between these PCA lines. (Fig. 6-12)

We now define the quantity $d\Theta = |\theta - \phi|$. This is computed at the location of the vertex seed. A PCA is then done to each of the tracks emanating from the 2D point and $d\Theta$ is then computed for new circles centered at 1 pixel increments along these lines away from the vertex seed. All points along the PCA which lie within a 40x40 region around the vertex seed are searched. When $d\Theta$ is minimized, we have found the optimal kink feature based on the local behavior of the ionization track. This minimization is illustrated in Fig. 6-13

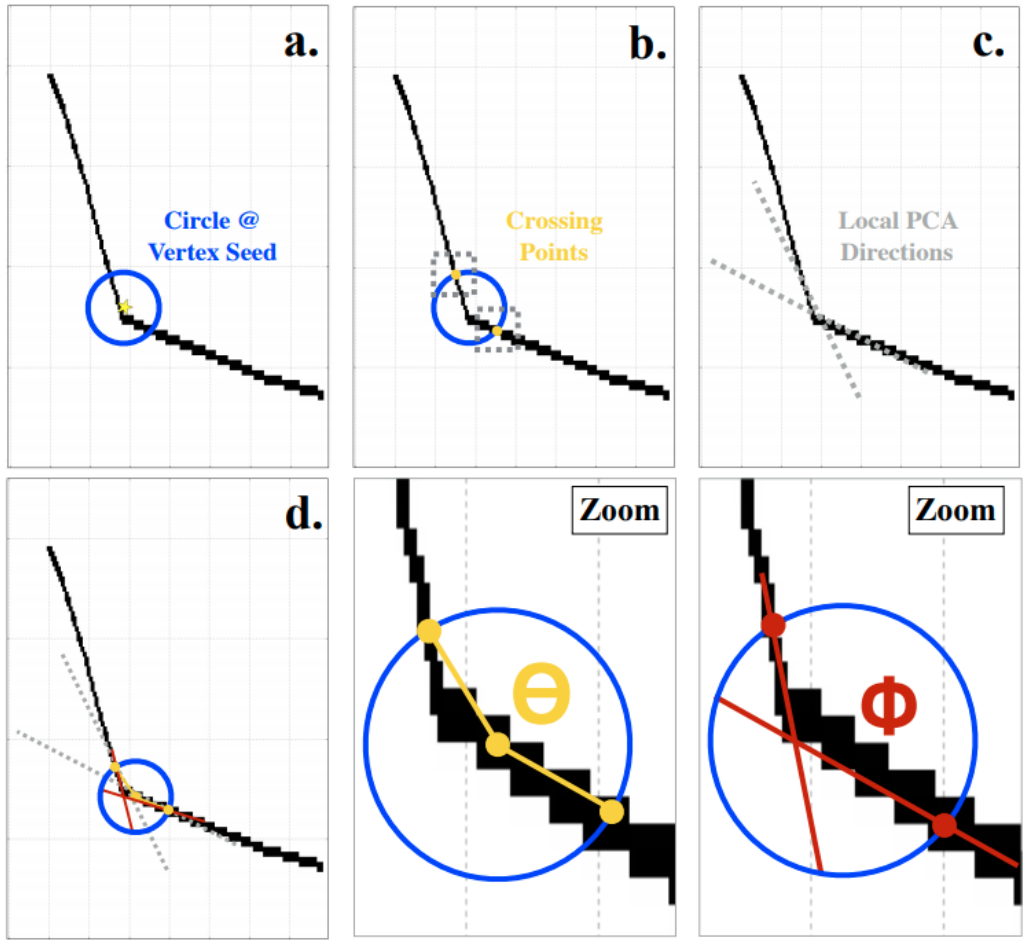

Figure 6-12: Method to take a vertex seed and define the two angular metrics, $\theta$ and $\phi$, which are used to identify 3D vertex candidates [44].

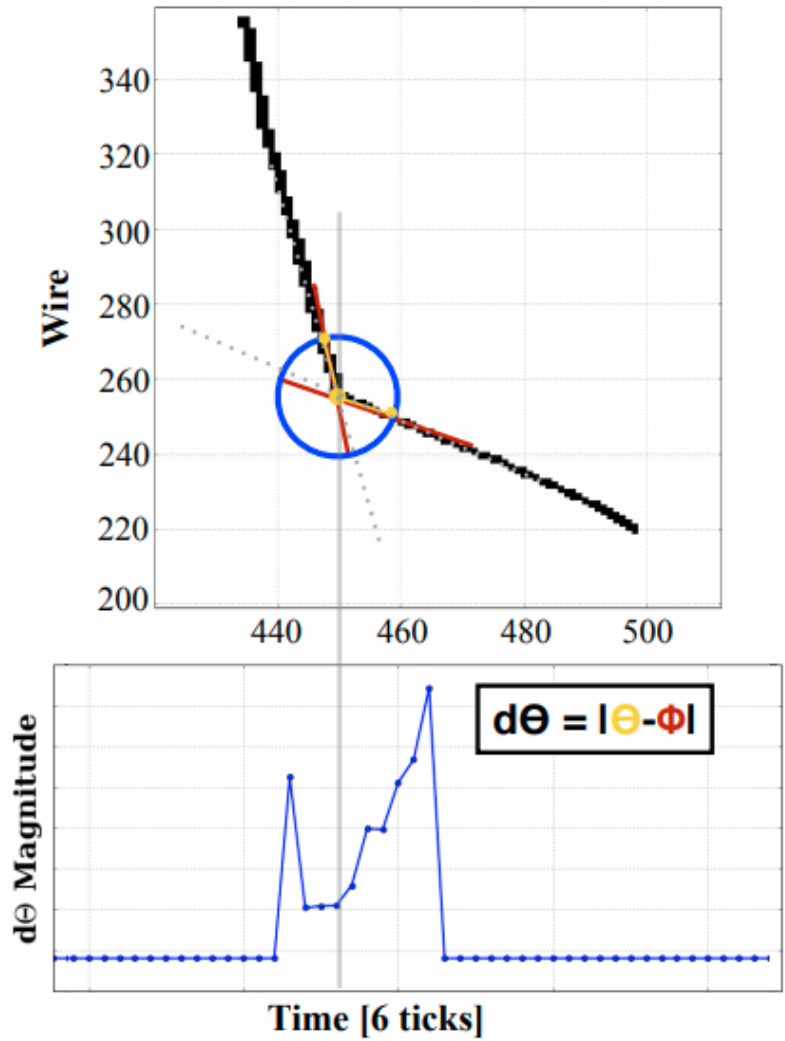

Figure 6-13: dΘ is computed at points around the vertex seed, and the minimum of dΘ corresponds to the optimal kink point [44].

We now have a set of refined 2D vertex points which have been refined from the initial seeds via minimization of the dΘ metric. This has been done on all three wire images. The final step is determine if a 3D consistent vertex can be built from these 2D points. Within the time resolution of our images, a vertex corresponding to a real physical event will always occur at the same time. So we first requiring that a 2D vertex point has a temporal match on one or both of the other images. The time provides the X coordinate of this vertex. The wire coincidence provides Y and Z.

A final 3D vertex is then refined from this point using a procedure analogous to the dΘ minimization done in 2D. A 4×4×4 cm$^3$ volume lattice with 0.5cm spacing is defined around the 3D vertex. For each point in this 3D space, the 2D projected point onto the wire image is calculated. If the projected point does not have nonzero charge on at least two of the 2D images, then the 3D point is rejected. If charge does exist on at least two 2D images, then dΘ is computed per plane and summed to yield the single metric dΘ$_3$ = $d\Theta_U + d\Theta_V + d\Theta_Y$. dΘ$_3$ is minimized over the 3D search volume to produce the optimal 3D vertex.

### 6.6.2 Shower-Track Vertex Finding

The shower track algorithm is similar to the track-track algorithm in that it too will search for an optimal 3D V shape. The shower-track algorithm differs in how it searches for the initial vertex seeds. Using the SparseSSNet pixel labeling, we have the track and shower images as illustrated in Fig. 6-10.

First, this algorithm looks at the track only image and identifies end points. This is done by locating the point which is farthest away from the cluster's barycenter. The second end is then found by locating the position on the track which is farthest from the first end. At each of these end points, a circle of radius 10 pixels is then drawn. This circle is then copied onto the shower image. The number of times the circle is pierced by shower pixels is measured. If this is zero, then this algorithm returns no shower-track vertex. If a single, continuous, cluster of shower pixels pierces the circle then a 2D vertex seed is formed. Continuity is required so as to prevent confusing an electron with a photon.

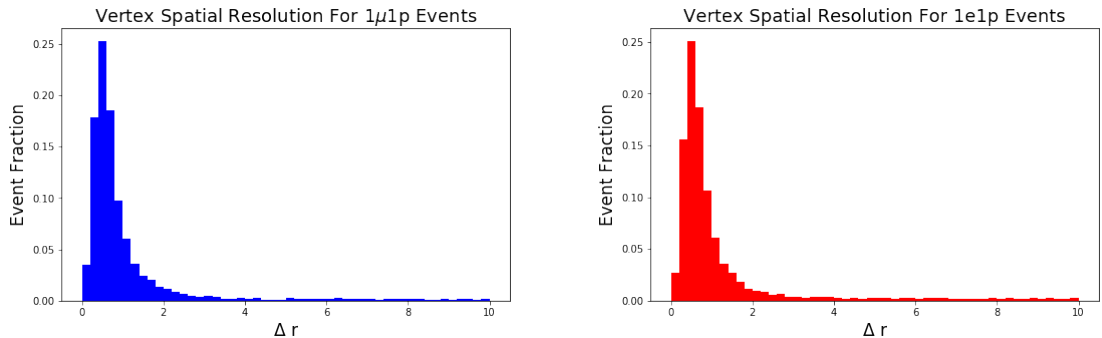

Figure 6-14: Distance between true interaction vertex and space charge corrected recosntructed vertex for a simulated sample of $1\mu 1p$ and 1e1p interactions.

To determine the best 3D vertex, it is still useful to perform the optimization of $d\Theta_3$ discussed in Section 6.6.1. This is important because SparseSSNet sometimes finds the very initial trunk of a shower to be track like. This would lead to some vertices which are systematically offset from the true vertex by $\mathcal{O}(cm)$. Optimizing for a V shape minimizes this problem.

### 6.6.3 Resolution & Efficiency

The efficiency and quality of vertex identification is characterized using a simulated sample of $1\mu 1p$ and $1e1p$ interactions. The primary metric of interest is the difference between the true and reconstructed vertex location, $\Delta r$. Electric field inhomogeneities within the detector will cause distortions in the apparent vertex location. This is known as the space-charge effect. A model of space charge is included within the simulated events themselves. So before computing $\Delta r$, the inverse of the space charge effect is applied to the reconstructed position to uncover the raw interaction location for comparison with truth. The resulting resolution is illustrated in Fig. 6-14. For both $1\mu 1p$ and $1e1p$ interactions, 68% of vertices are within 3 pixels ($\sim$1cm) of the true interaction location. A $1\ell 1p$ interaction that occurs within the detector fiducial volume will, by this stage of the analysis, produce a correct vertex with probability illustrated in Fig 6-15.

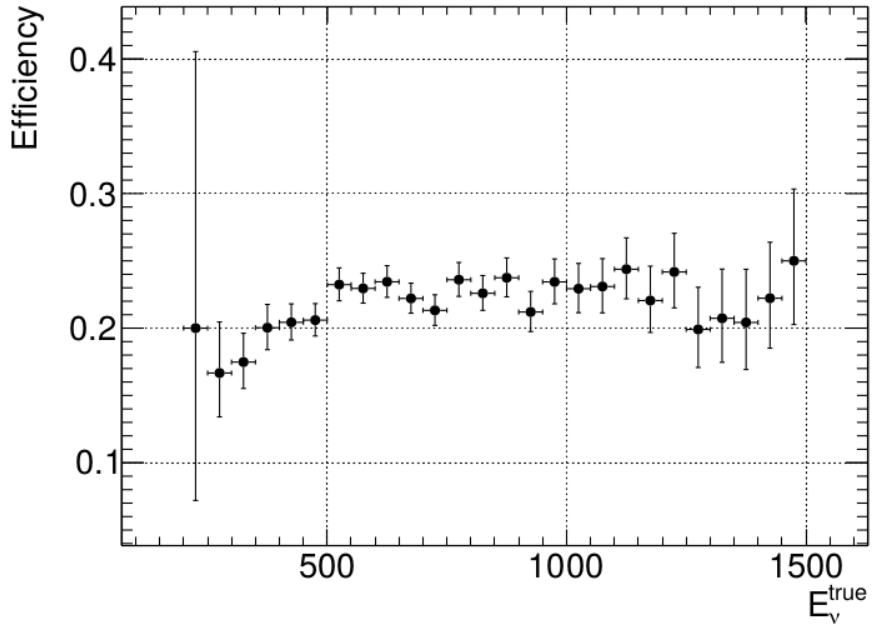

Figure 6-15: The probability, as a function of true neutrino energy, that a $1\ell1p$ interaction which occurs within the detector fiducial volume will yield a good vertex.

## 6.7 Track Reconstruction

At this point we have reconstructed vertices which represent points at which a neutrino interaction may have happened. We now must reconstruct the final state daughter particles. This full 3D reconstruction [45] identifies all points in the detector associated with an ionization track. This naturally provides the angular information for the daughter particle. In conjunction with range-based energy deposition splines for liquid argon, this also provides the deposited energy. Ionization density along the reconstructed track provides particle idenficiation. Ultimately, this reconstruction yields the 4-momentum of all track-producing particles (muons, protons, $\pi^{\pm}$). This algorithm is also applied to particles that yield electromagnetic showers (e,$\gamma$) and yields reliable angular information. Energy information for these particles, however, is handled by a separate algorithm discussed in Section 6.8.1.

### 6.7.1 Track Finding

The track finding algorithm uses the full ADC image. The image is preprocessed with a simple threshold of 15 ADC on all three wire plane images. These are set to zero. Reconstruction then begins at the previously identified 3D vertex point. A stochastic search identifies new 3D points by exploring the region about the vertex which contain 3D consistent points with active charge on multiple wire planes. New 3D points away from the vertex then seed this algorithm again. This is iterated until all 3D points along a track have been found.

The procedure to identify neighboring points, given an initial 3D seed, proceeds as follows:

1. Set seed to the last 3D point found. At the beginning this is the vertex, in later iterations it will be the previous 3D track point.

2. Generate a search region in which to find new 3D charge. The geometry of the search region depends on the track position:

    - *Within 5 cm of vertex*: Generate a set of random 3D points within a sphere of radius $r_{search} = 2 + 4e^{-L/5}$ cm, where L is the distance between the vertex and seed. The purpose of this decaying function is to permit an initially wider search around the vertex. Once a track is found, however, a narrower radius is desirable to prevent jumping from a track to adjacent charge.

    - *More than 5 cm from vertex*: Generate a set of random 3D points within a cone which faces forward in the direction of the previous 10 cm. If the track is <10 cm then use a cone with opening angle $30^o$. The radius of the cone is given by $r_{search}$. Switching to a cone permits faster tracking of straight tracks vs. a sphere.

3. Within the search sphere / cone, project all random 3D points onto the 2D wire images. A point must either project onto either 3 nonzero pixels or 2 nonzero pixels + 1 region with known inactive wires.

4. A 3D point from the random search region is stored as part of the track if it mapped to the largest total ADC value of any random point in the search region.

We have now identified a set of neighboring points to the seed. Any point that is at least one wire displaced from the seed (0.3 cm) is added to a proto-track, a cloud of 3D points corresponding to 3D consistent active pixels. The point within this cloud which is the farthest from the original seed then becomes the next seed and the process repeats for as long as new points are added.

Once this iteration is complete, the proto-track consists of an unordered set of points. Further, the stochastic nature of the search means that even nearby points may zig-zag around the track width rather than following a straight line. The next step is then to order the points. Starting at the vertex the next point is that which minimizes Eq. 6.3. The distances $L_1$ and $L_2$ and the angles $\theta$ and $\phi$ are defined as in Figure 6-16. The numerical parameters of this metric have been tuned to optimize the smoothness of tracks.

$$5L_1 + 0.1L_2 + 2(2 - cos\theta) - 10(2 - cos\phi) \tag{6.3}$$

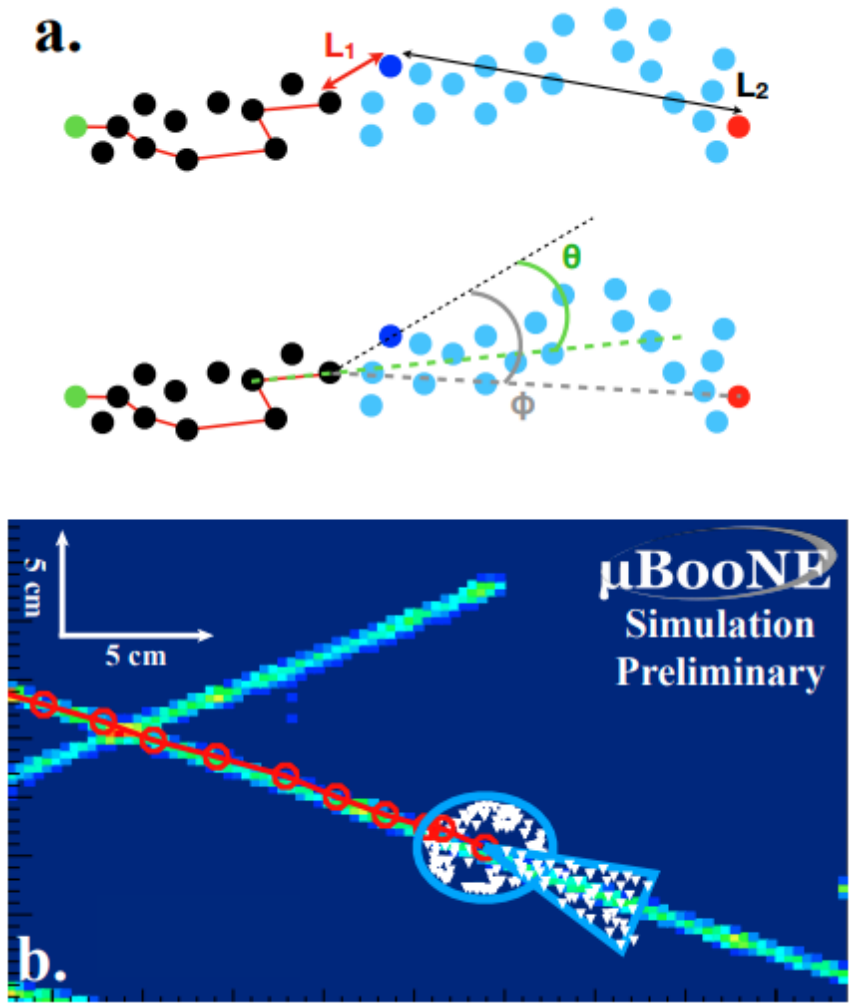

Figure 6-16: (a) Each dot represents a 3D point in the proto track. Black dots havce been sorted already, blue havec not. The green dot represents the interaction vertex whiel the red is the end of the track. The red line illustrates the path found within the points via a minimizing Eq. 6.3 at each point. $L_1$ is the distance to the prior selected 3D point, $L_2$ is the distance to the end of the proto track. $\theta$ is the angle from the last two sorted points to the current point, and $\phi$ is the angle between the candidate, prior sorted poitn, and end of the proto track. (b) Illustration of how stochastic searching is used to generate a cloud of 3D consistent points.

After doing this for all points in the cloud, there now exists a logical path to follow from vertex to end. However, there will still be substantial zig-zag along the width of the track and so a second-order smoothing is required. First, a new set of points is created by performing a rolling average of each point and the two neighbors. We then loop over this new set of points, moving from the vertex to each points nearest neighbor. The nearest neighbor must satisfy several criteria

- The nearest neighbor must be closer to the end point than the previous point.

- The distance from the previous point must not exceed 5 cm (this usually indicates that you have jumped from the track to background charge).

- The point must deviate by more than 0.5 cm from the line between points n-1 and n+1.

Once the first track has been identified, the entire procedure is iterated until there are no more tracks to identify. To prevent the algorithm from finding the same track repeatedly, or to find multiple tracks that contain overlapping charge, the pixels associated with previously found tracks are masked from the images before iterating. Specifically, if the 3D points are within 2cm of the vertex, then pixels within a 3 pixel cylinder around the projected track are masked. If more than 2cm from the vertex then the masking cylinder has a radius of 6 pixels. This smaller cylinder near the vertex is used to ensure that tracks which partially overlap at the very beginning due to their finite width aren't missed.

### 6.7.2 Track Ionization Density

Once 3D tracks have been reconstructed, ionization charge density can be calculated at different positions along the track. Ionization densities are a crucial ingredient in performing particle identification in a LArTPC because they permit discrimination between highly ionizing protons and minimum ionizing leptons.

For all 3D points on a track, their 2D projections on any given wire plane allow us to compute the mean ADC value across planes for that track position.

$$\mu_{ion} = \frac{1}{3n} \sum_{i=1}^{3} \sum_{j=1}^{n} \left(\frac{dQ}{dx}\right)_{i,j} \tag{6.4}$$

Where $\mu_{ion}$ is the mean ionization density for a track, and $(dQ/dx)_{i,j}$ is the pixel intensity of 2D projection onto plane i of 3D track point j. The ability of this metric to discriminate minimum and highly ionizing tracks is illustrated in Fig. 6-17 using a sample of simulated $1\mu 1p$ interactions.

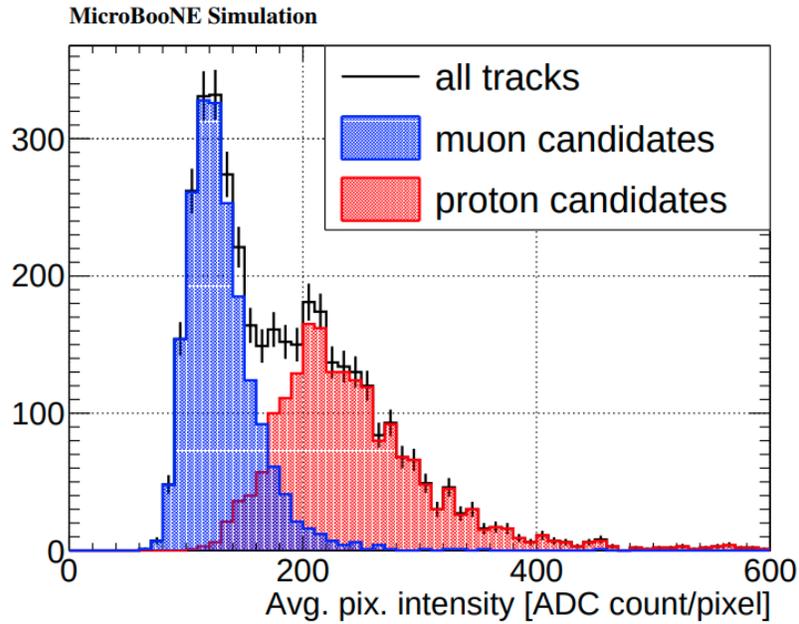

Figure 6-17: Illustration of the discrimination power of $\langle dQ/dx \rangle$ as computed by the 3D track reconstruction. In a simulated sample of $1\mu 1p$ interactions the difference in ionization density between the reconstructed muon and proton tracks is clear.

### 6.7.3 Angular Estimation and Resolution

Once 3D tracks have been reconstructed, the 3D points must be converted to a track direction before analytic kinematic quantities can be calculated. To obtain an estimate of the initial particle direction all 3D points with 15 cm of the vertex are averaged. A vector which points from the vertex to this mean location defines the track direction vector, $\vec{d}$. The angles $\theta$ and $\phi$ as defined in detector spherical coordinates, illustrated in Fig. 5-6, are then calculated from this vector as in Eqs. 6.5 & 6.6

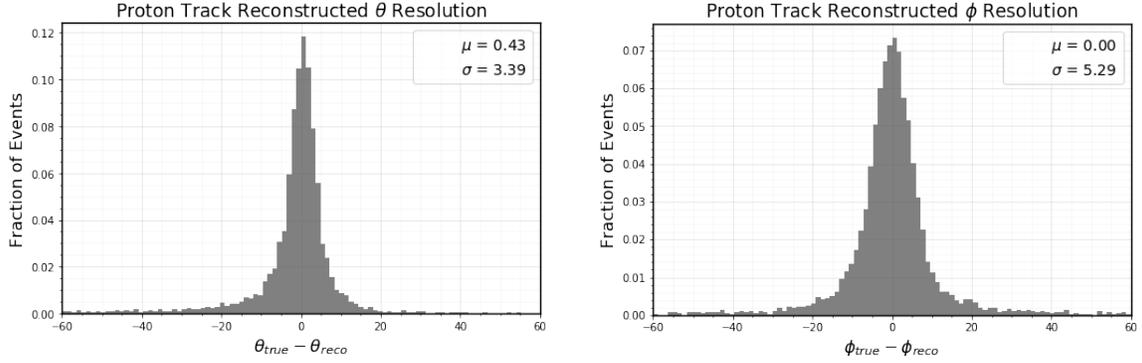

Figure 6-18: Angular resolution achieved for protons from true $1\mu1p$ interactions for which both particles were reconstructed by the 3D track reconstruction. A resolution of $3.4^o$ in $\theta$ and of $5.3^o$ in $\phi$ is achieved.

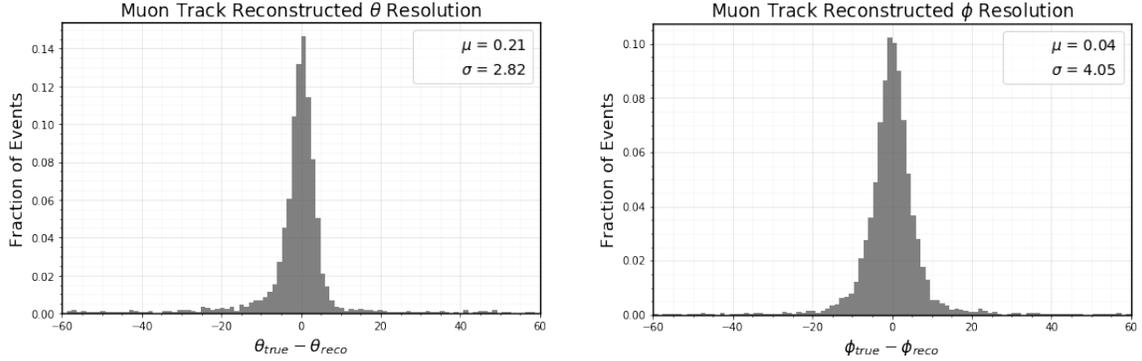

Figure 6-19: Angular resolution achieved for muons from true $1\mu1p$ interactions for which both particles were reconstructed by the 3D track reconstruction. A resolution of $2.8^o$ in $\theta$ and of $4.1^o$ in $\phi$ is achieved.

$$\theta = cos^{-1}\frac{d_z}{|d|} \quad (6.5)$$

$$\phi = atan2(d_y, d_x) \quad (6.6)$$

The angular resolution on the target signal events is characterized using a simulated sample of $1\mu1p$ and $1e1p$ interactions and are shown is Figs. 6-18 - 6-20

### 6.7.4 Energy Estimation and Resolution

The energies of particles that produce ionization tracks can be directly known from their lengths, assuming a given particle identification. This is obtained from the known stopping power of muons and protons in liquid argon. Electrons, which produce showers rather than tracks, are handled with a calorimetric rather than range-based algorithm, which is discussed in Section 6.8.1. Particle identification is not explicitly performed until later in the analysis, so the energy corresponding to both a muon and proton is computed for all tracks. The range splines used to derive energy from length for protons and muons are illustrated in Fig. 6-21 [42].

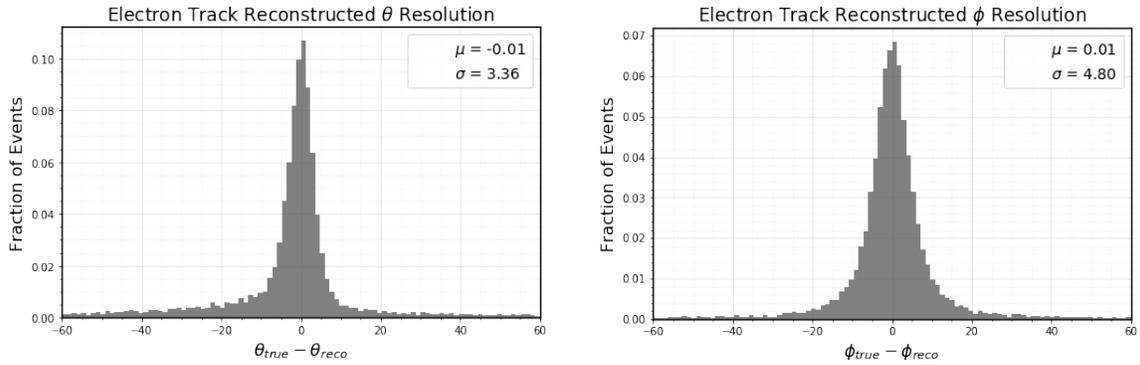

Figure 6-20: Angular resolution achieved for electrons from true 1e1p interactions for which both particles were reconstructed by the 3D track reconstruction. A resolution of $3.4^o$ in $\theta$ and of $4.8^o$ in $\phi$ is achieved.

The conversion from length to energy is highly accurate. When the reconstruction does reach the end of a track it typically estimates the length very accurately. However, significant underestimation of the track energy can occur if the reconstruction halts prematurely before reaching the end of the tracks. This can occur because the particle exited the detector active volume, making it impossible to track. It can also occur if the track crosses a region with unresponsive wires on one of the planes, making it difficult to match, and thus stopping early. This leads to two visible populations visible in Figs. 6-22- 6-23. Many events cluster in a tight peak at zero corresponding the well reconstructed interactions. A second distribution corresponding to significantly underreconstructed tracks is also visible. The selections discussed in Section 7 will seek to isolate only those well reconstructed interactions.

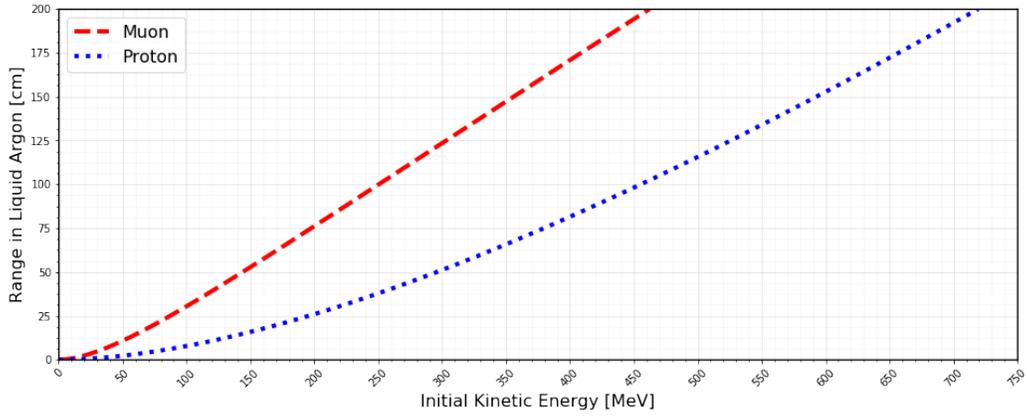

Figure 6-21: The conversion from reconstructed 3D length to deposited energy is achieved using the known stopping power of liquid argon for protons and muons.

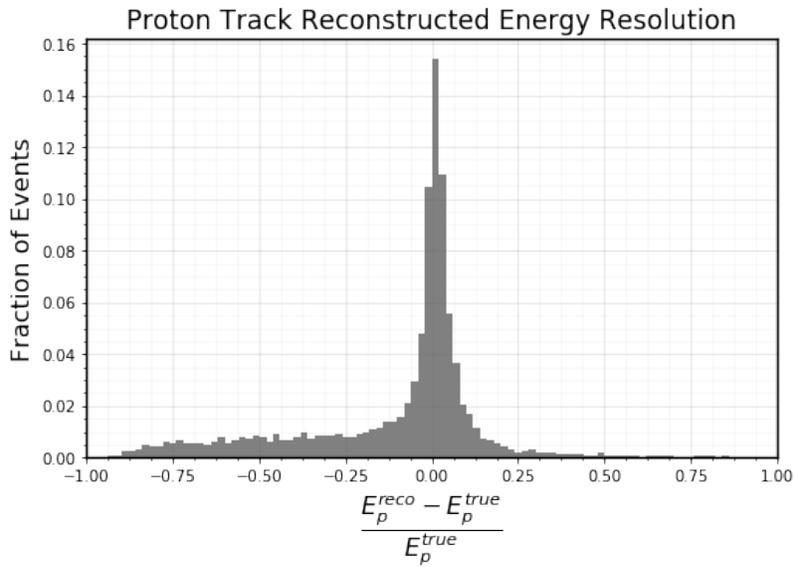

Figure 6-22: Energy resolution on protons from true $1\mu 1p$ interactions for which both particles were reconstructed by the 3D track reconstruction.

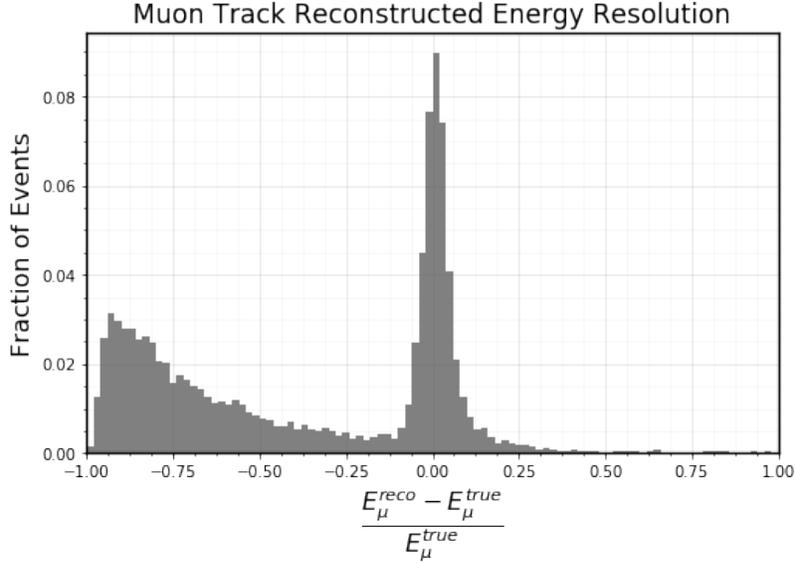

Figure 6-23: Energy resolution on muons from true $1\mu1p$ interactions for which both particles were reconstructed by the 3D track reconstruction.

## 6.8 Shower Reconstruction

Electrons will produce electromagnetic showers and thus need a fundamentally different algorithm to reconstruct their energies. Instead of a range based algorithm such as is used for muons and protons, shower reconstruction is a calorimetric algorithm which attempts to identify all charge related to the shower and then convert this to a total energy [51].

This algorithm also seeks to find a second shower if a first shower is reconstructed. This is done to minimize the potential of $\pi^0$ backgrounds.

### 6.8.1 Shower Finding

Shower finding is performed independently on each of the three 2D wire images. The first step is to use the SparseSSnet pixel labeling to keep only shower-like features. Only pixels with a shower score $> 0.5$ are retained, and all others are zeroed out. Further, pixels with an ADC intensity $< 10$ are masked. A 30 cm long isosceles template triangle is then placed extending from the vertex with an initial angle of 0.3 radians. An angular scan is performed to identify the optimal direction for the triangle. This is chosen by the angle for which the triangle contains the most charge. Once this has been found, a radial scan that moves the peak of the triangle away from the vertex is done. This explores the possibility of a gap near the vertex. The size of the gap is chosen to be the distance that maximizes the enclosed number of active pixels. For an electron, this gap will typically be zero. For a photon it may be larger since the charge is displaced from the vertex. Once the optimal initial direction and gap have been found, a 2-parameter scan is performed in which both the length and opening angle of the triangle is varied to optimize the amount of active pixels that can be enclosed.

Once a first cluster has been identified, these shower pixels are masked from the image and the process is repeated to determine if a second shower is present. The only difference is that the size of the initial triangle is 60 cm rather than 30 cm. Situations with a second shower are frequently $\pi^0$ decays, which are a potentially significant background to this analysis. To tag and remove these, it will be useful to determine if the two showers are consistent with arising from a $\pi^0$ decay. To do this, the invariant mass of the two showers needs to be computed. To achieve this, the 2D clusters for the first and second showers are matched across all 3 planes to determine the direction in 3D. Both the first and second shower finding are run independently on all three 2D wire images.

The triangle clustering has now yielded a total ADC charge for the showers on each plane. Converting this to energy is done using a simulated set of $1e1p$ interactions with known electron energy and which were vertexed correctly. A linear fit is calculated to give the conversion from ADC to energy. This is illustrated in Fig. 6-24

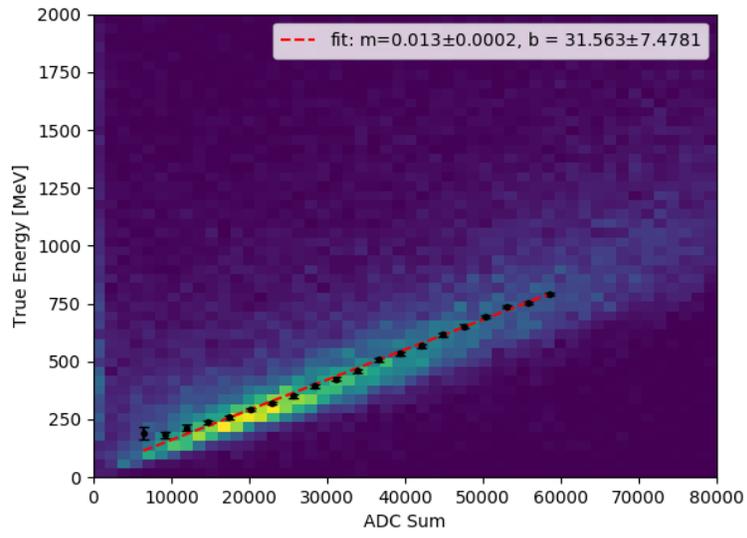

Figure 6-24: Shows the linear relationship between true electron energy and enclosed ADC charge for electrons from a simulated 1e1p sample. This best fit line is used to convert ADC $\rightarrow$ energy.

### 6.8.2 Energy Resolution

Showers are significantly more difficult to reconstruct well than tracks. As a result, resolution on electromagnetic cascades from electrons cannot match that of muons or protons. The resolution for electron showers is characterized using a $1e1p$ simulated sample for neutrino's of energy <800 MeV, the signal region in which the MiniBooNE excess appears. Within this energy region the electron shower resolution is 24%.

As with the muons, there are situations such as exiting particles and unresponsive wires that contribute to the populations with poorer resolution. Those events with sharper resolution will be targeted during selection.

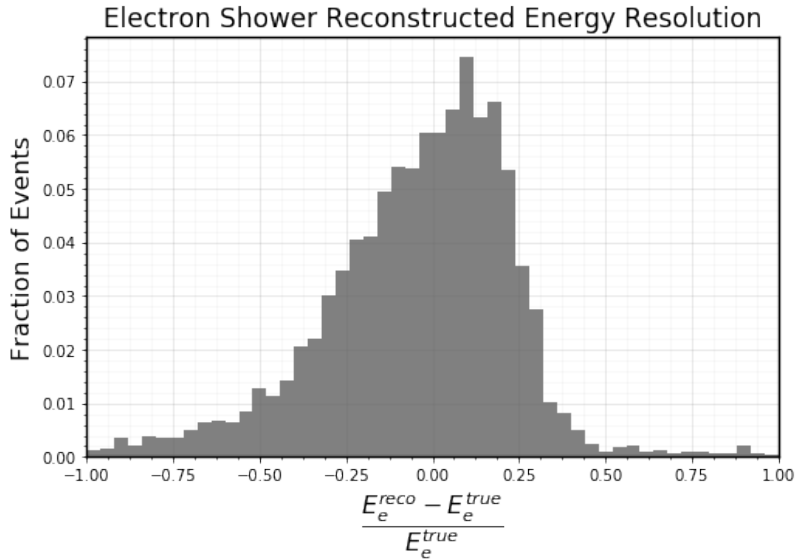

Figure 6-25: Energy resolution on electrons from true 1e1p interactions witih $E_\nu^{true} <$ 800 MeV for which both particles were reconstructed by the 3D track reconstruction.

### 6.8.3 Data Driven Shower Reconstruction Checks

We can cross-check the ADC-to-MeV conversion using two "standard candles" with known values. The Michel electron cut-off should appear at 53 MeV and the $\pi^0$ mass should peak at 135 MeV. As described below, we can obtain calibration test-points separately for data and Monte Carlo. To do this, a $\pi^0$ and Michel selection are applied to the following samples

- Beam neutrino simulation, simulates the portion of on-beam data containing a neutrino interaction;

- Data cosmic ray activity, corresponds to the portion of on beam data containing only cosmic background;

- 4.4×$10^{19}$ POT of on beam data made available as an unbiased open box.

The simulation, in conjunction with the cosmic data, allows a full predicted spectrum. Comparison of this prediction with on-beam data allows us to characterize the level of agreement in our shower reconstruction calibration between simulation and data.

### $\pi^0$ Shower Reconstruction Check

To select $\pi^0$, the following cuts are applied:

1. Two showers are reconstructed on the collection plane with >35 MeV

2. Both collection plane showers overlap spatially with showers found on one other plane by at least 50%

3. Two different shower in the collection plane cannot overlap with the same shower in a different plane

4. The higher energy shower must have energy > 80 MeV

5. The ADC sum in a sphere 5 cm around the the vertex is > 250 ADC

6. The more energetic of the two showers must reconstruct < 1.5 radians from the beam direction

7. The angle between the two showers is < 2.5 radians

8. Invariant mass <200 MeV

9. Require a proton to be reconstructed. Together with the two photons compute the invariant mass of the $\Delta^+$. This must be <1400 MeV

The result is a predicted spectrum, derived from combining the beam neutrino simulation with the cosmic data. Once properly normalized to reflect the same POT as the data sample, this should then agree with the selected data spectrum. Deviations in shape and position of the simulated vs data mass peaks will characterize the goodness of the simulation driven ADC→Energy calibration. We find excellent agreement between simulation and observation, as illustrated in Fig. 6-26

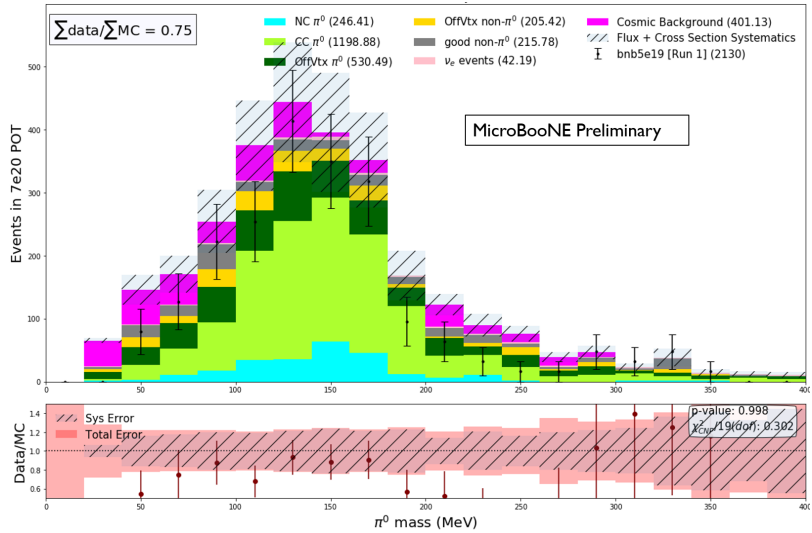

Figure 6-26: $\pi^0$ selection applied to a combined prediction derived from simulated neutrino interactions plus cosmic data. This is compared with a selected set of on beam data. "Off vertex" means that the reconstructed vertex is > 5cm from the true interaction location.

The $\chi^2$ in Eq. 6.7,

$$\chi^2 = \sum \left( \frac{m_{\pi^0}^{true} - m_{\pi^0}^{reco}}{\sigma} \right)^2, \quad (6.7)$$

is then minimized vs the $ADC \to$E conversion factor. $\sigma$ is decided by the width of a guassian fit to the simulation spectrum and is set to 30. This is done independently for the data spectrum and simulated spectrum to determine which values are favored by each. These values, in addition to those derived from the Michel check, are illustrated in Fig. 6-28

**Michel Shower Reconstruction Checks**

To select Michels, the following cuts are applied to the beam neutrino simulation and to the cosmic ray data:

1. Two particles attached to the vertex are reconstructed

2. The longest particle is >100cm in length

3. The shorter particle is < 30cm in length

4. The longest particle is <20% shower-pixel tagged

5. The shorter particle is >80% shower-pixel tagged

In the case of the Michel cross-check, we fit the shower charge sum ($ADC$) spectrum shown in Fig. 6-27, to the following five-parameter function:

$$f(x = \frac{ADC}{ADC_{cutoff}}; \sigma, N) = N \int_0^1 (3y^2 - 2y^3) \frac{1}{\sqrt{2\pi\sigma^2}} \exp \frac{-(x-y)^2}{2\sigma^2} dy$$

$$\sigma = y\sqrt{r_1^2 + \frac{r_2^2}{yADC_{cutoff}} + \left(\frac{r_3}{yADC_{cutoff}}\right)^2}$$
(6.8)

Here, $f(x)$ parameterizes the true Michel spectrum convoluted with a Gaussian representing charge resolution. $N$ is a normalization parameter. $\{r_1, r_2, r_3\}$ represent contributions to the charge resolution corresponding to a constant noise term, a statistical charge-counting term, and a radiative photon term, respectively.

We first fit $f(x)$ to the selected Michel spectrum from simulation and data. This is done by minimizing the $\chi^2$ over all five parameters $\{ADC_{cutoff}, N, r_1, r_2, r_3\}$.

$$\chi^2(ADC_{cutoff}, N, r_1, r_2, r_3) = \sum_{i \in ADC\ bins} \left(\frac{(O_i - f(x = \frac{ADC_i}{ADC_{cutoff}}; N, r_1, r_2, r_3))}{\sigma_{i,stat.}^2}\right)^2$$
(6.9)

where $O_i$ is the number of observed Michel events in ADC sum bin $i$ and $\sigma_{i,stat.} = \sqrt{O_i}$ is the Poisson error.

Next, the parameters $\{N, r_1, r_2, r_3\}$ were fixed at the minimum $\chi^2$ values (respectively for data/MC) and a 1D scan over $ADC_{cutoff}$ was performed, this time only including the tail of the $ADC$ spectrum in the $\chi^2$ (corresponding to an $ADC > 3000$ cut). Figure 6-27 shows the observed Michel ADC spectra in data and MC, along with their respective best fits from the 1D scan. Using Wilks' theorem for one fit parameter, we can find the $1\sigma$ interval on $ADC_{cutoff}$ by requiring $\chi^2(ADC_{cutoff}) - \min\{\chi^2; ADC_{cutoff}\} \leq 1$.

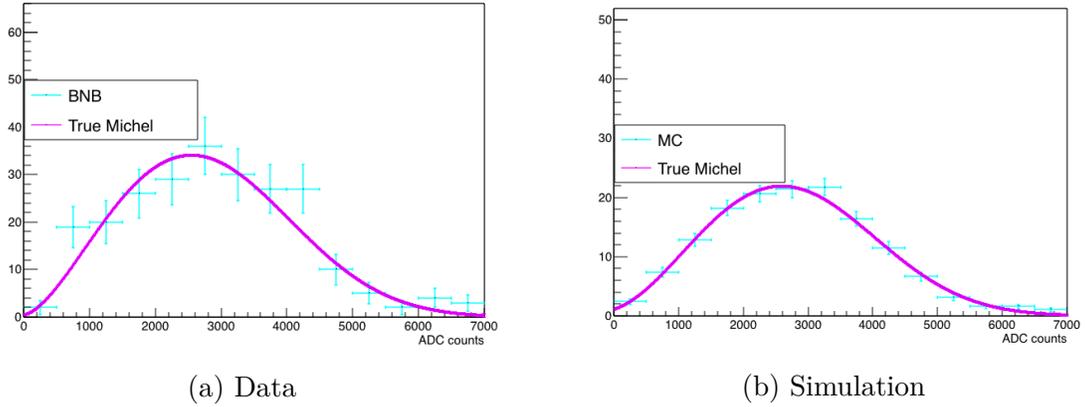

Figure 6-27: Observed michel spectra in data/MC (cyan) and the corresponding best fit parameterization of 6.8 (pink)

In Fig. 6-28, the resulting test points are indicated with statistical and systematic uncertainties added in quadrature. In principle, one expects perfect agreement between the test-points and calibration line for simulation. These test-points are shown on Fig. 6-28 in red, and one sees the expected agreement with the blue calibration line within $\pm 1\sigma$. The data test-points then allow a check that the ADC-to-MeV conversion found using the Monte Carlo applies as well as the data. These are shown as orange test-points on Fig. 6-28, and the agreement with the Monte Carlo test-points and the line are within the uncertainties as expected.

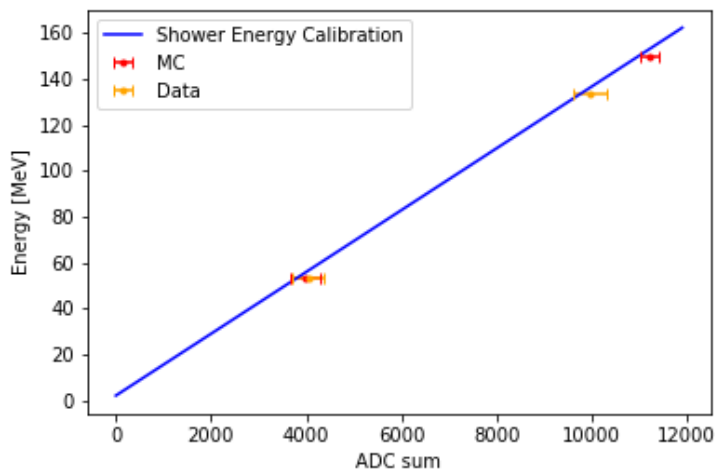

Figure 6-28: The four test points mentioned in the text (from the michel/$\pi_0$ samples for data & simulation) compared with the ADC-to-MeV calibration line derived from simulation. Error bars denote statistical and systematic errors. One can see $1\sigma$ agreement between all test points and the calibration line

## 6.9 Particle Identification with Neural Nets

In addition to SparseSSNet, which determines only if pixels are showers or tracks, another neural network is used. The Multi Particle Identification Network (MPID) analyzes 2D wire images and determines the probability that each of five particles, proton, muon, electron, photon, and charged pion, are present in the image [52]. Except for the use of the vertex to define the analyzis region, MPID is independent from other aspects of the analysis chain and thus provides a valuable cross check and selection tool.

### 6.9.1 Network Training & Architecture

The MPID network is trained on simulated images which contains only one of the five training particles. Particles are generated at a random location within the TPC active volume. For 80% of the training samples the kinetic energy is drawn uniformly from 60-400 MeV for protons and 30-1000 MeV for all other particles. The remaining 20% is low energy biased and draws from 30-100 MeV for all particles. The direction of the particles is generated isotropically.

The events are filtered at truth level to eliminate those which contain inelastic scattering. After the position, direction, and initial momentum are determined the particle is tracked through the argon using GEANT4. The MicroBooNE detector simulation is then used to produce 3 2D wire plane images as in a genuine event. Final training is performed with 60,000 events per each particle type while 20,000 per particle type are reserved for validation.

The network architecture uses ten convolutional layers to abstract the information from the 2D input images. An average pooling layer is applied after every second convolutional layer to extract higher level features. Two final fully connected layers combine all features derived from the convolutional layers. This outputs a vector of five floating point numbers corresponding to the five classes. A sigmoid is applied to yield a score from 0-1 corresponding to the class probability. This does not, however, ensure that the sum probability of all classes is one, rather each class is evaluated for probability independently. A schematic of the network architecture is shown in Fig. 6-29 and a visual example of an input and output is shown in Fig. 6-30.

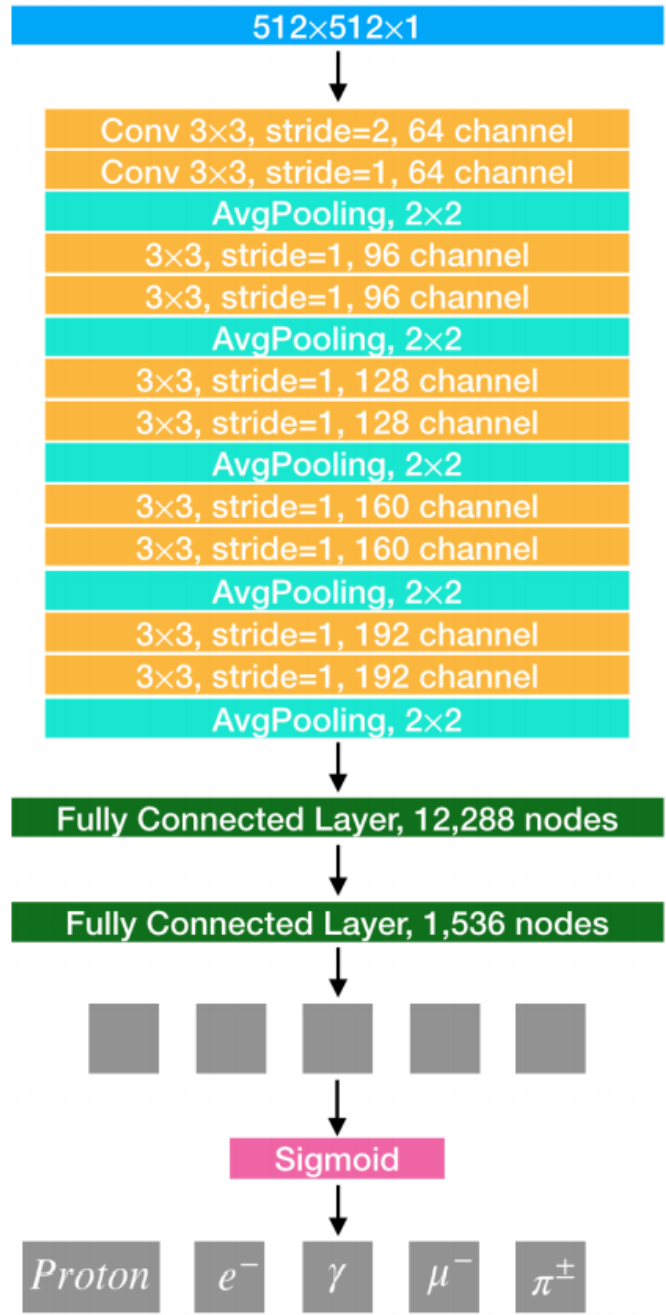

Figure 6-29: Schematic of the MPID network. Input is a 2D image and the output is a vector of 5 particle probability scores [52].

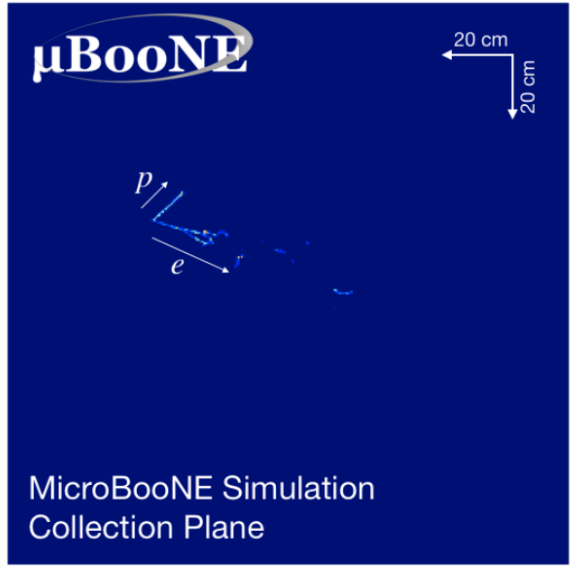

|  | $p^+$ | $e^-$ | $\gamma$ | $\mu^-$ | $\pi^\pm$ |
|---|---|---|---|---|---|
| MPID Score | 0.97 | 0.98 | 0.10 | 0.02 | 0.03 |

Figure 6-30: The MPID network will take as input a 2D image such as this simulated $1e1p$ interaction and yield a vector of five scores. As seen here, the proton and electron probabilities are high while other absent particles are lower [52].

## 6.9.2 Accuracy

While the MPID is trained on individual particle simulation, the ultimate metric of interest to this analysis is the ability of the network to identify particles within $1e1p$ and $1\mu 1p$ interactions. Using a simulated set of both $1e1p$ and $1\mu 1p$ interactions, the efficiencies and particle misidentification rates are summarized in Figs. 6-31 & 6-32

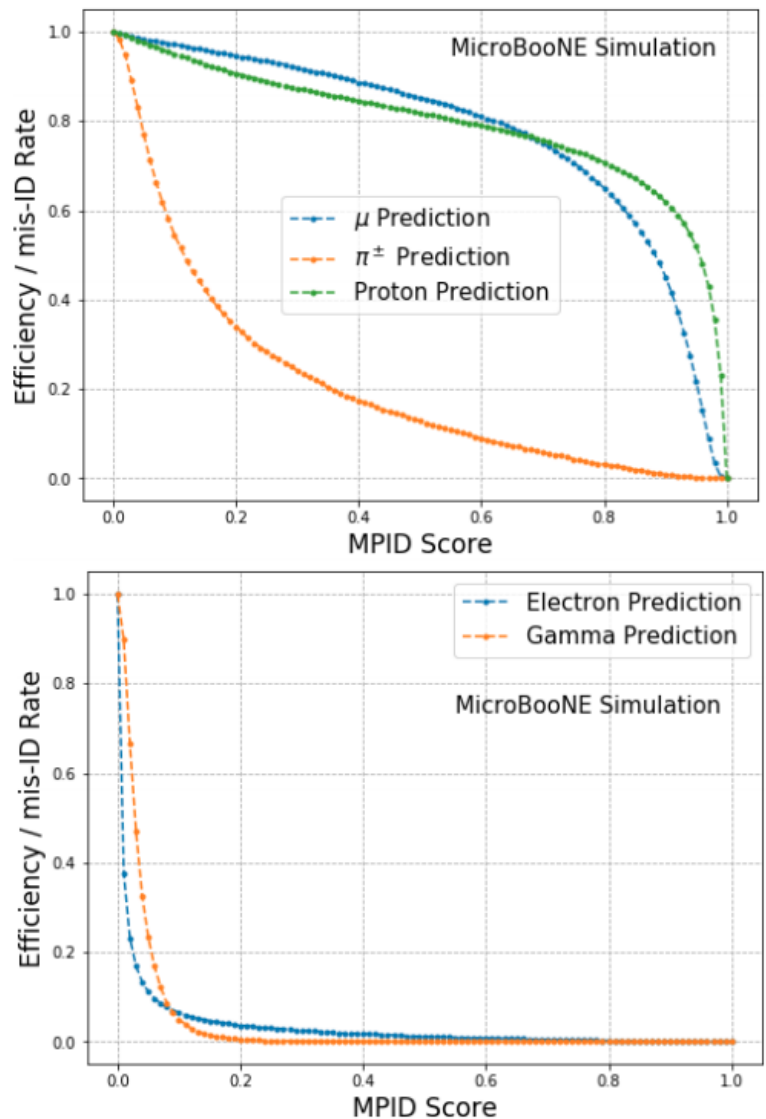

Figure 6-31: Efficiency (misidentification of wrong particle) curves for the MPID network on particles identified within a simulated sample of $1\mu 1p$ interactions [52].

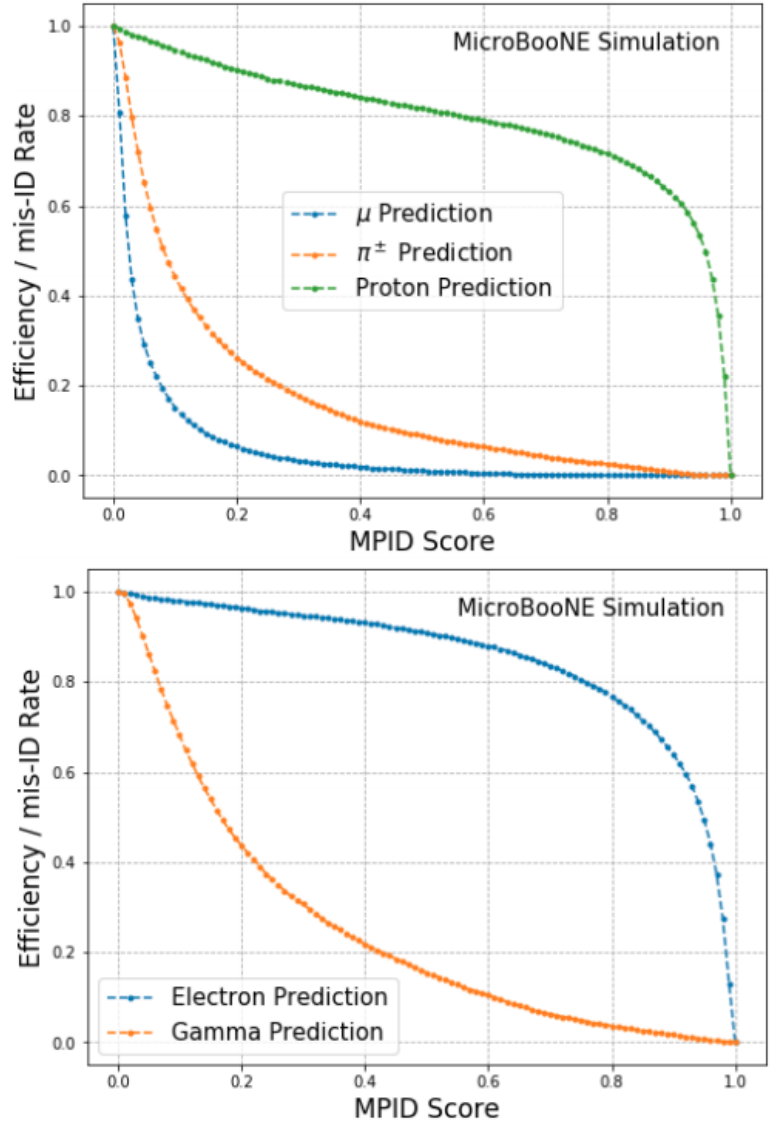

Figure 6-32: Efficiency (misidentification of wrong particle) curves for the MPID network on particles identified within a simulated sample of $1e1p$ interactions [52].

## Chapter 7

# Neutrino Event Selection

The final step in this analysis is to take advantage of the reconstructed information discussed in prior chapters to extract the CCQE $1\ell 1p$ signal from among the cosmic and non-signal neutrino backgrounds. Parallel selections are used to identify $1\mu 1p$ and $1e1p$ interactions. They follow broadly similar procedures, but differ in certain details. A mixture of cuts and machine learning in the form of boosted decision trees (BDT) are used to produce the final set of selected events.

## 7.1 General Overview of $1\ell 1p$ Event Selection

This section introduces the overall concepts behind the event selection. Here we will provide an overview of the selection criteria that will be expanded upon throughout the remaining text. Specifically, sections 7.5 and 7.6 provide explicit step by step selection procedures for the $1e1p$ and $1\mu 1p$ selections. In order to maximize the ability of the $1\mu 1p$ sample to constrain systematics on the $1e1p$ these selections are kept as similar as possible. However there will be some necessary differences. In addition to the overview provided here, and the detailed discussion in 7.5 and 7.6, we further provide Tables 7.2 and 7.3 which offer a side by side overview of the similarities and differences between the cuts and variables used in both selections.

In this section, we will broadly discuss useful cuts and their motivations, we will discuss the use of BDTs, which are employed in both the selections.

The selections will make use of variables calculated during 2D and 3D reconstruction as well as by the SparseSSNet and MPID neural networks. We describe the event selection in three tiers. The first tier is called the preselection cuts. The second combines the preselection with a BDT score requirement. The third combines preselection, BDT score, and PID requirements. These three sets of requirements are considered the "Event Selection Requirements" and are used for all comparisons that follow, unless otherwise noted.

**Preselection Cuts**

Preselection cuts downselect the sample that has passed the wirecell cosmic tagger with at least one in-time cROI. Here, we discuss the motivation behind the cuts.

We use the requirement of "two-prongs" from the reconstruction code to assure an event is consistent with $1e1p$ or $1\mu1p$. In this case, a "prong" is simply a reconstructed object emanating from the vertex. At this point, we remain agnostic about any PID associated with these prongs. The strict two-prong requirement results in a sample that is highly pure in 1 lepton and 1 proton events at the truth level, as we will discuss in the details of selected events. We use an "orthogonality cut" to divide the events into candidates for the two non-overlapping samples, $1e1p$ and $1\mu1p$. This cut makes use of the pixel-labeling from SparseSSnet and separates selected events into those that do or do not have a shower-like particle present.

We place several spatial requirements on the interaction vertex and reconstructed particles. In both the $1e1p$ and $1\mu1p$, the vertex must be reconstructed within a defined fiducial volume that excludes interactions near the edge of the detector as well as regions of the detector with significant dead wires. Beyond this, events are required to be "well-contained," *i.e.* to not pass close to or across the edge of the active volume. Containment is required for two reasons: to reduce energy smearing from poorly reconstructed particles and to remove incoming cosmic rays. This comes at the cost of removing exiting muons from neutrino interactions, leading to a low-energy bias to the analysis. This bias is acceptable because the LEE signal is also at low energy. The containment cut assures collection of most charge from the lepton or proton.

Lastly, initial reconstruction-quality cuts are applied. As discussed later, we will require an event to be "boostable" to the nucleon rest frame. To this end, in the pre-selection, we require $0 < \beta < 1$ for the boost, which eliminates mis-reconstructed background events. We also place minimum requirements on particle energy and opening angle in line with our definition of $1\ell 1p$ events.

**BDT Cuts**

The second tier adds the "BDT cuts." The specific variables used in each selection are discussed in Sections 7.5 and 7.6. A side by side comparison of the variables used in each BDT are listed in Table 7.3. In this section, we provide a brief overview of our approach to the BDTs.

The $1e1p$ and $1\mu 1p$ analysis each make use of BDTs. The $1e1p$ BDT is trained with a simulated sample of CCQE $1e1p$ $\nu_e$ interactions as the signal class. The $1\mu 1p$ BDT is trained using a simulated sample of CCQE $1\mu 1p$ interactions as the signal class. Both samples use a set of data cosmic interactions plus a beam simulation which contains all the non-signal $\nu$ interactions which also constitute a background. In both cases, the BDT training samples employ "well-reconstructed CCQE" events only as the target sample. Here "well-reconstructed" means events where the reconstructed energy is reconstructed within 20% of the true neutrino energy. By only showing the BDT well reconstructed events it better learns meaningful patterns which enables strong rejection backgrounds and only selects events with patterns indicative of CCQE $1\ell 1p$ interactions with reliable energy reconstruction.

In fact, $1e1p$ and $1\mu 1p$ each have two associated BDTs that have identical input variables, but are trained on different sample sets, one for Run 1 and one for post-Run 1 data. Changes to detector electronics were made to MicroBooNE after the first complete data running cycle. This leads to some qualitative changes in images and it merits training seprate BDTs for the run periods.

We make use of the python package XGBoost to produce the BDTs. The BDTs are regularized to reduce the potential for spurious results.

**PID Cuts**

The third tier of cuts are the "PID cuts" that are applied after the preselection and BDT cuts. These remove backgrounds that survive the second tier. These backgrounds are already significantly lower than the irreducible intrinsic background, but because they have relatively large systematic errors, it is preferable to cut them. e.g. $\pi^0$ are one of the most pernicious backgrounds to the $1e1p$ selection and as such an explicit cut on the presence of a two-shower interactions is made.

We also apply a set of cuts based on MPID output. These cuts differ between the $1\mu 1p$ and $1e1p$ selections as each has different backgrounds that persist beyond the BDT. These are summarized in Table 7.2.

In principle, these PID variables could be incorporated have been incorporated directly into the BDTs. We may choose to do so in a future analysis. However, at present, we chose to apply them them separately and explicitly, so that we can study the effect of varying the cuts.

## 7.2 More Details on Reconstructed Quantities Used

As discussed previously, the reconstruction chain provides many reconstructed quantities available which are leveraged for the final selection of $1\ell 1p$ signal events. This information broadly falls into three categories.

- Ionization patterns which look at charge deposition behavior.

- Kinematic variables use the reconstructed energy and angles to check for physical or non-physical kinematic behavior.

- Neural network output. The MPID and SparseSSNet provide information about the particles and charge deposition patterns within an image.

## 7.2.1 Ionization Patterns

The event images provide information about the ionization density of the objects we have reconstructed. For each reconstructed particle, the total charge deposited per track length is computed, providing a mean ionization density $\mu_{ion}$

The magnitude of this ionization density provides a simple yet effective particle ID within an event. We have limited ourselves to 1 lepton-1 proton events, so the proton is identified by choosing the track with higher ionization density. The magnitude of the ionization asymmetry is also useful. Signal interactions will have a proton attached to a lepton. In a typical case, this involves a minimum ionizing lepton attached to a proton that is roughly twice as ionizing. It is atypical for backgrounds to show a similar pattern. We capture this asymmetry using the following variable dubbed $\eta$:

$$\eta = \left| \frac{\mu_{ion}^p - \mu_{ion}^\ell}{\mu_{ion}^p - \mu_{ion}^\ell} \right|$$

## 7.2.2 Neural Network Output

**MPID**

The ability to visually gauge what particles are present in a reconstructed event is a major strength of an image based analysis. Using the MPID neural network described in Section 6.9 we can check that a candidate event contains a proton, the proper lepton, and no undesired particles that would typify a background. A cosmic ray background, for instance, will typically be identified as containing a muon.

There are two different types of score reported by the MPID. They differ in how the input image was filtered. The *interaction score* feeds the MPID only the portions of the image which are attached to the vertex. As a result this produces a score which cleanly captures only attached particles. However it can miss relevant detached particles such as $\pi^0$ decays which convert away from the vertex. For situations in which detached information may be useful, the *image score* returns probabilities associated with the entire image around the vertex.

The MPID is a 2D quantity and as a result in addition to the two-fold score discussed above, each of these two also has three different scores, one derived from the U,V, and Y plane images. Unless explicitly specified otherwise, the MPID score used by this analysis is that derived from the Y plane.

**SparseSSNet**

The ability to identify pixels which are shower like vs track like has already been extensively leveraged during reconstruction. It is further useful during selection. SparseSSNet provides a pixel by pixel shower vs track label. At the selection level this is used in aggregate. In each of the 2D images the clusters attached to the vertex are identified and the fraction of each that is shower like is analyzed. A $1\mu 1p$ interaction, for instance, will have a proton and muon track only and should virtually never contain a particle with a high shower fraction. The $1e1p$ on the other hand should contain one particle with a very small shower fraction and one with a very large fraction.

As with the MPID, a given event has one score per particle and one per wire plane image. Unless explicitly specified otherwise, the score used is that derived from the Y plane.

### 7.2.3 Kinematic Variables

Kinematic variables provide the strongest source of signal identification. Insofar as only contained events are considered, once the particles are identified a LArTPC enables full 4-momentum reconstruction from the energies and angles. From here, virtually any useful kinematic quantity can be reconstructed. The most useful of these and their associated definitions are summarized in Table 7.1.

| Variable Name | Definition |
|---|---|
| **Base Variables** | |
| $E_p$ | Energy of proton determined from range |
| $E_\mu$ | Energy of muon determined from range in detector |
| $E_e$ | Energy of electron determined from deposited charge |
| $m_\ell, m_n, m_p$ | Masses of the lepton, neutron and proton |
| $\cos\theta_p$ | $p_p^z/|\vec{p}_p|$ |
| $\cos\theta_\ell$ | $p_\ell^z/|\vec{p}_\ell|$ |
| $\phi_p$ | $\mathrm{atan2}(p_p^y, p_p^x)$ |
| $\phi_\ell$ | $\mathrm{atan2}(p_\ell^y, p_\ell^x)$ |
| $P_p = (E_p, \vec{p}_p)$ | Reconstructed 4-momentum of the proton |
| $P_\ell = (E_\ell, \vec{p}_\ell)$ | Reconstructed 4-momentum of the lepton |
| $E_b$ | Binding Energy for argon; the analysis assumes $B = 28.5$ MeV |
| **Definitions Related to Neutrino Energy** | |
| $E_\nu^{range}$ * | $E_p + E_\ell - (m_n - E_b)$ |
| $E_\nu^{QE-p}$ | $\dfrac{E_p(m_n - E_b) + \frac{1}{2}(m_\ell^2 - (m_n - E_b)^2 - m_p^2)}{(m_n - E_b) + |\vec{p}_p|\cos\theta_p - E_p}$ |
| $E_\nu^{QE-\ell}$ | $\dfrac{E_\ell(m_n - E_b) + \frac{1}{2}(m_p^2 - (m_n - E_b)^2 - m_\ell^2)}{(m_n - E_b) + |\vec{p}_\ell|\cos\theta_\ell - E_\ell}$ |
| $\Delta^{QE}$ | $\sqrt{\left(E_\nu^{QE-p} - E_\nu^{QE-\ell}\right)^2 + \left(E_\nu^{QE-p} - E_\nu^{range}\right)^2 + \left(E_\nu^{QE-\ell} - E_\nu^{range}\right)^2}$ |
| | * Unless explicitly noted otherwise, any reference to $E_\nu$ refer to $E_\nu^{range}$ as reconstructed in the laboratory frame. |
| **Event Kinematics** | |
| $Q^2$ | $2E_\nu(E_\ell - P_\ell^z) - m_\ell^2$ |
| Hadronic Mass ($m_{had}$) | $E_\nu - E_\ell$ |
| Björken's Scaling x ($x_{Bj}$) | $Q^2/2m_n m_{had}$ |
| Björken's Scaling y ($y_{Bj}$) | $m_{had}/E_\nu$ |
| Opening angle | $\cos^{-1}(\hat{p}_\ell \cdot \hat{p}_p)$ |
| $p_T$ | $\sqrt{(p_\ell^x + p_p^x)^2 + (p_\ell^y + p_p^y)^2}$ |
| $p_L$ | $p_p^z + p_\ell^z$ |
| $\alpha_T$ | $\cos^{-1}\left(-\dfrac{\vec{p}_T^{\,\ell} \cdot \vec{p}_T}{|\vec{p}_T||\vec{p}^{\,\ell}|}\right)$ |
| $\phi_T$ | $\cos^{-1}\left(-\dfrac{\vec{p}_T^{\,\ell} \cdot \vec{p}_T^{\,p}}{|\vec{p}_T^{\,p}||\vec{p}^{\,\ell}|}\right)$ |

Table 7.1: Kinematic variables derived from the reconstruction code used in this analysis.

## 7.3 Quasi-elastic Scattering Consistency

Beginning with the simplest picture, a neutrino may exchange a W with a neutron and produce the analogous charged lepton and proton in a final state. $\nu_\ell + n \to \ell^- + p$ This is termed quasi-elastic because the lepton's rest mass must be produced. While an analogous process can occur via a Z exchange, this will not produce our sought after 1-lepton 1-proton topology.

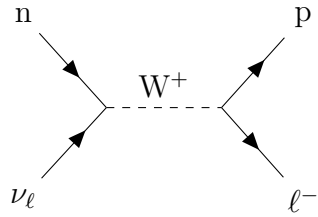

Figure 7-1: Feynman diagram illustrating charged current quasi-elastic scattering of a $\nu$ off of a nucleon.

The elastic kinematics associated with CCQE processes not only enables a ready computation of the initial neutrino energy, but it provides several methods by which this can be done. Crucially, this will require the knowledge that the neutrino momentum was aligned with the z-axis. Consider the following:

$$(P_\nu - P_p)^2 = (P_\ell - P_n)^2 \tag{7.1}$$

$$E_\nu = E_p + E_\ell - m_n \tag{7.2}$$

$$\vec{p}_\nu = \vec{p}_p + \vec{p}_\ell \tag{7.3}$$

Where Eq.7.1 is one of the Mandelstam variables - a Lorentz invariant - and Eqs.7.2 and 7.3 enforce energy and momentum conservation. From 7.1

$$(p_\ell - p_n)^2 = (p_\nu - p_p)^2$$

$$E_\ell^2 - |\vec{p}_\ell|^2 + E_n^2 - 2E_\ell E_n = E_\nu^2 - |\vec{p}_\nu|^2 + E_p^2 - 2E_\nu E_p + 2\vec{p}_\nu \cdot \vec{p}_p - |\vec{p}_p|^2$$

$$m_\ell^2 + m_n^2 - 2E_\ell m_n = m_p^2 - 2E_\nu E_p + 2\vec{p}_\nu \cdot \vec{p}_p \quad (7.4)$$

$$m_\ell^2 + m_n^2 - 2E_\ell m_n = m_p^2 - 2E_\nu E_p + 2E_\nu |\vec{p}_p| cos\theta_p$$

$$E_\nu = \frac{E_\ell m_n + \frac{1}{2}(m_p^2 - m_n^2 - m_\ell^2)}{E_p - |\vec{p}_p| cos\theta_p}$$

Using the fact that the neutrino momentum was aligned with the z-axis, and that $|\vec{p}_\nu| \approx E_\nu$, Eq.7.3 becomes $E_\nu = |\vec{p}_p|cos\theta_p + |\vec{p}_\ell|cos\theta_\ell$. Subtracting this from Eq.7.2 yields $E_p - |\vec{p}_p|cos\theta_p = m_n + |\vec{p}_\ell|cos\theta_\ell - E_\ell$. Substituting this into the denominator above yields an equation solely in terms of the reconstructed lepton. Furthermore, the above procedure would proceed equally well in terms of either final state particle $p \leftrightarrow \ell$. This yields two equations, either of which can be used to obtain the initial neutrino energy in terms of the 4-momentum of the proton or lepton. We will call these $E_\nu^{QE-p}$ and $E_\nu^{QE-\ell}$ hereafter.

$$E_\nu^{QE-\ell} = \frac{E_\ell m_n + \frac{1}{2}(m_p^2 - m_n^2 - m_\ell^2)}{m_n + |\vec{p}_\ell|cos\theta_\ell - E_\ell} \quad (7.5)$$

$$E_\nu^{QE-p} = \frac{E_p m_n + \frac{1}{2}(m_\ell^2 - m_n^2 - m_p^2)}{m_n + |\vec{p}_p|cos\theta_p - E_p} \quad (7.6)$$

And in fact we have already seen a third equation, 7.2, which requires no directional information but requires that we have reconstructed both the lepton and proton. Hereafter we will refer to this as $E_\nu^{range}$.

$$E_\nu^{range} = E_p + E_\ell - m_n \quad (7.7)$$

Because these three formulae in principle reconstruct the same quantity, it will prove useful to consider the following quantity which captures their level of mutual agreement. The smaller the value of this parameter, the better the agreement between the quasi-elastic energies. We will hereafter refer to this as the quasi-elastic scattering consistency.

$$\Delta^{QE} = \sqrt{\left(E_\nu^{QE-p} - E_\nu^{QE-\ell}\right)^2 + \left(E_\nu^{QE-p} - E_\nu^{range}\right)^2 + \left(E_\nu^{QE-\ell} - E_\nu^{range}\right)^2} \quad (7.8)$$

In principle, these three formulae are exactly equivalent for the quasi-elastic scattering of a neutrino off of a free neutron. In other words, the idealized quasi-elastic scattering consistency value is $\Delta^{QE} = 0$. However, we must account for the fact that the neutron exists within a nuclear environment. This introduces several complications.

Firstly, the neutron is not free. A certain amount of energy must be expended to remove it from the nucleus. In reality this is not a fixed value within a given nucleus. However the differences are small $\mathcal{O}(10 \text{ MeV})$. So at the energy scales of interest to this analysis we approximate it as a fixed value capturing the mean removal energy. A value of $E_b = 28.5$ MeV is used. This acts as a reduction to the neutron's effective scattering mass $m_n \to m_n - E_b$ in Eqs. 7.5-7.7.

Secondly, we have assumed that the momenta of the outgoing lepton and proton which are measured exactly reflect those which emerged from the scatter. Because of the potential for final state interactions within the nucleus this will not always be true. We do not introduce a correction to account for this. Rather we will proceed under the assumption that these final state interactions are small. There will be a population of interactions for which this is false. Our equations will accordingly show tension with each other in situations such as this. The decision to neglect final state interactions is thus a tacit shift in the sought after signal to be quasi-elastic interactions which have small final state interactions.

Finally, we deduced these equations under the assumption that the struck nucleon was at rest. Because of Fermi motion this is also not necessarily true. An approach to help minimize this is discussed in the following section.

## 7.4 Strengthening Variables With Boosting to the Nucleon-at-Rest Frame

Variables associated with quasielastic scattering are generally derived in a picture in which a neutrino scatters off of a bound nucleon at rest. Using these formulae with values reconstructed in the lab frame works to a first approximation, but doing so neglects the initial Fermi motion of the nucleon which was struck and thus introduces tacit approximations. If these formulae could be properly computed in the rest frame of the nucleon then their power and accuracy can be increased.

The Fermi motion is random, and a priori cannot be known. However, our signal of interest comprises only quasielastic interactions. If final state interactions are small, we then have sufficient reconstructible information to estimate the Fermi momentum if the momenta of all final state particles are reconstructed and so long as the initial direction of the neutrino beam is known. This follows from a consideration of the kinematics of the scattering.

$$\begin{aligned}
\vec{p}_\nu + \vec{p}_{fermi} &= \vec{p}_{final} \\
\vec{p}_\nu + \vec{p}_{fermi} &= \sum_{particles} \vec{p} \\
\langle 0,0,p_\nu \rangle + \langle p_f^x, p_f^y, p_f^z \rangle &= \langle p_p^x, p_p^y, p_p^z \rangle + \langle p_\ell^x, p_\ell^y, p_\ell^z \rangle \\
\to \vec{p}_f &= \langle p_p^x + p_\ell^x, p_p^y + p_\ell^y, p_p^z + p_\ell^z - p_\nu \rangle \\
\to \vec{p}_f &\approx \langle p_p^x + p_\ell^x, p_p^y + p_\ell^y, p_p^z + p_\ell^z - E_\nu \rangle
\end{aligned} \quad (7.9)$$

This allows us to estimate all three components of the Fermi momentum using reconstructible quantities. This, combined with the energy of the struck nucleon $E_n$, can then be used to define the boost vector $\vec{\beta}$. The value of $E_n$ in this case is the off shell mass given by Eq. 7.10 in which $m_n$ is the on shell nucleon mass, $E_b$ is the removal energy, and $T_f$ is the final state nuclear recoil kinetic energy.

$$E_n = m_n - E_b - T_f$$
$$\text{use} T_f \ll (m_n - E_b) \tag{7.10}$$
$$\approx m_n - E_b$$

The boost vector $\beta$ is then given by Eq. 7.11

$$\vec{\beta} = \frac{\vec{p}_f}{E_N} \tag{7.11}$$

It then becomes possible to boost all reconstructed 4-momenta into the struck nucleon rest frame. These boosted vectors can then be used to compute derived kinematic quantities which will be more powerful in this frame.

$$\begin{pmatrix} E' \\ p'_x \\ p'_y \\ p'_z \end{pmatrix} = \begin{pmatrix} \gamma & -\gamma\beta_x & -\gamma\beta_y & -\gamma\beta_z \\ -\gamma\beta_x & 1 + \frac{\gamma-1}{\beta^2}\beta_x^2 & \frac{\gamma-1}{\beta^2}\beta_x\beta_y & \frac{\gamma-1}{\beta^2}\beta_x\beta_z \\ -\gamma\beta_y & \frac{\gamma-1}{\beta^2}\beta_x\beta_y & 1 + \frac{\gamma-1}{\beta^2}\beta_y^2 & \frac{\gamma-1}{\beta^2}\beta_y\beta_z \\ -\gamma\beta_z & \frac{\gamma-1}{\beta^2}\beta_x\beta_z & \frac{\gamma-1}{\beta^2}\beta_y\beta_z & 1 + \frac{\gamma-1}{\beta^2}\beta_z^2 \end{pmatrix} \begin{pmatrix} E \\ p_x \\ p_y \\ p_z \end{pmatrix} \tag{7.12}$$

For example, in the prior section we saw that for CCQE interactions we can compute the initial neutrino energy in three distinct ways. However the formulae assumed we were in the nucleon rest from. Performing this boost minimizes the detrimental impact of that assumption. As illustrated in Fig. 7-2, for a simulated CCQE sample boosting provides reconstructed energies which are in tighter agreement with each other.

## Correlations Between Energy Reconstruction Equations With and Without Boosting

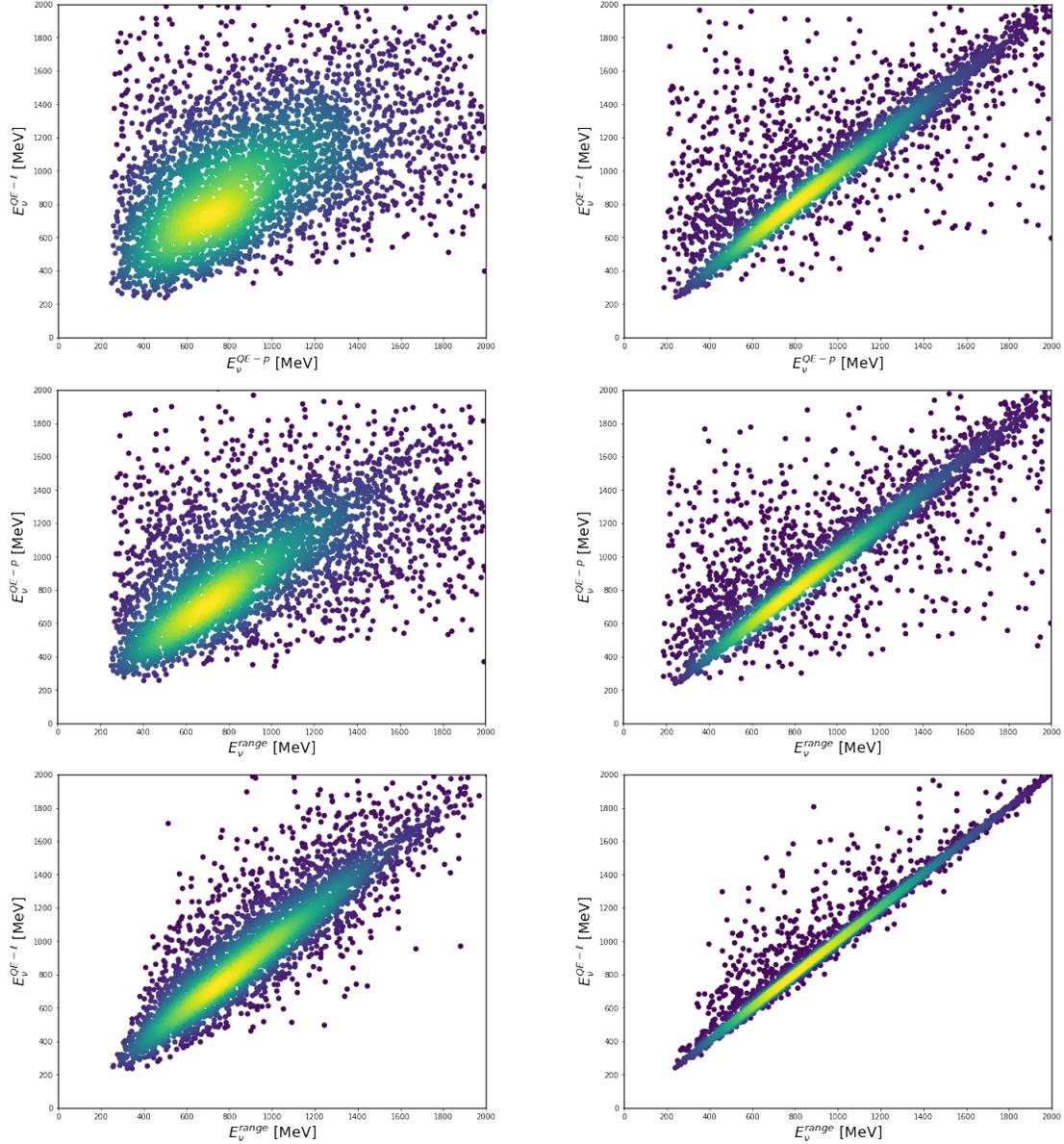

Figure 7-2: Correlations between the three different equations used to compute the reconstructed neutrino energy, $E_\nu^{range}$, $E_\nu^{QE-p}$, & $E_\nu^{QE-\ell}$. These are computed using truth level kinematic values for a simulated sample of CCQE interactions with 1 proton and 1 muon in the final state. These are illustrated prior to (left) and after (right) boosting into the frame in which the struck nucleon was at rest.

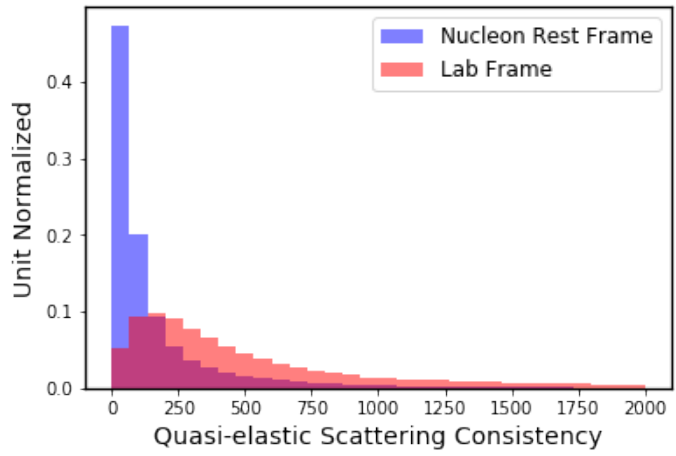

Figure 7-3: Improvement produced in $\Delta^{QE}$ by boosting for a simulated sample of CCQE $1\mu 1p$ interactions.

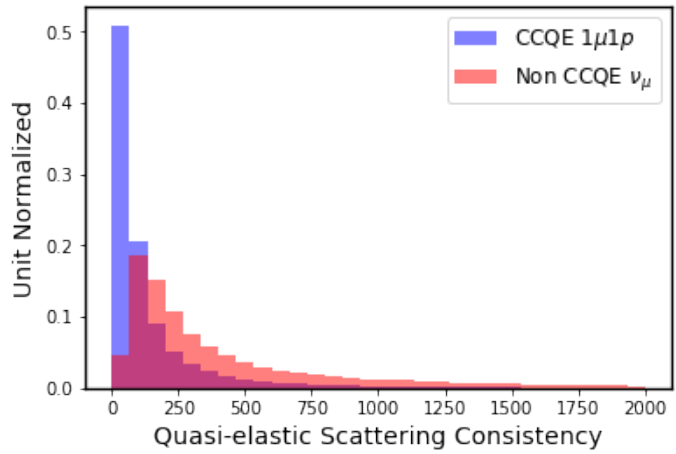

Figure 7-4: Comparison of the boosted $\Delta^{QE}$ variable for a simulated sample of CCQE $1\mu 1p$ interactions compared to other $\nu_\mu$ interactions that would constitute background.

## 7.5 $1e1p$ Selection

As discussed previously, the $1e1p$ selection comprises three stages. The exact cuts and procedures employed in each of these stages are detailed here.

### 7.5.1 Pre Selection

The preselection is the initial set of post reconstruction cuts applied. These cuts enforce some basic consistency with our anticipated signal patterns and ensure that variables which will be used in the upcoming BDT are always well defined. e.g. we will require two particles to be present, certain BDT variables are only defined for a two particle situation. These cuts are enumerated below

- Exactly two prongs are reconstructed, both must be >5 cm in length.

- The vertex must lie inside a fiducial volume. This is defined by a 10cm buffer on all boundaries of the active volume. A region on the collection plane with a large number of dead wires is also omitted. Specifically, the fiducial volume is defined by:

$$10 < x < 246.25$$
$$-106.5 < y < 106.5$$
$$10 < z < 700 \text{ or } 740 < z < 1026.5$$

- The event must reconstruct as contained. This is enforced by looking at how close both of the prongs approach to the edge of the detector. A prong may not be tracked to within 15 cm of the edge of the active volume. A prong also may not enter the region between $y = \frac{1}{\sqrt{3}}z - 117$ and $y = \frac{1}{\sqrt{3}}z - 80$ which corresponds to a region of dead induction plane wires leading to an increased probability of poorer reconstruction.

- The electron must reconstruct with energy > 35 MeV.

- The proton must reconstruct with energy > 60 MeV.

- The 3D opening angle between the particles must be $> 0.5$ radians.

- A physical boost to the neutron rest frame with $0 < \beta < 1$ must be possible. The computation is detailed in Section 7.4.

- The reconstructed electron shower energy on all three planes must match within a certain tolerance. Specifically

$$\frac{\sqrt{(E_e^U - E_e^V)^2 + (E_e^U - E_e^Y)^2 + (E_e^V - E_e^Y)^2}}{E_e^Y} < 2 \qquad (7.13)$$

- The proton must reconstruct as forward going, $\theta_p < \pi/2$

These initial cuts eliminate 98.9% of cosmic background vertices, 98.5% of $\nu_\mu$ background vertices, and retain 59% of CCQE $1e1p$ interactions below 800 MeV (the LEE regime) that were vertexed within the fiducial volume. Of the 41% of signal interactions that are cut at this stage, 29% are eliminated because they were uncontained. An illustration of these efficiencies for this cut and the subsequent two stages of cut vs true energy is illustrated in Fig. 7-7.

## 7.5.2 BDT Cut

The BDT is used to further discriminate against the remaining backgrounds is trained using the subset of the following samples which has already passed the precuts.

1. Cosmic data

2. Simulated $\nu_\mu$ with cosmic background data overlaid

3. Simulated CCQE $1e1p$ interactions

The CCQE $1e1p$ simulation provides the training signal class. Only vertices within this sample which were located on the true interaction point, and which reconstructed with energy resolution better than 20%. Cosmics and $\nu_\mu$ conglomerated provide the background class. Both background and signal samples were split 50/50 into

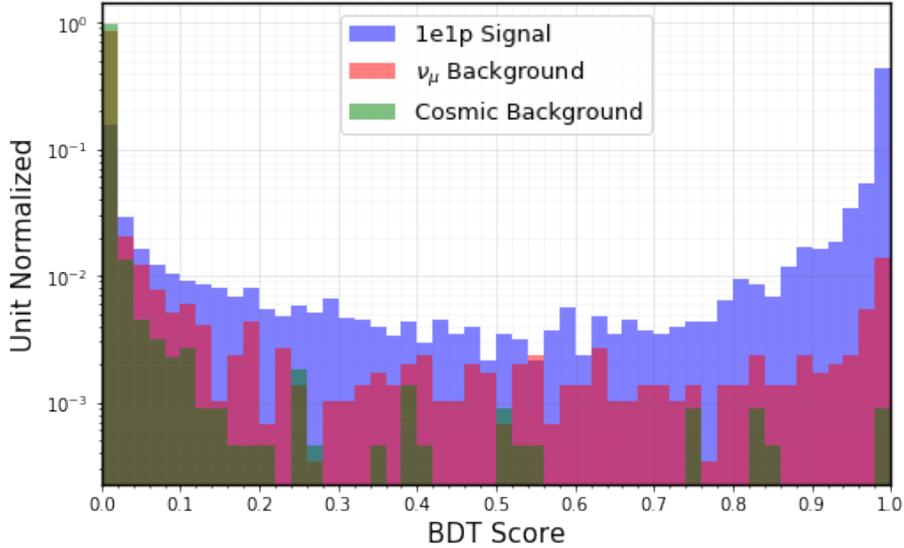

Figure 7-5: The signal probability for validation samples of signal, $\nu_\mu$ background, and cosmic background as predicted by the trained selection BDT.

training / validation samples. The hyperparameters of the BDT training were tuned to maximize signal / background in the region of BDT confidence > 80%. This tuning was subject to the constraint that the accuracy on the training vs validation samples may not differ by more than 2%

The variables used to train this BDT are summarized in Table 7.3. The relative importance of each variable is summarized in Fig. 7-6. Once trained, the BDT can take a tuple of these variables for any reconstructed event and output a probability that the event is the signal class. The results of this as derived from the validation sample are illustrated in Fig. 7-5

There is an inherent trade off between purity and efficiency as harsher cuts on this probability are used. The cut position was chosen via optimization for sensitivity to the low energy excess signal and is placed at 0.7. This increases the rejection of cosmic vertices to 99.99%, $\nu_\mu$ backgrounds are rejected at 99.9%, and 45.9% of CCQE $1e1p$ interactions below 800 MeV (the LEE regime) that were vertexed within the fiducial volume are retained.

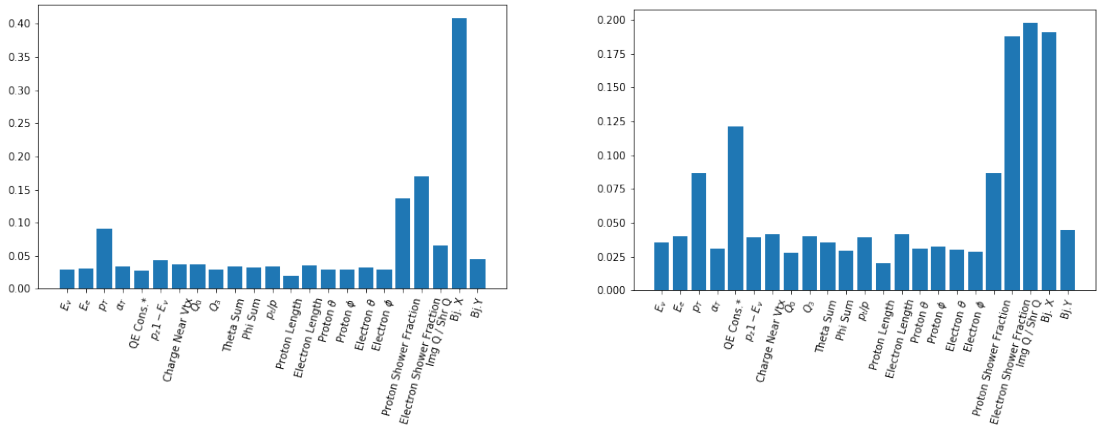

Figure 7-6: Importance of different training variables for the $1e1p$ BDTs. Training associated for simulations prior to detector electronics upgrades (left) and with training after upgrades (right).

### 7.5.3 PID Cuts

The final tier of the selection cuts specifically targets the existence of particles which should never be present in our signal. To do this, we use results from the MPID (Section 6.9) and from the $\pi^0$ reconstruction (Section 6.8.1). The following cuts are applied.

1. If two showers are reconstructed, compute the invariant mass. If this is $> 50$ MeV then veto the event.

2. Require the MPID have a muon score from the interaction image which is $<0.2$. This cut is only used for events with an electron candidate above 100 MeV.

3. Require that the MPID have an electron and photon score from the whole image such that $\frac{\text{electron score}}{\text{photon score}} > 2$

The first cut explicitly checks for a feature that reconstructs like a $\pi^0$. There is no situation in which our desired signal should have this signature. The final cut approaches the $\pi^0$ problem with a different tool, the MPID. Recall that the whole image score uses charge both attached to and distinct from the vertex. It will thus see the presence of detached photons from a $\pi^0$. If, when analyzing the whole image, the MPID is not at least twice as confident that an electron is present rather than than a photon, the event is eliminated. The MPID muon cut removes any CC$\nu_\mu$ interactions which have persisted. It also assists with the rare case in which a cosmic muon pierces close enough to the vertex so as to be clustered along with the interaction.

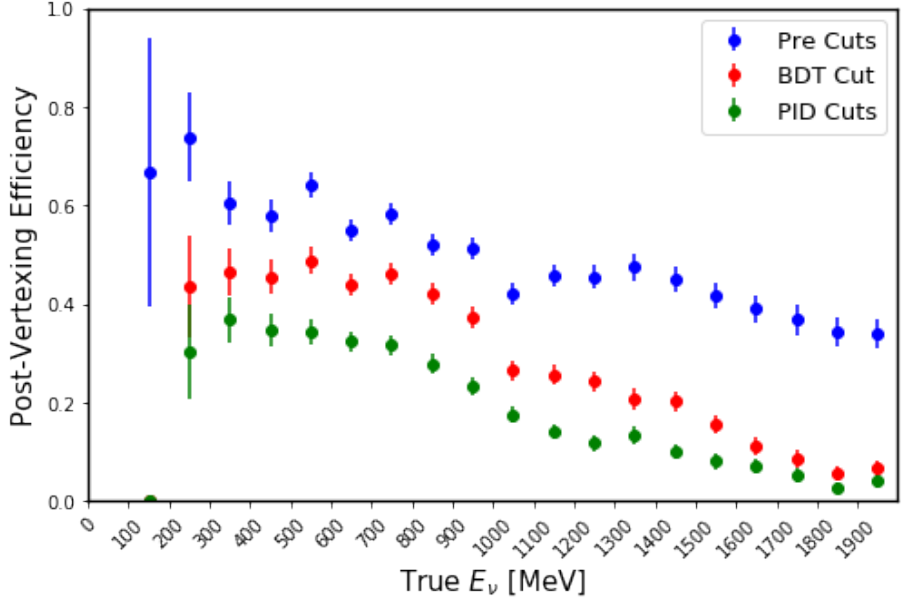

Figure 7-7: The efficiencies of the three tiers of selection cuts on a sample of simulated CCQE, $1e1p$ interactions in the LEE region of interest ($< 800$ MeV). Efficiencies are illustrated vs true $E_\nu$

The addition of these cuts successfully eliminates 100% of cosmics within available samples giving us confidence that actual rejection is >99.999%. $\nu_\mu$ backgrounds are rejected at >99.99%. 33.1% of CCQE $1e1p$ interactions below 800 MeV (the LEE regime) that were vertexed within the fiducial volume are retained. If, at this stage, a single event contains more than one reconstructed event which has passed all cuts, that with the higher BDT score is selected for that event.

## 7.6   $1\mu 1p$ Selection

### 7.6.1   Pre Selection

The preselection is the initial set of post reconstruction cuts applied. These cuts enforce some basic consistency with our anticipated signal patterns and ensure that variables which will be used in the upcoming BDT are always well defined. e.g. we will require two particles to be present, certain BDT variables are only defined for a two particle situation. These cuts are enumerated below

- Exactly two prongs are reconstructed, both must be >5 cm in length.

- The vertex must lie inside a fiducial volume. This is defined by a 10cm buffer on all boundaries of the active volume. A region on the collection plane with a large number of dead wires is also omitted.

$$10 < x < 246.25$$
$$-106.5 < y < 106.5$$
$$10 < z < 700 \text{ or } 740 < z < 1026.5$$

- The event must reconstruct as contained. This is enforced by looking at how close both of the prongs approach to the edge of the detector. A prong may not be tracked to within 15 cm of the edge of the active volume.

- The muon must reconstruct with energy > 35 MeV.

- The proton must reconstruct with energy > 60 MeV.

- The 3D opening angle between the particles must be > 0.5 radians.

- A physical boost to the neutron rest frame with $0 < \beta < 1$ must be possible. The computation is detailed in Section 7.4.

- The proton must reconstruct as forward going, $\theta_p < \pi/2$

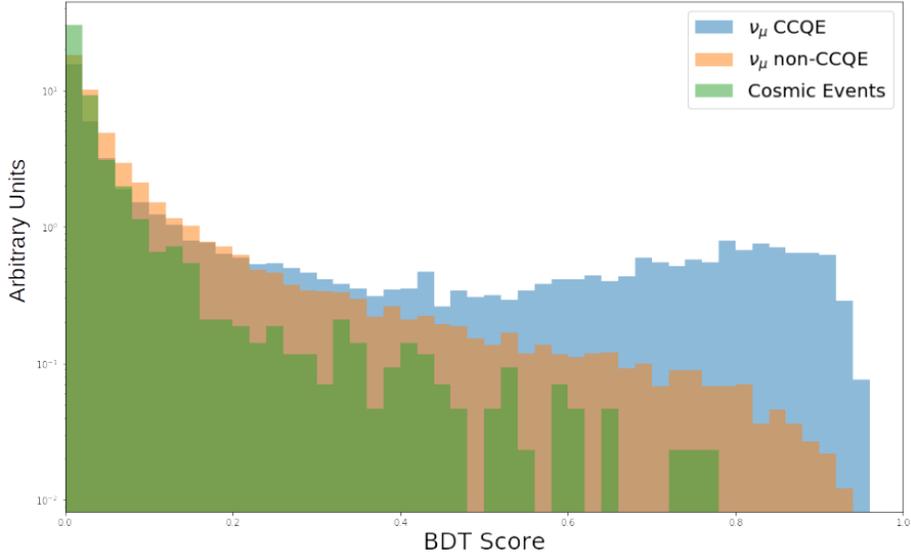

Figure 7-8: The signal probability for validation samples of signal, $\nu_\mu$ background, and cosmic background as predicted by the trained selection BDT.

### 7.6.2 BDT Cut

The BDT used to further discriminate against the remaining backgrounds is trained using the subset of the following samples which has already passed the precuts.

1. Cosmic data

2. Simulated $\nu_\mu$ with cosmic background data overlaid

The subset of the $\nu_\mu$ sample which is true CCQE, $1\mu1p$, located on the true interaction point, and has a reconstructed energy within 20% of the true value forms the signal training sample. Cosmics and all other $\nu_\mu$ interactions form the background.

The variables used to train this BDT are summarized in Table 7.3. The relative importance of each variable is summarized in Fig. 7-9 Once trained, the BDT can take a tuple of these variables for any reconstructed event and output a probability that the event is the signal class. The results of this as derived from the validation sample are illustrated in Fig. 7-8. Events which have a score >0.4 will be retained for this selection.

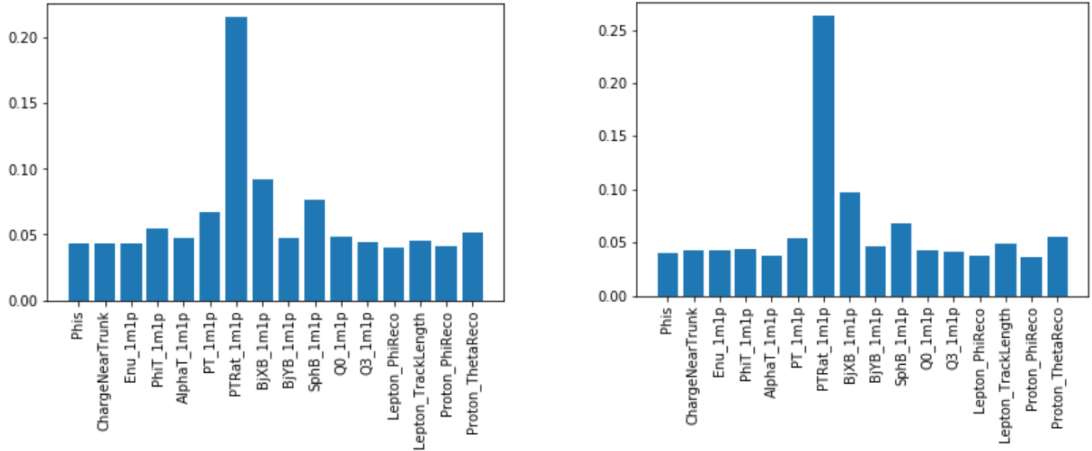

Figure 7-9: Importance of different training variables for the $1\mu 1p$ BDTs. Training associated for simulations prior to detector electronics upgrades (left) and with training after upgrades (right).

### 7.6.3 PID Cut

The final tier of the selection cuts uses only one PID cut which is distinct from any of those used in the $1e1p$ selection. It targets backgrounds at the lowest energies, $< 400$ MeV. Using the MPID network, if an event reconstructs with a neutrino energy less than 400 MeV then a proton probability $>80\%$ is required.

| Name/Description of Code | Requirement for $1\mu 1p$ | Requirement for $1e1p$ |
|---|---|---|
| **Cuts/Selections Applied in External Code** | | |
| Common Optical Filter | > 20 PMT hits in 2 adjacent 6-tick bins in beam window 6-tick bins in beam window | Same |
| WC tagger | Mask cosmic-associated charge | Same |
| **Preselection Cuts for Event Selection** | | |
| Run Quality Requirements: | | |
| Good Run Flag | True | Same |
| Beam Quality Flag | True | Same |
| Vertex Requirements: | | |
| $1\ell 1p$ consistent | Two prongs of 5cm length, no additional prongs | Same |
| Edge fiducial | >10 cm from active volume edge | Same |
| No dead region | outside $z = 700$ to $740$ cm | Same |
| Containment Requirements: | | |
| Edge containment | closest approach >15 cm from edge | Same |
| Efficient region | N/A | outside $y = (1/\sqrt{3}) * z - 117$ to $y = 1/\sqrt{3} * z - 80$ cm. |
| Particle Energy Requirements: | | |
| Lepton energy | > 35 MeV | Same |
| Proton energy | > 60 MeV | Same |
| Analysis Orthogonality Cut | Max Shower Fraction < 0.2 | Max Shower Fraction > 0.2 |
| Other Basic Quality Requirements: | | |
| Opening Angle | > 0.5 radians | Same |
| Proton $\theta$ | < $\pi/2$ | Same |
| Boostable | $0 < \beta < 1$ | Same |
| Shower 3-view energy consistency | N/A | Inconsistency < 140% |
| Vertex with highest BDT score | $\nu_\mu$ BDT | $\nu_e$ BDT score |
| **BDT Requirements for Event Selection** | | |
| $\nu_\mu$ BDT > 0.4 | $\nu_e$ BDT > 0.9 | |
| **PID Requirements for Event Selection** | | |
| $\pi^0$ mass | N/A | $\pi^0$ mass < 50 MeV |
| muon MPID image score | N/A | < 20% if $E_e$ > 100 MeV |
| $\gamma/e$ image score ratio | N/A | < 2 |
| proton MPID interaction score | > 80% if $E_\nu$ < 400 MeV | N/A |

Table 7.2: Three categories of requirements used in the $1\mu 1p$ and $1e1p$ selections.

| Variable | Used in 1$\mu$1p BDT | Used in 1e1p BDT |
|---|---|---|
| Variables Used in BDTs, Based on Ionization | | |
| Charge within 5 cm of vertex | Yes | Yes |
| Shower charge in event image / shower charge clustered as electron | No | Yes |
| proton shower fraction | No | Yes |
| Electron shower fraction | No | Yes |
| Variables Used in BDTs, Related to Energy Measurements | | |
| Neutrino Energy | Yes | Yes |
| Energy of electromagnetic shower | No | Yes |
| Lepton length | Yes | Yes |
| Proton length | No | Yes |
| $p_z - E_\nu$ | No | Yes |
| Variables Used in BDTs, Related to 2-Body Scattering Consistency | | |
| Bjorken's $x$ | Yes * | Yes * |
| Bjorken's $y$ | Yes * | Yes * |
| QE Consistency | Yes * | Yes * |
| $Q_0$ | Yes | Yes |
| $Q_3$ | Yes | Yes |
| Variables Used in BDTs, Related to Transverse Momentum | | |
| $\alpha_T$ | Yes | Yes |
| Event $p_T$ | Yes | Yes |
| Event $p_T/p$ | Yes | Yes |
| $\phi_T$ | Yes | No |
| Variables Used in BDTs, Related to Angles | | |
| Proton $\phi$ | Yes | Yes |
| Proton $\theta$ | Yes | Yes |
| Lepton $\phi$ | Yes | Yes |
| $\phi_p - \phi_\ell$ | Yes | Yes |
| $\theta_p + \theta_e$ | No | Yes |
| Variables Useful for Comparison, Not Used in Either BDT | | |
| $\eta$ (Normalized average ionization difference) | No | No |
| Maximum Shower Pixel Fraction | No | No |
| Minimum Shower Pixel Fraction | No | No |
| Opening Angle | No | No |
| $x$ Vertex | No | No |
| $y$ Vertex | No | No |
| $z$ Vertex | No | No |

Table 7.3: Summary of variables investigated. Variables specifically used in the 1$\mu$1p BDTs and the 1e1p BDT are noted. The analysis has been designed substantial variable overlap. If a * appears, the variable is used in the boosted frame. The mathematical definitions of many of these variables appear in Table 7.1.

| Metric | value |
|---|---|
| Average muon energy resolution | 2.7% |
| Average muon angle resolution | 3° |
| Average proton energy resolution | 2.6% |
| Average proton angle resolution | 4° |
| Average $\nu_\mu$ energy resolution | 1.8% |
| Average Electron energy resolution | 16.4% |
| Average Electron angle resolution | 3.7°% |
| Average $\nu_e$ energy resolution | 16.5% |

Table 7.4: Metrics for energy reconstruction for events passing the $1\mu1p$ and $1e1p$ selections

# Chapter 8

# Systematics

While the sensitivity of our analysis will be statistics limited, we are still subject to various important systematic uncertainties. These fall into three categories:

- Uncertainties about the beam
- Uncertainties about the $\nu - Ar$ interaction cross section
- Uncertainties about the detector

It is further useful to group these systematics into two categories. Reweightable and non-reweightable. These are so called based on whether or not the systematic introduces new types of events (non-reweightable) or simply alters the probability distribution of events that would already exist (reweightable). Flux, cross section, and Geant4 related systematics are typically reweightable. Detector systematics, however, generally produce new types of events which cannot be weighted into existence from the central value. These must be simulated using designated samples.

For all of these uncertainties, the ultimate metric of interest is the correlations between analysis bins and samples. To capture these correlations a covariance matrix is built. This is defined as in Eq. 8.1:

$$\mathcal{M}_{ij} = \frac{1}{\mathcal{N}} \sum_{k}^{\mathcal{N}} \left( N_i^{CV} - N_i^k \right) \left( N_i^{CV} - N_i^k \right) \tag{8.1}$$

Where $\mathcal{M}_{ij}$ is the ij$^{th}$ element of the covariance matrix, $\mathcal{N}$ is number of different systematic variations explored. $N_i^{CV}$ is the number of events in the $i^{th}$ bin of the central value prediction/ $N_i^k$ is the number of events in the $i^{th}$ bin for the $k^{th}$ variation.

Because we often must deal with simulations and data sets of variable size, it is often useful to rewrite the covariance matrix in a fractional form as in Eq. 8.2:

$$\mathcal{F}_{ij} = \frac{\mathcal{M}_{ij}}{N_i^{CV} N_j^{CV}} \qquad (8.2)$$

## 8.1 Flux Systematics

Systematics associated with the beam flux are reweightable. [29] Flux systematics are all linked to the properties of the hadron that decayed to produce the neutrino initially. These fall into several categories:

1. Hadron production at the target of $\pi^{\pm}, K^{\pm}, \& K_L^0$

2. Modeling of the focusing horn

3. Secondary hadron interactions within the target.

Where the production of $\pi^+$ is the largest as it is the primary source of neutrinos within the beam, especially in the low energy regions of interest to this analysis. An illustration of the impact on the $1e1p$ and $1\mu 1p$ flux is illustrated in Fig. 8-1

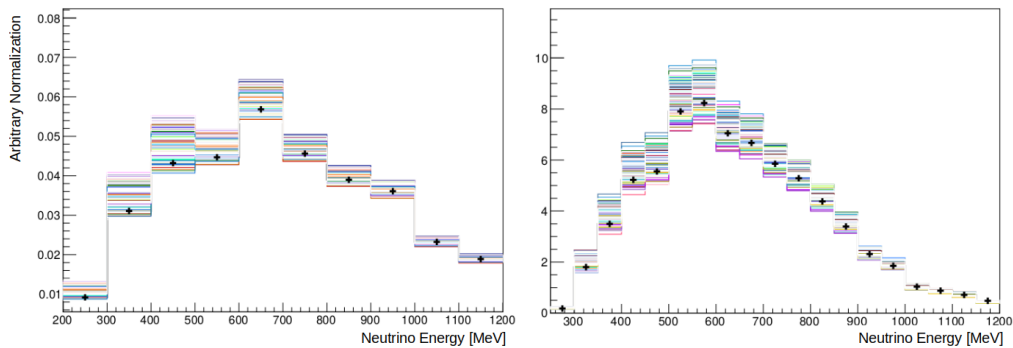

Figure 8-1: The impact on $1e1p$ flux (left) and $1\mu 1p$ flux (right) for variations in the $\pi^0$ production rate on the BNB target. This is the leading flux systematic.

## 8.2  Cross Section Systematics

Cross section systematics are all linked to uncertainties related to the interaction of the neutrino with the argon nucleus. [36] This too is a reweightable systematic in which the reweighting is carried out as a function of the relevant truth level parameters. This reweighting is primarily carried out using architecture provided within GENIE, augmented by some additional MicroBooNE specific code to reweight parameters not supported explicitly by GENIE.

This machinery provides a variety of "knobs" that can be dialed to alter the interaction model. These include knobs related to CCQE, NC elastic, MEC, resonant, coherent, non-resonant pion, and DIS channels. There are also knobs associated with final state interactions (FSI) which impact the interactions of the daughter particles after the initial neutrino interaction but before exiting the nucleus. The majority of these parameters are varied in a correlated way to produce a single knob which we refer to as the GENIE ALL knob. As it encapsulates a large number of correlated variations, it will be the leading order systematic. Several other have only two discrete recommended parameters and are individually varied. These include:

- **RPA_CCQE**: Strength of the RPA effect

- **AxFFCCQEshape**: Shape of the CCQE cross section due to the axial form factor

- **VecFFCCQEshape**: Shape of the CCQE cross section due to the vector form factor

- **DecayAngMEC**: Angular distribution of decaying nucleon pairs in MEC events

- **Theta_Delta2Npi**: Fraction of MEC events in the $\Delta$-like MEC channel

- **NormCCCOH**: Normalization of CC coherent pion production

- **NormNCCOH**: Normalization of NC coherent pion production

- **ThetaDelta2Rad**: Angular distribution variation for radiative $\Delta$ decays

An illustration of the impact on the $1e1p$ and $1\mu 1p$ reconstructed energy spectra is illustrated in Fig. 8-2.

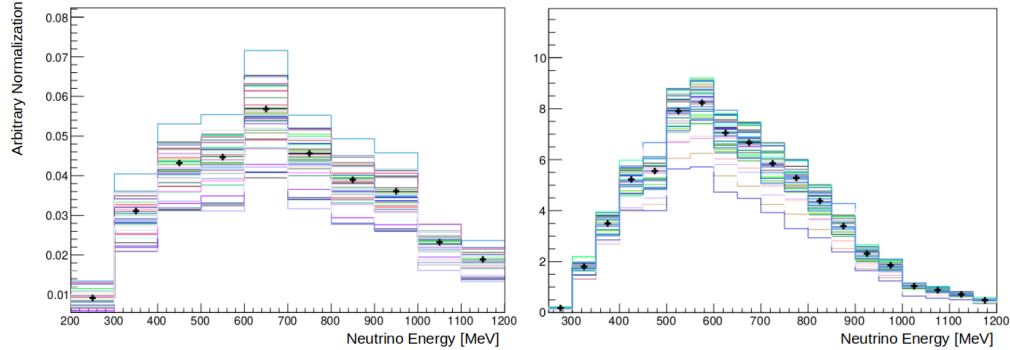

Figure 8-2: The impact on $1e1p$ (left) and $1\mu 1p$ (right) selected energy spectra for variations in the GENIE ALL cross section systematics knob. This is the leading interaction systematic considered. The 100 variation universes used in this systematic analysis are illustrated.

As of the time of writing this thesis, the collaboration had not decided on the implementation of several systematic uncertainties related to the MEC events. We do not expect a significant change to the results of this analysis when these are added, because MEC events end up representing less than 5% of the selected sample.

## 8.3 Hadron Re-interaction Systematics

Hadron re-interaction systematics are linked to uncertainties in the behavior of daughter hadrons within the liquid argon. This is a reweightable systematic linked to daughter particle truth information. Geant4 describes the trajectories of the daughter particles through the argon and secondary interactions may occur. Of particular interest are variations in the re-interaction behavior of protons and charged pions. Given the exclusive topology used by this analysis, we find this source of uncertainty to be significantly subdominant.

## 8.4 Detector Systematics

Detector systematics cannot be reweighted from existing events because certain variations can introduce entirely new types of events which does no exist in the absence of the systematic. As a result, different simulations with the variations explicitly incorporated into the detector simulation are used. [53] The primary systematics fall into three categories:

1. Wire Modification (WireMod): These produce changes in the wire waveforms. They are motivated and informed by discrepancies between data and simulation in the detector response. WireMod samples are a family of different variations which are variously parameterized in terms of x, y, z, $\theta_{XZ}$, $\theta_{YZ}$ and the dE/dx of deposited charge.

2. Other TPC Variations: These capture modifications to the drift simulation itself, not to the wire response. This includes modifications to the space charge modeling.

3. Optical Variations: These capture changes to the light simulation. Variations in absolute light yield and the Rayleigh scattering length are included.

Each of these detector variation samples is identical up to the indicated detector systematic. That is, all of them contain the same GENIE generated interactions and the same Geant4 propagation simulations.

Due to limitations in computation resources, the detector systematic simulation samples available to this analysis are themselves statistically limited. Therefore when evaluating detector systematics on a given distribution a single bin evaluation is performed. This allows sufficient statistics to characterize the systematic uncertainty on normalization alone. However, we believe this to be sufficient as studies on the energy smearing introduced by the detector systematics indicate that they are significantly smaller than the bin widths used, leaving shape uncertainties subdominant to normalization uncertainties. Applying this procedure yields a 9.1% detector systematic on the $1\mu 1p$ selected results and a 1.9% for the $1e1p$.

## 8.5 Systematics Results

All systematics discussed previously in this chapter: flux, cross section, hadron reinteraction, and detector systematics treated as as a flat effect on normalization uncertainty have been implemented for both $1\mu1p$ and $1e1p$ selected samples. Using the binning that will be used in the final sensitivity analysis, discussed more in Section. 9, we compute the full fractional covariance matrix including both selections. This is illustrated in Fig. 8-3. The correlations visible in the off diagonal blocks are crucial as it is these which will be used to constrain the systematics on the final $1e1p$ prediction prior to estimating signal sensitivity. The impact of this is evident in Table. 8.1 which breaks down the systematic uncertainties by category and shows the reduction in uncertainty due to the $1\mu1p$ constraint.

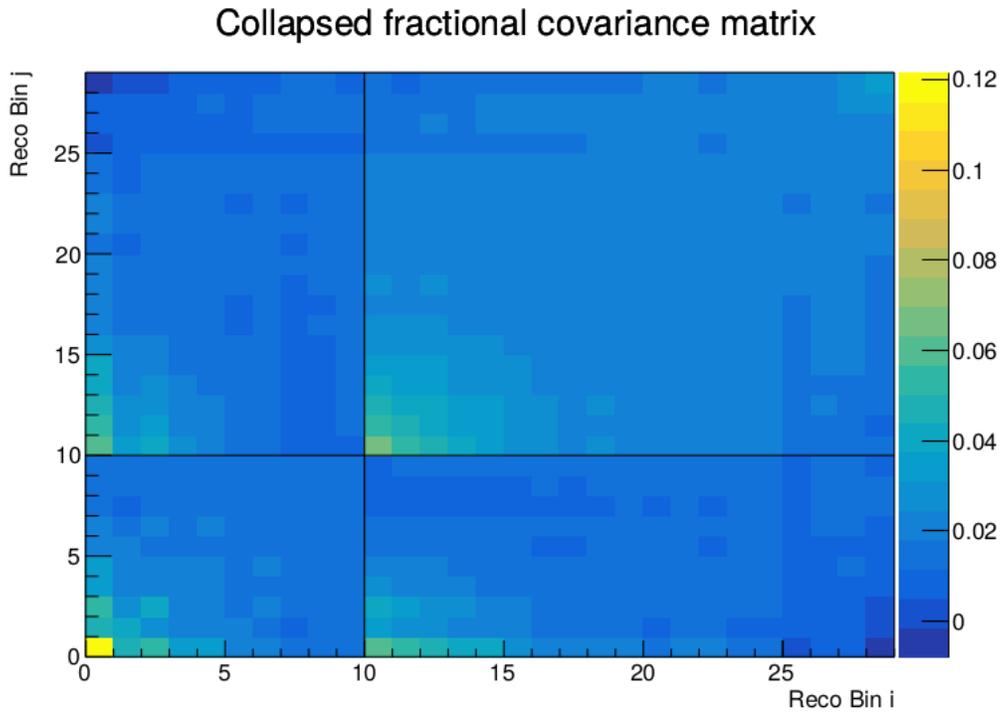

Figure 8-3: Fractional covariance matrix for flux, cross section, hadron re-interaction, and flat detector systematics. Captures both $1e1p$ and $1\mu1p$ selections. The $1e1p$ block is in the lower left, the $1\mu1p$ in the upper right. The off diagonal blocks capture the correlations between the two.

| 1e1p bin | Flux | XSec | Geant4 | Detector | Total | Total Constrained |
|---|---|---|---|---|---|---|
| 200–300 MeV | 0.230 | 0.261 | 0.015 | 0.019 | 0.349 | 0.241 |
| 300–400 MeV | 0.111 | 0.167 | 0.059 | 0.019 | 0.209 | 0.159 |
| 400–500 MeV | 0.129 | 0.149 | 0.011 | 0.019 | 0.198 | 0.130 |
| 500–600 MeV | 0.080 | 0.123 | 0.026 | 0.019 | 0.150 | 0.106 |
| 600–700 MeV | 0.067 | 0.121 | 0.016 | 0.019 | 0.141 | 0.098 |
| 700–800 MeV | 0.053 | 0.118 | 0.026 | 0.019 | 0.133 | 0.103 |
| 800–900 MeV | 0.053 | 0.119 | 0.015 | 0.019 | 0.133 | 0.096 |
| 900–1000 MeV | 0.052 | 0.106 | 0.009 | 0.019 | 0.120 | 0.090 |
| 1000–1100 MeV | 0.054 | 0.105 | 0.005 | 0.019 | 0.120 | 0.088 |
| 1100–1200 MeV | 0.058 | 0.113 | 0.018 | 0.019 | 0.129 | 0.096 |

Table 8.1: Fractional uncertainties for $1e1p$ events due to flux, cross section, hadron re-interactions, and detector systematics. A flat 1.9% detector uncertainty has been used. The last two columns provides the total fractional systematic uncertainty before and after the correlations with the $1\mu1p$ sample are used to constrain.

# Chapter 9

# Data Validations & Anticipated Sensitivity

MicroBooNE is performing a blind analysis. At the time of writing this thesis, the "box" has not been opened to give access to the data. However, there are open data sets, as well as a method of accessing the signal-region using blind-safe plots. These are used to demonstrate that the analysis is on a firm ground.

After providing the data validations, at the end of this chapter, we present the final result, which is the predicted energy spectrum and sensitivity of this analysis to the MiniBooNE Low Energy Excess.

## 9.1 Explanation of Data-Simulation Comparison Plots

The primary goal of the following sections in this chapter is is to benchmark the agreement between data and simulation using various available data samples. Here we briefly describe the method used to present the uncertainties and to characterize the agreement between a predicted spectrum and observation.

In general, the plots will be presented in two frames. The top frame shows the data overlaid on the simulated prediction. The bottom frame will present the ratio of data-to-expectation.

The top frame will generally show stacked plots with statistical errors shown on data points, but without statistical errors on the prediction. When considering the results of the top frame, keep in mind that In certain situations, especially at higher purity, the statistical errors on the simulation are not necessarily negligible (statistical error on the simulation will be included in the lower frame). The contributions to the stacked plots are indicated by the label. The uncertainty bars on data points in the top frame are 68% Poisson confidence intervals.

The ratio plots in the lower frame will contain shaded bands which indicate the flux, cross-section, and detector systematic uncertainties. The systematic uncertainties were discussed in detail in Section 8. The points on the ratio plot illustrate the 68% credible interval for the data-to-prediction ratio with a flat prior. This credible interval accounts for the statistical uncertainty in <u>both</u> data and in the Monte Carlo used to produce the prediction. In the case of empty bins, the upper bar illustrates the same 34% upper coverage as other points, but because the lower bar extends fully to 0 these are technically 84% total coverage intervals.

We willll use the Combined Neyman-Pearson $\chi^2$ ($\chi^2_{CNP}$) formula [55] to compare the data to expectation:

$$\chi^2_{CNP} = \sum_i \begin{cases} \dfrac{(\mu_i - M_i)^2}{\frac{3}{1/M_i + 2/\mu_i}} & M_i \neq 0 \\[2ex] \dfrac{(\mu_i - M_i)^2}{\frac{\mu_i}{2}} & M_i = 0 \end{cases} \qquad (9.1)$$

It has been found that this $\chi^2$ definition has a significantly lower bias on the best-fi parameters than the traditional Neyman's or Pearson's $\chi^2$, and so has been adopted as the metric of choice for the MicroBooNE Collaboration.

## 9.2  $1\mu 1p$ Data Comparison Plots

High statistics data comparisons are available for Runs 1-3 via a selection filter which processes data and only retains events which are selected by this analysis as $1\mu 1p$, thus preserving blindness. The availability of this filter in conjunction with the fact that $\nu_\mu$ interactions are significantly less rare allows us to make high statistics validation plots to compare with simulated predictions. These plots provide valuable checks on our understanding of the analysis and on our understanding of uncertainties. Further, the $1e1p$ reconstruction and selection overlaps significantly with the $1\mu 1p$ and as such good agreement in the $1\mu 1p$ sector bolsters confidence in our understanding of the $1e1p$. A set of comparisons is provided.

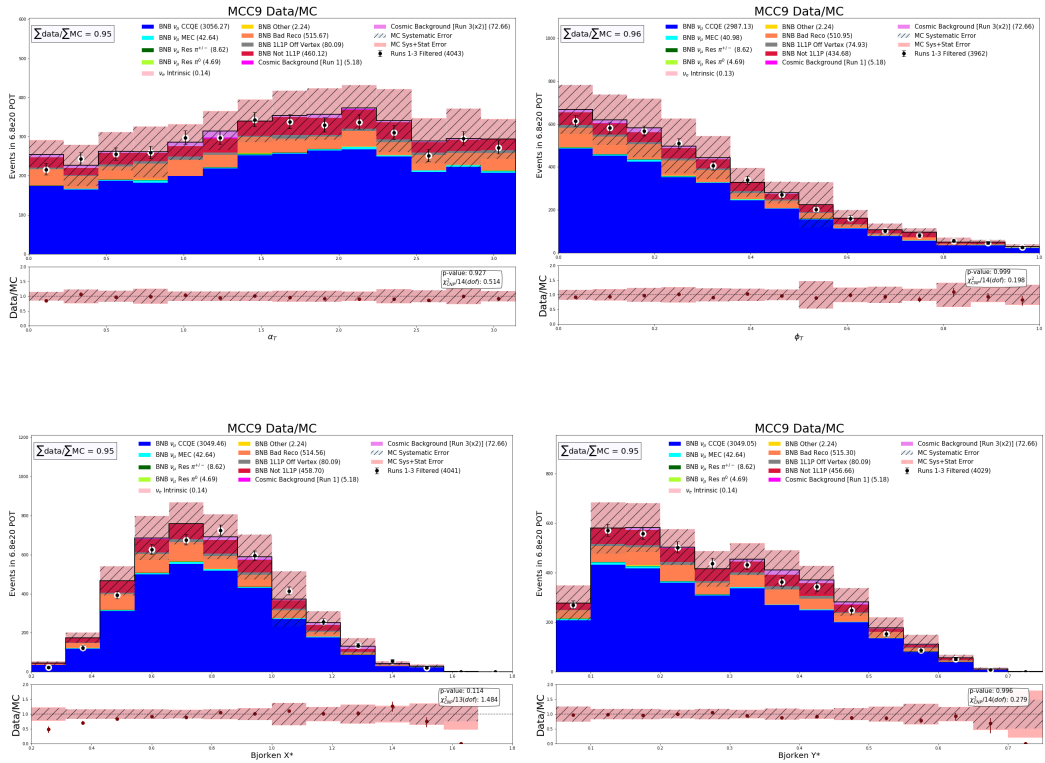

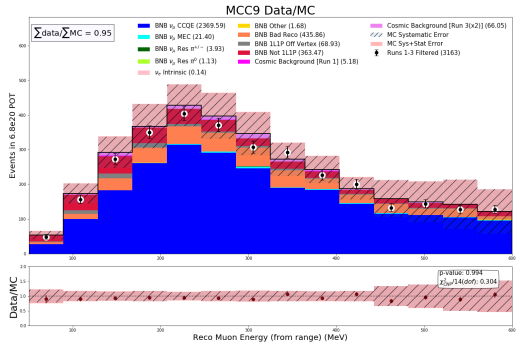
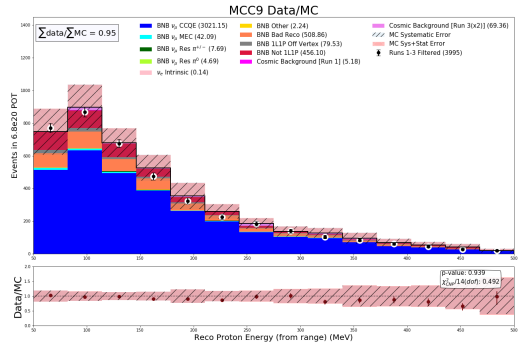
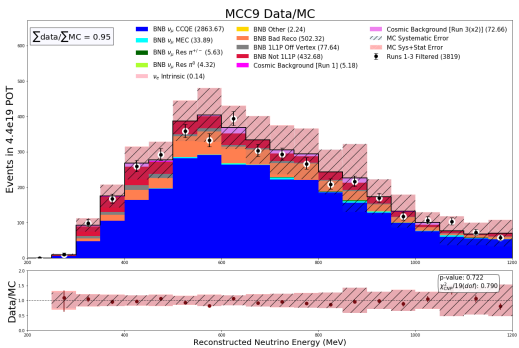
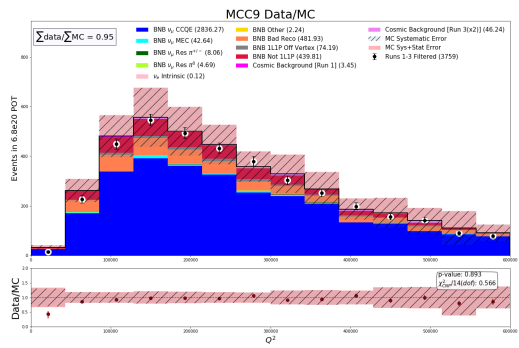
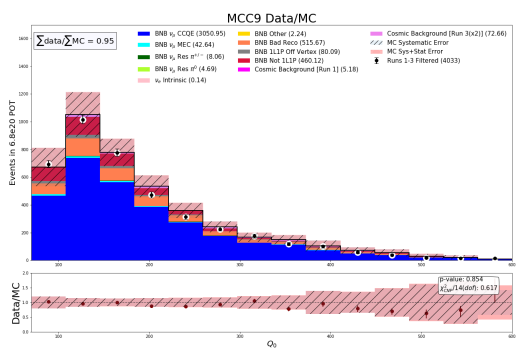
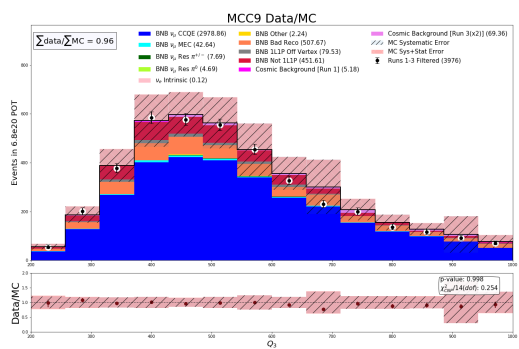
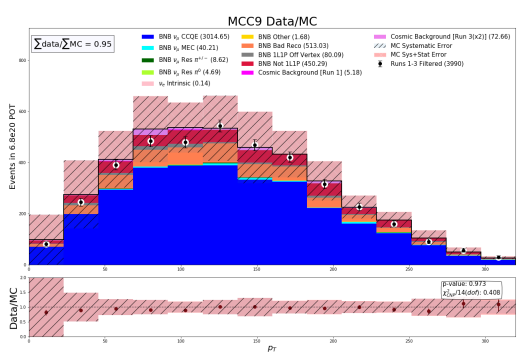
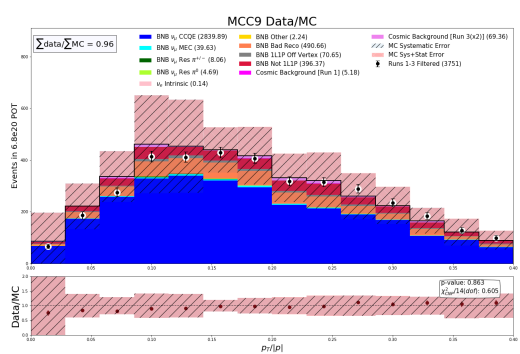

## 9.3   $1e1p$ Data Comparison Plots

Only a small fraction ∼7.5% of the total data has been approved by the MicroBooNE collaboration to be unblinded to a $1e1p$ selection as of the writing of this thesis. As a result, our ability to provide higher statistics validations are more limited than for the $1\mu1p$. We do, however, have two available validations that give us confidence in the efficacy of the reconstruction and selection discussed in this thesis. These are:

1. A high energy sideband

2. Blind safe histograms

Both of these samples are defined in such a way as to avoid premature sensitivity to the presence of a MiniBooNE excess, while still providing sufficient data statistics to validate reconstruction variables used in this analysis. The high energy sideband is a filtered sample which provided events from Runs 1-3 which reconstructed with a neutrino energy > 700 MeV. Because the overwhelming bulk of sensitivity to a low energy excess derives from the < 400 MeV region, this sample cannot reveal the presence of an excess with any significance.

The blind safe histograms access information spanning the entire energy range. These histograms are generated by running the $1e1p$ selection over all data first. Based on analyzing simulation, we have an expectation that we would select $N_{exp}$ in the absence of an excess. If we ran this filter and observed N events, we could compare to $N_{exp}$ and learn if there was an excess. So to avoid this, we compute $N_{low-90}$ which is the 90% lower Poisson confidence value givcen $N_{exp}$. When running the filter, only this fixed number of events are kept. This is done at random. Analyzers do not know how many events were actually selected. Further, analyzers do not get unrestricted access to these events. An excess could be spotted even in the absence of full normalization information because the shape of certain spectra, like neutrino energy, differs between a universe with and without a low energy excess. So in advance of running this filter, a predetermined set of variables are identified which are insensitive to the presence of a low energy excess. Only histograms of these variables are made. A summary of the the goodness of agreement for both comparison data sets is provided in Figs. 9-1 & 9-2.

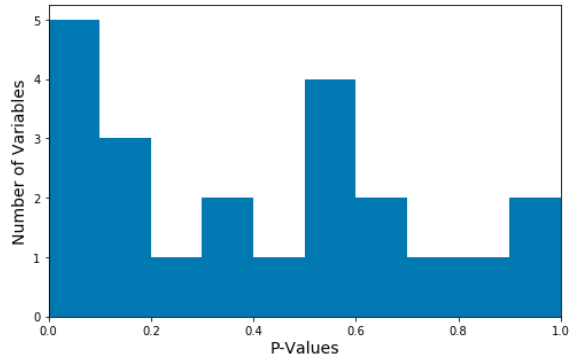

Figure 9-1: Summary of p-values for goodness of agreement between prediction and observation of analysis variables for the blind plots.

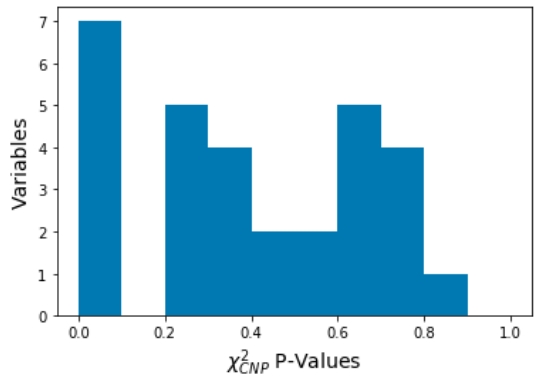

Figure 9-2: Summary of p-values for goodness of agreement between prediction and observation of analysis variables for the high energy sideband plots.

## 9.3.1 High Energy Sideband

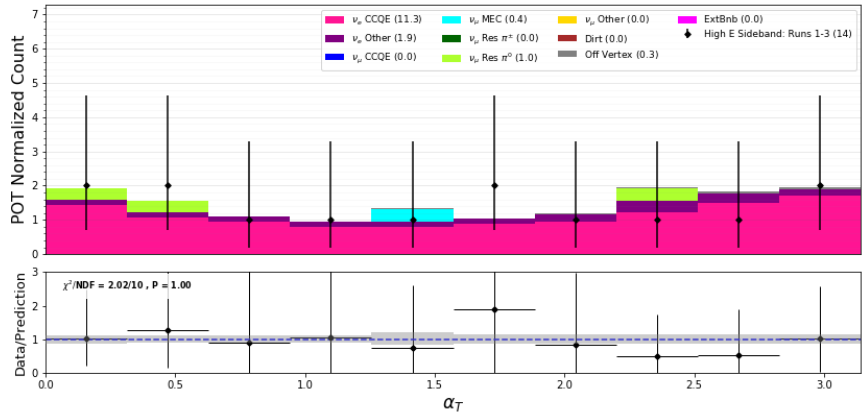

Figure 9-3: $1e1p$ High Energy Box $\alpha_T$.

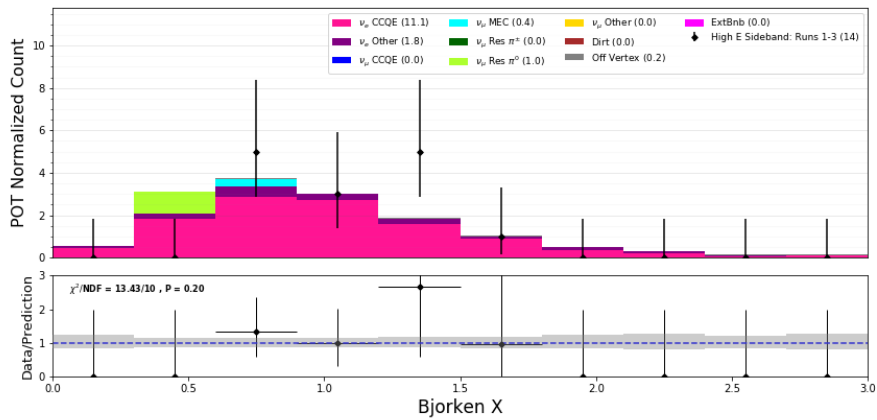

Figure 9-4: $1e1p$ High Energy Box $x_{Bj}$.

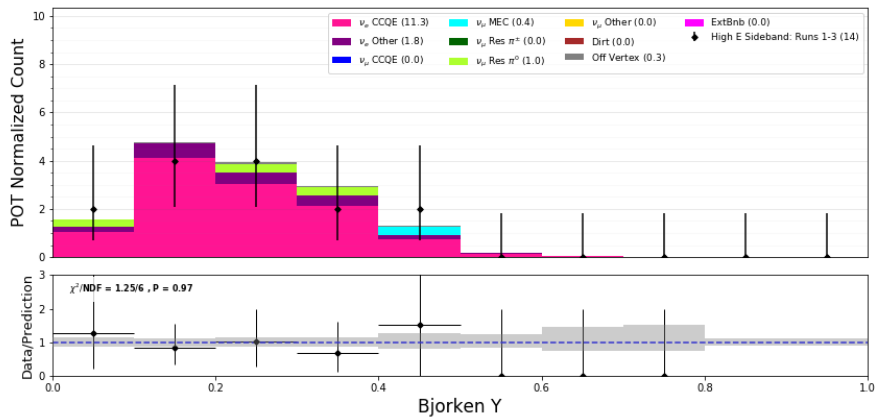

Figure 9-5: $1e1p$ High Energy Box $y_{Bj}$.

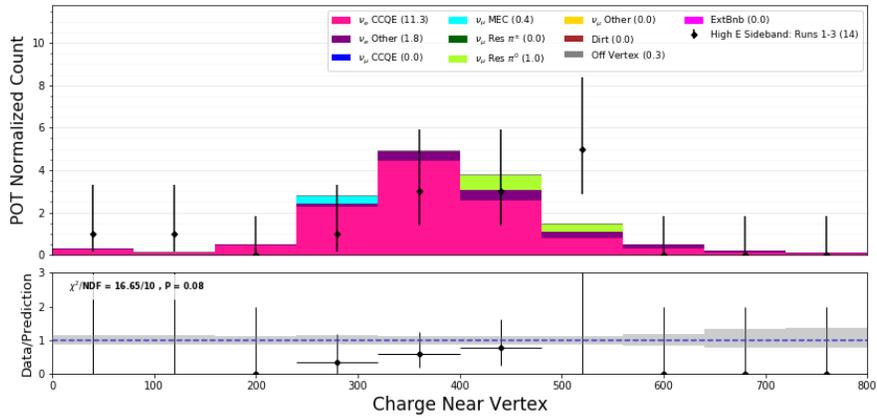

Figure 9-6: $1e1p$ High Energy Box Charge Near Trunk.

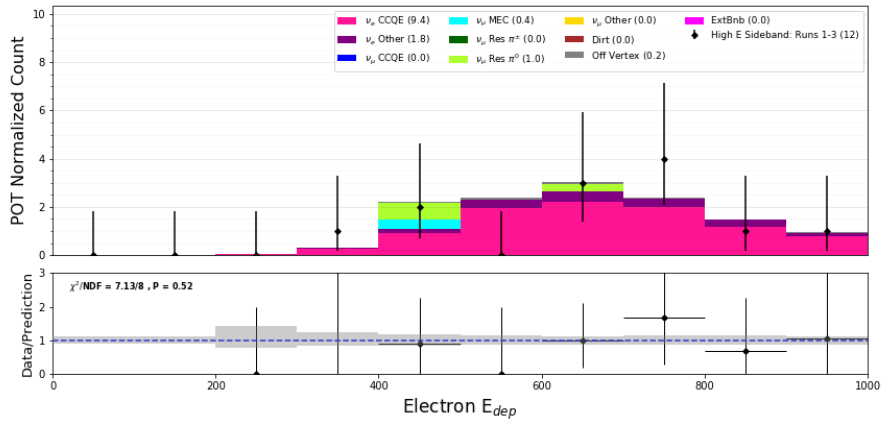

Figure 9-7: $1e1p$ High Energy Box $E_e$.

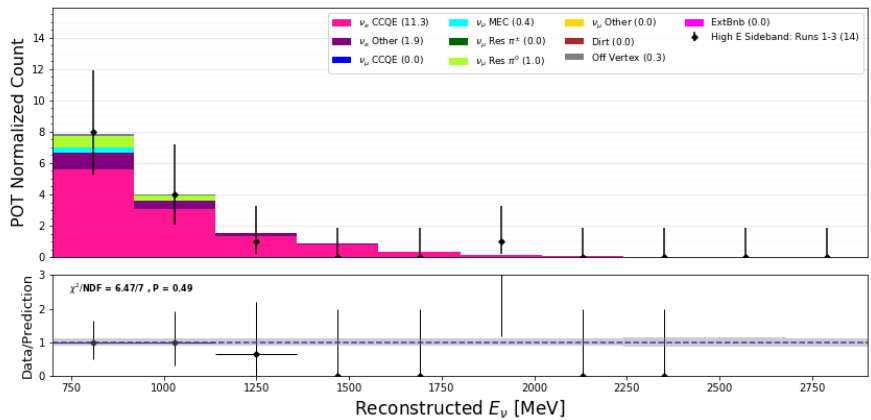

Figure 9-8: $1e1p$ High Energy Box $E_\nu$.

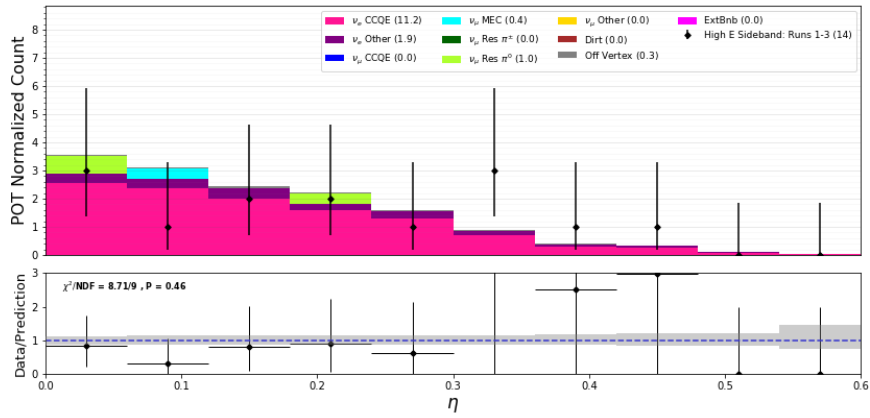

Figure 9-9: 1e1p High Energy Box $\eta$.

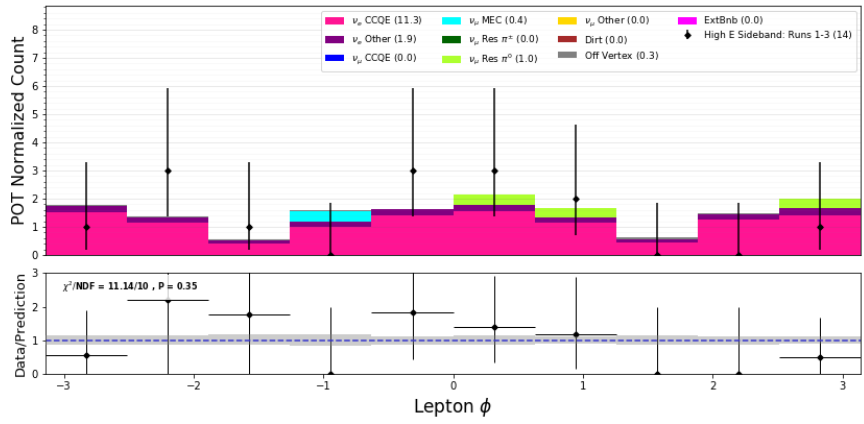

Figure 9-10: 1e1p High Energy Box Lepton $\phi$.

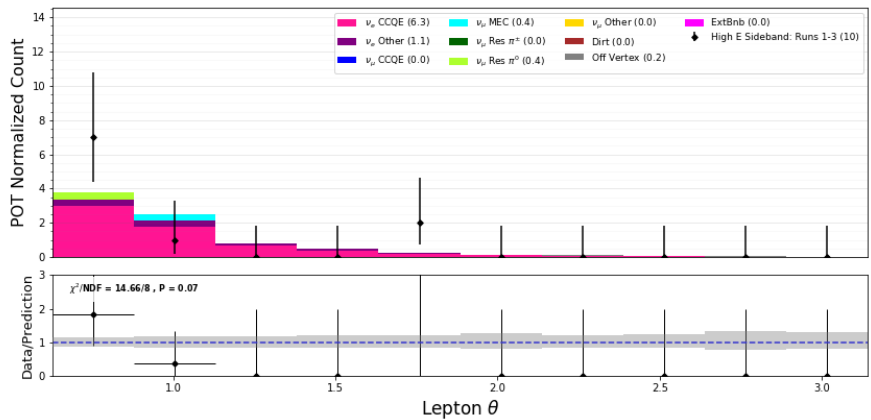

Figure 9-11: 1e1p High Energy Box Lepton $\theta$.

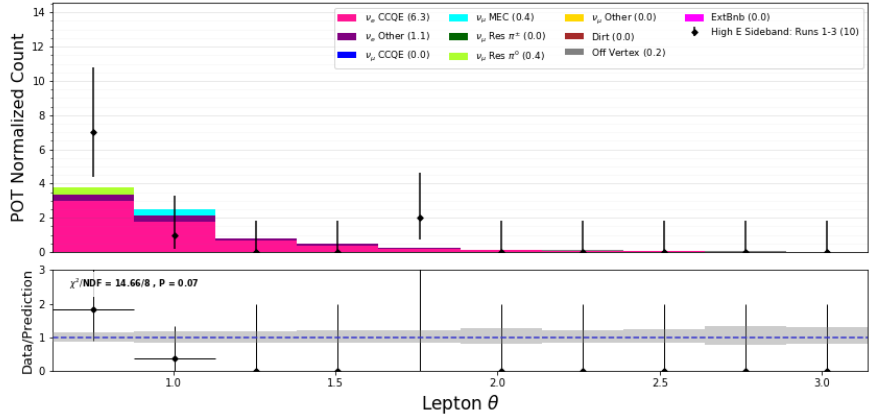

Figure 9-12: $1e1p$ High Energy Box Lepton $\theta$.

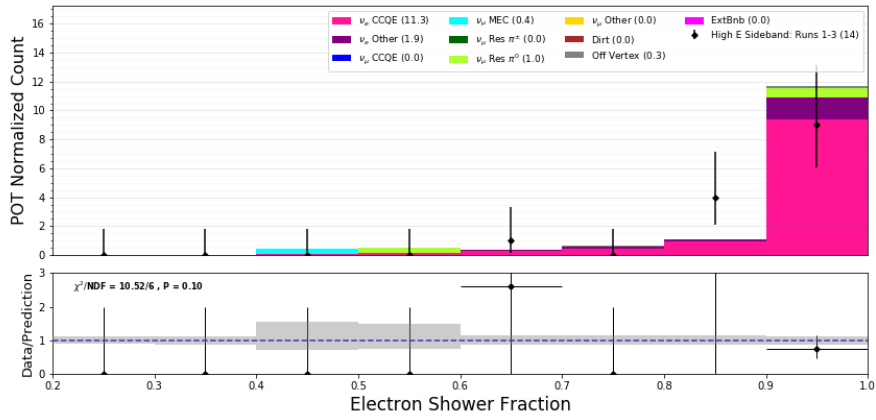

Figure 9-13: $1e1p$ High Energy Box Maximum Shower Fraction.

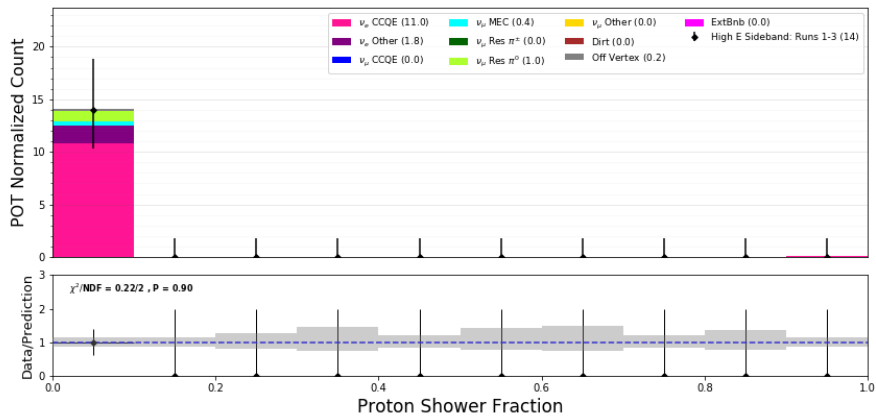

Figure 9-14: $1e1p$ High Energy Box Minimum Shower Fraction.

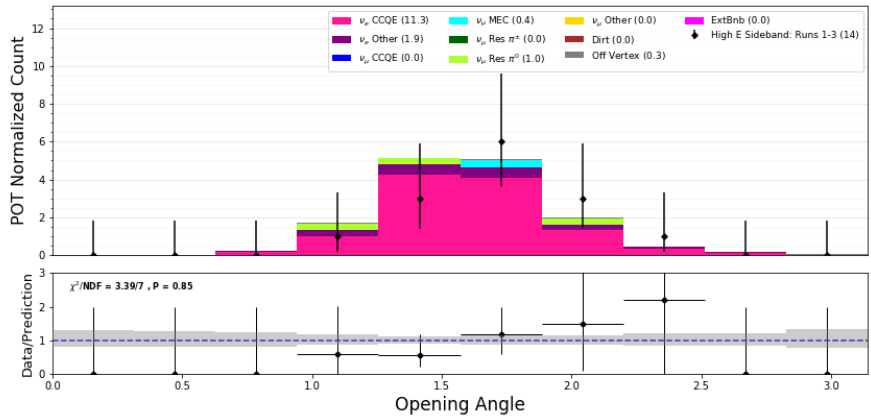

Figure 9-15: $1e1p$ High Energy Box Lepton Opening Angle

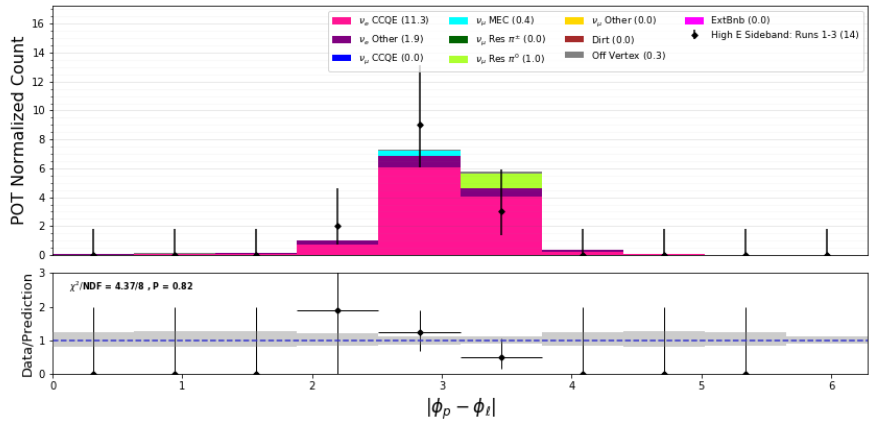

Figure 9-16: $1e1p$ High Energy Box Lepton Difference in $\phi$.

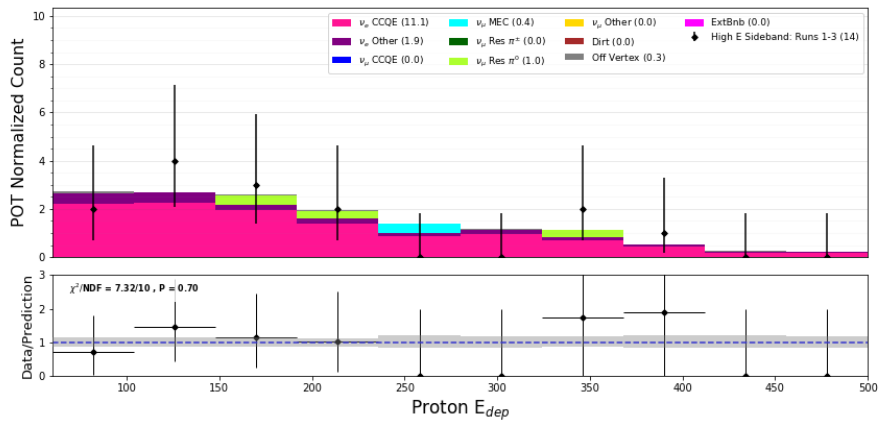

Figure 9-17: $1e1p$ High Energy Box Lepton Proton Energy.

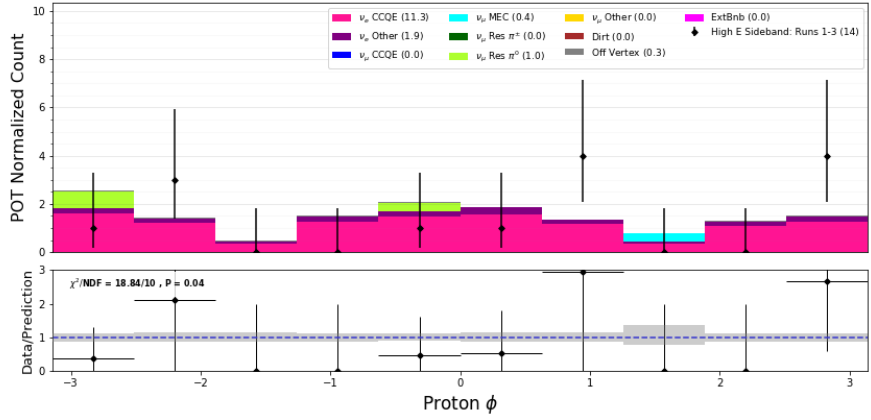

Figure 9-18: $1e1p$ High Energy Box Proton $\phi$.

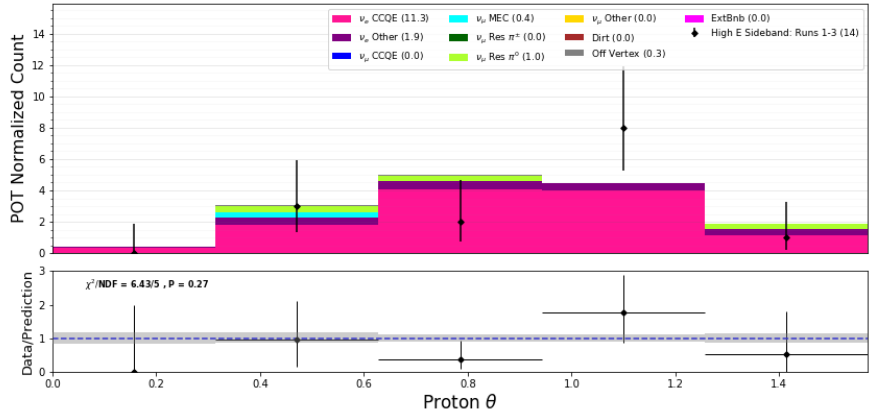

Figure 9-19: $1e1p$ High Energy Box Proton $\theta$.

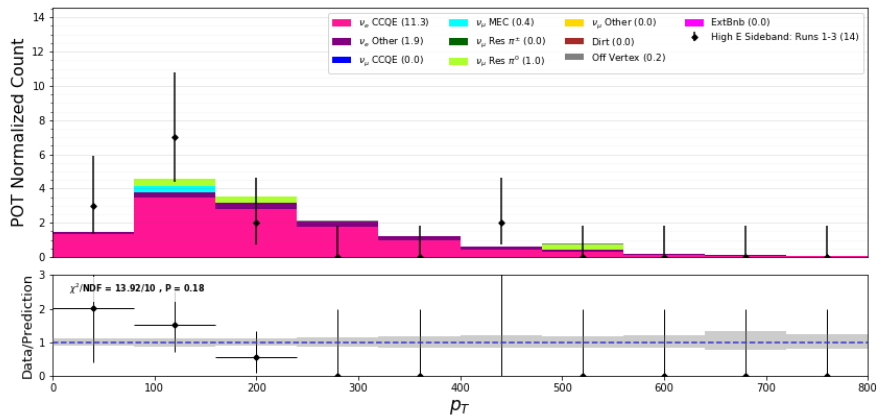

Figure 9-20: $1e1p$ High Energy Box Event $p_T$.

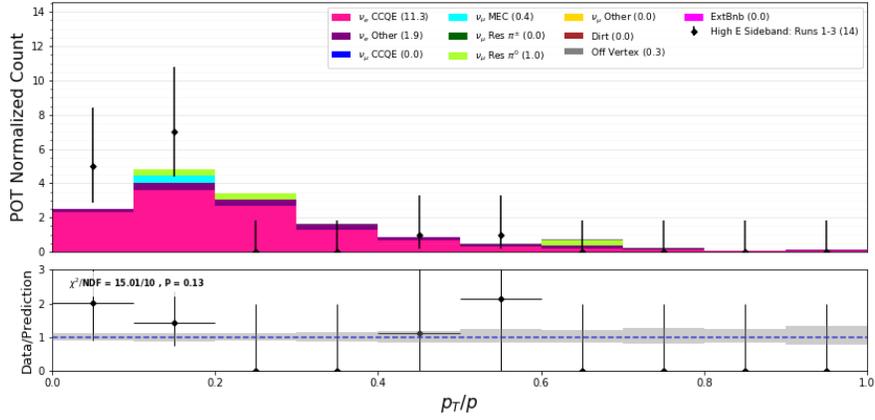

Figure 9-21: $1e1p$ High Energy Box Event $p_T/|p|$.

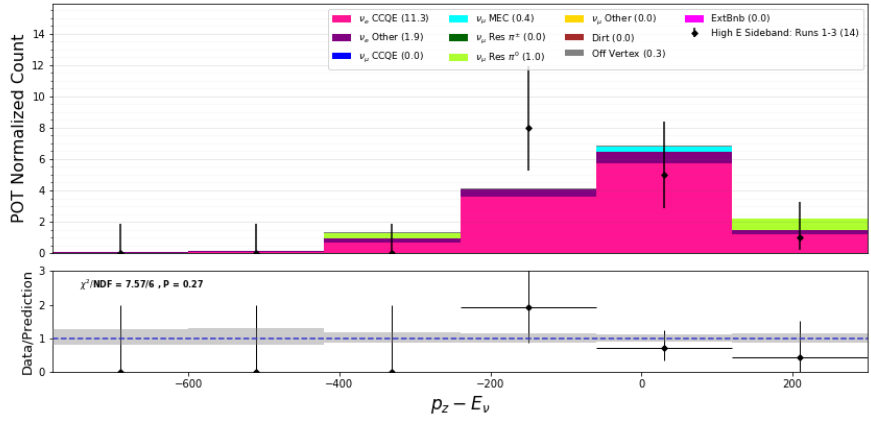

Figure 9-22: $1e1p$ High Energy Box Fermi $z$ Momentum Variable.

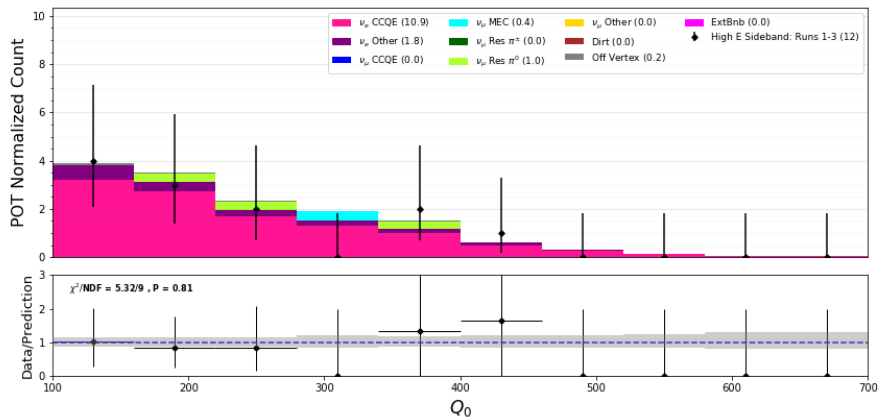

Figure 9-23: $1e1p$ High Energy Box $Q^0$.

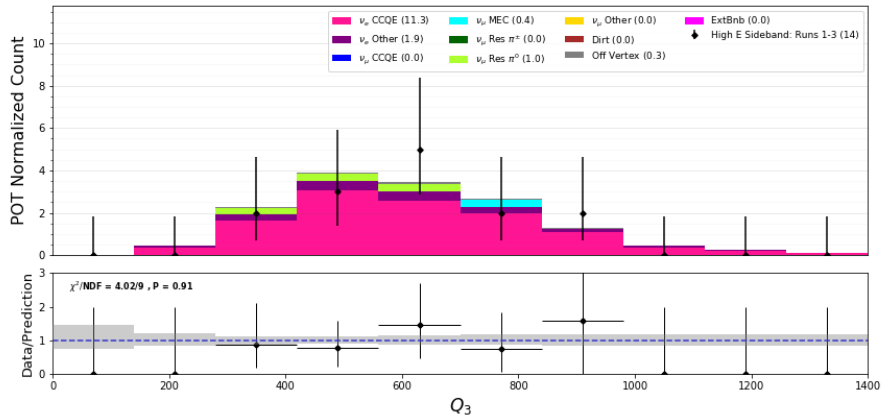

Figure 9-24: $1e1p$ High Energy Box $Q^3$.

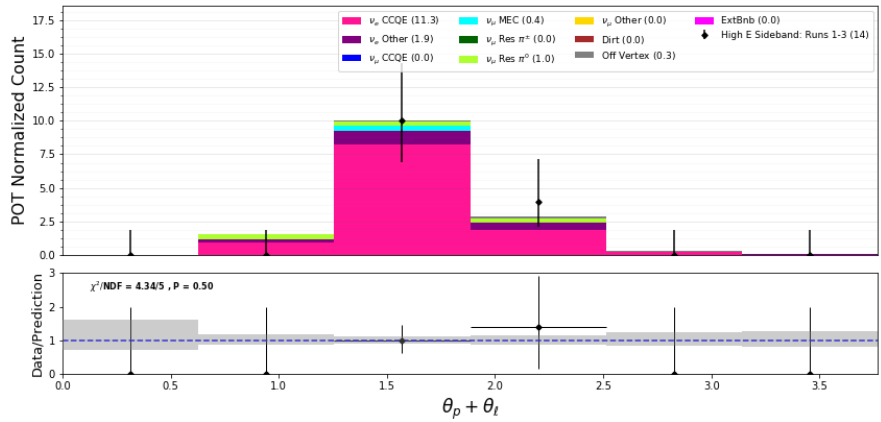

Figure 9-25: $1e1p$ High Energy Box Sum of $\theta$s

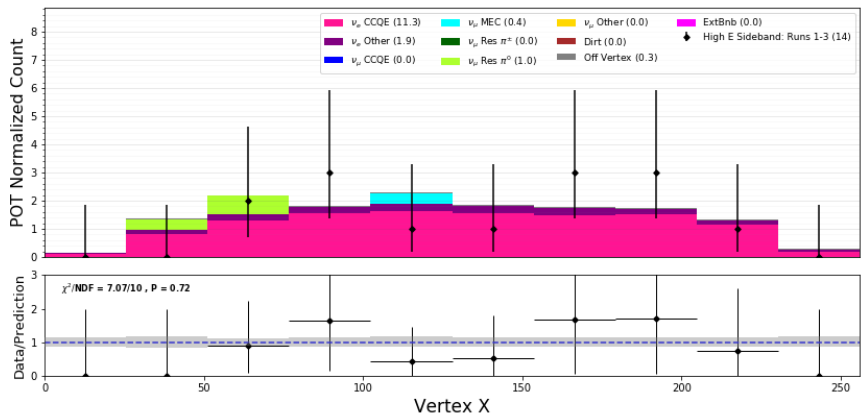

Figure 9-26: $1e1p$ High Energy Box $x$ vertex position.

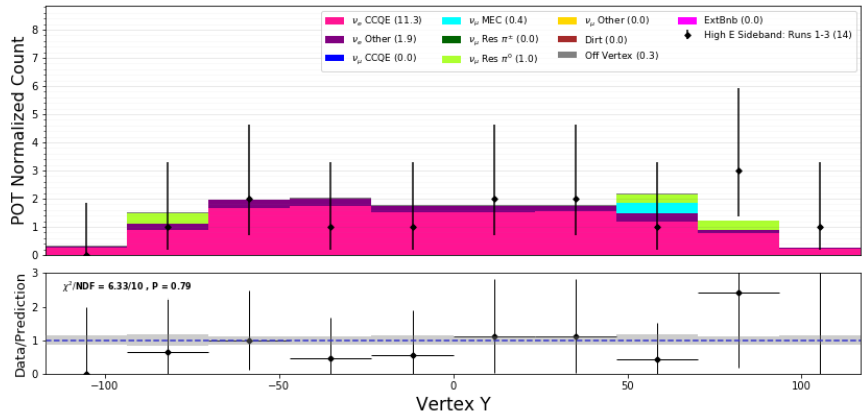

Figure 9-27: $1e1p$ High Energy Box $y$ vertex position.

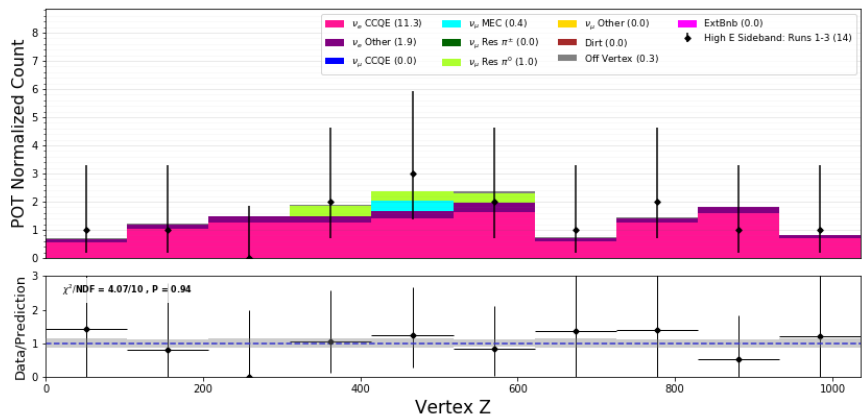

Figure 9-28: $1e1p$ High Energy Box $z$ vertex position

## 9.3.2 Blind Safe Histograms

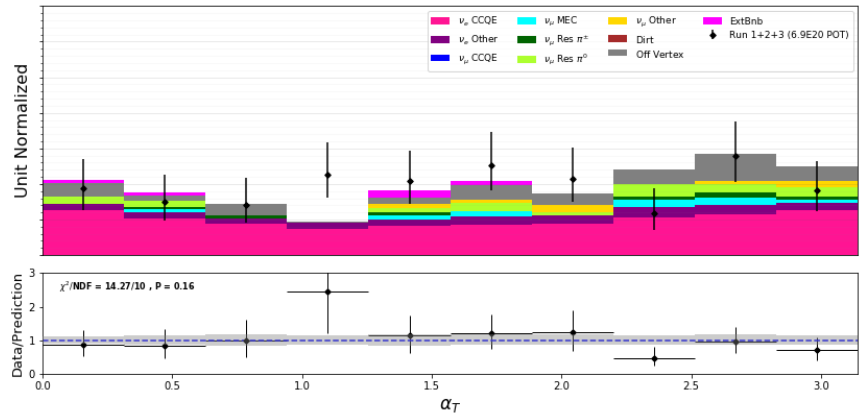

Figure 9-29: Blind plot: $\alpha_T$.

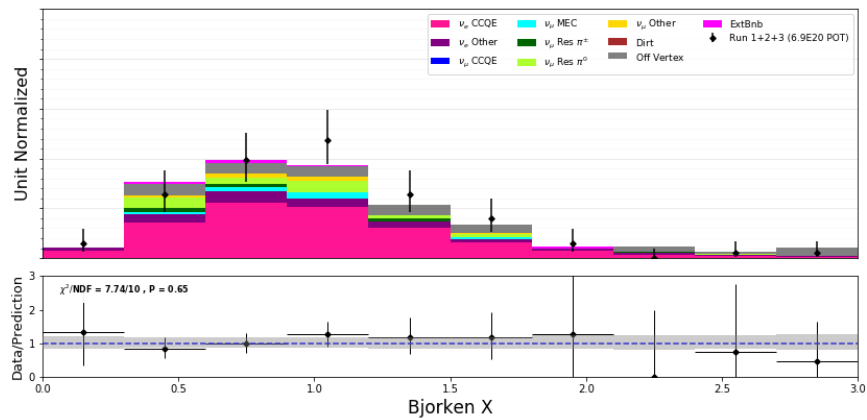

Figure 9-30: Blind plot: Bjorken X

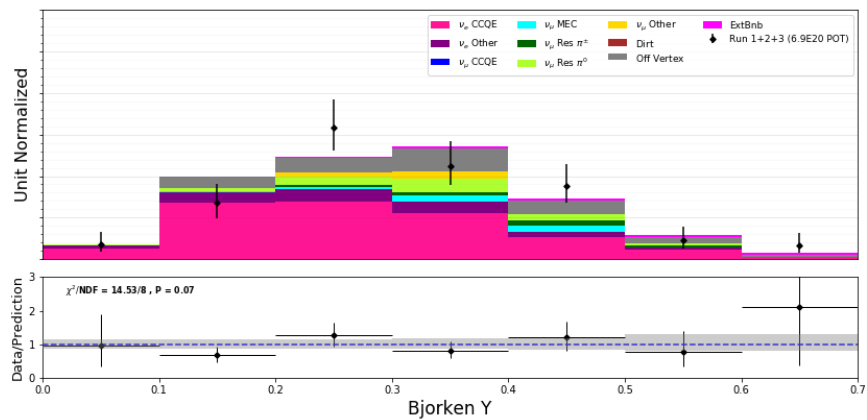

Figure 9-31: Blind plot: Bjorken Y

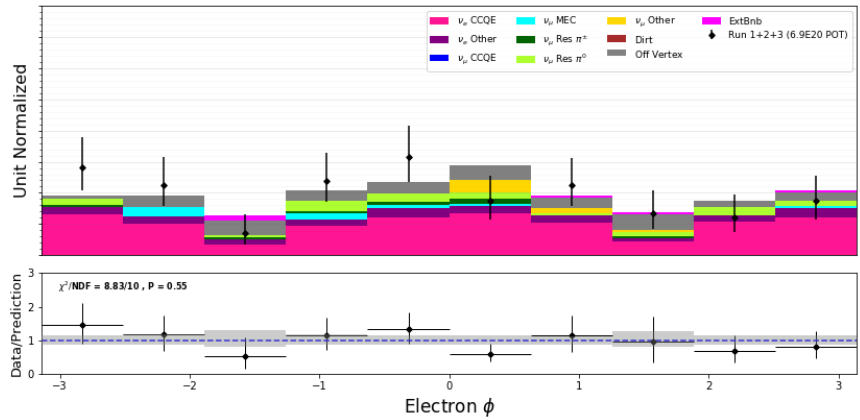

Figure 9-32: Blind plot: Electron $\phi$

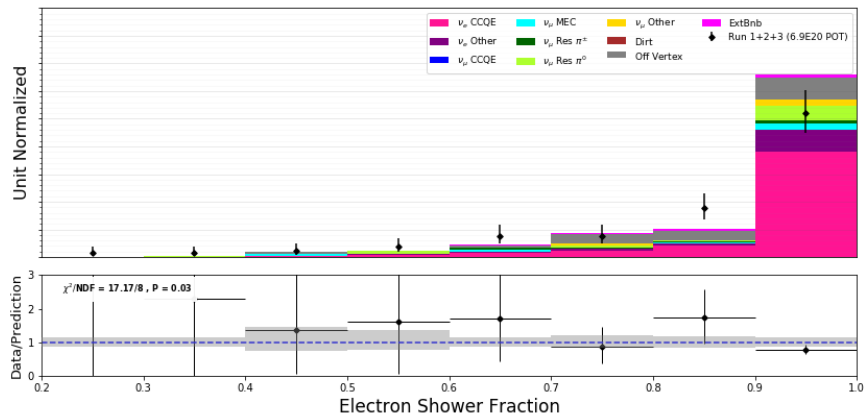

Figure 9-33: Blind plot: Electron Shower Fraction

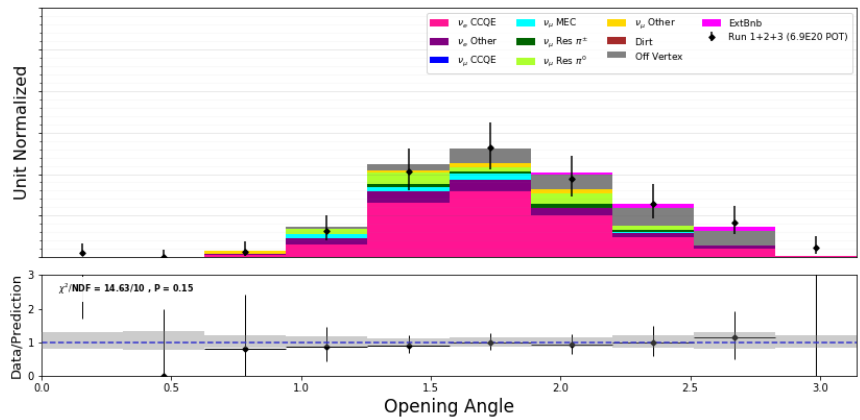

Figure 9-34: Blind plot: 3D Opening Angle

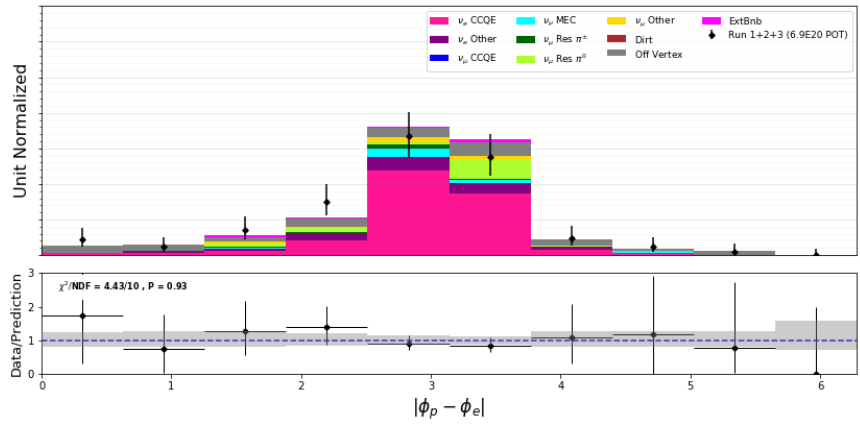

Figure 9-35: Blind plot: Absolute difference in $\phi$, $|\phi_p - \phi_e|$

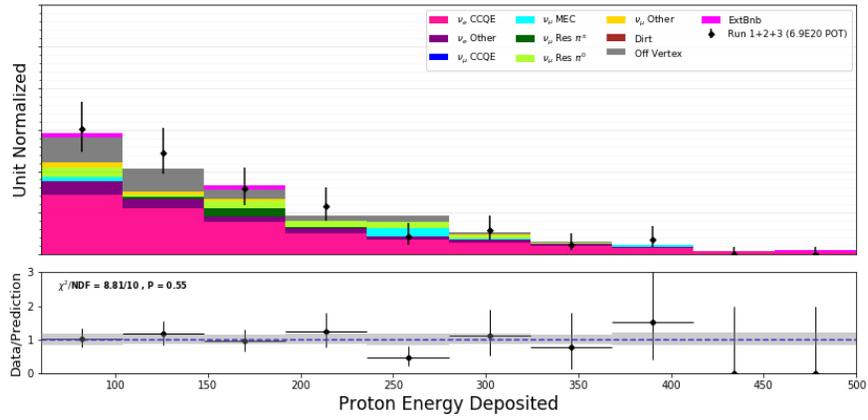

Figure 9-36: Blind plot: Proton deposited energy

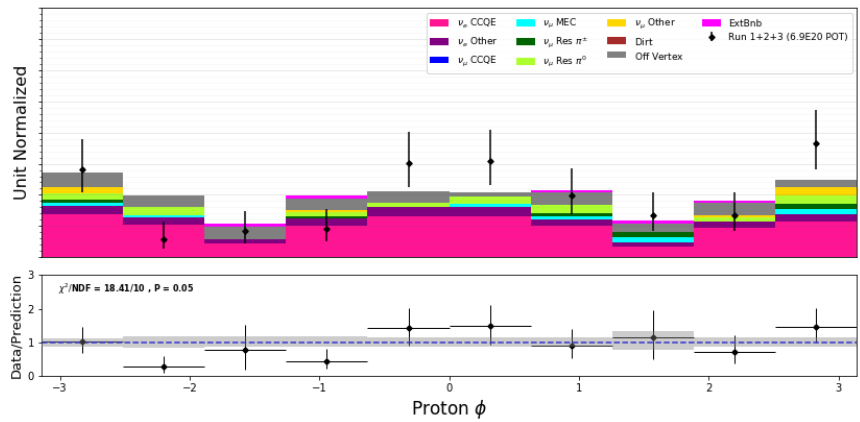

Figure 9-37: Blind plot: Proton $\phi$

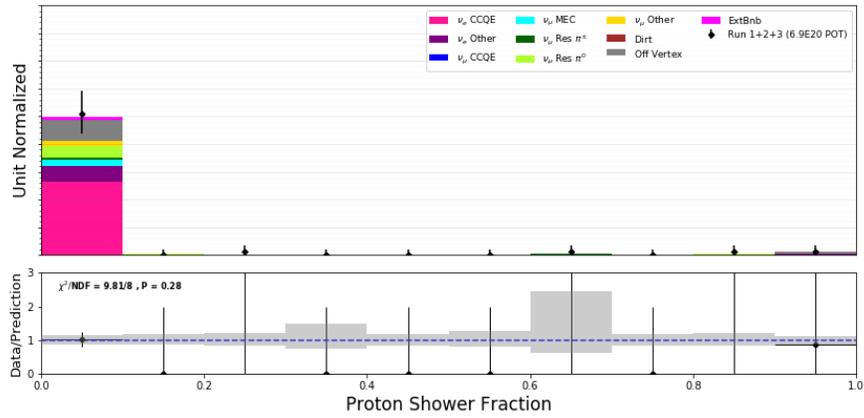

Figure 9-38: Blind plot: Proton Shower Fraction

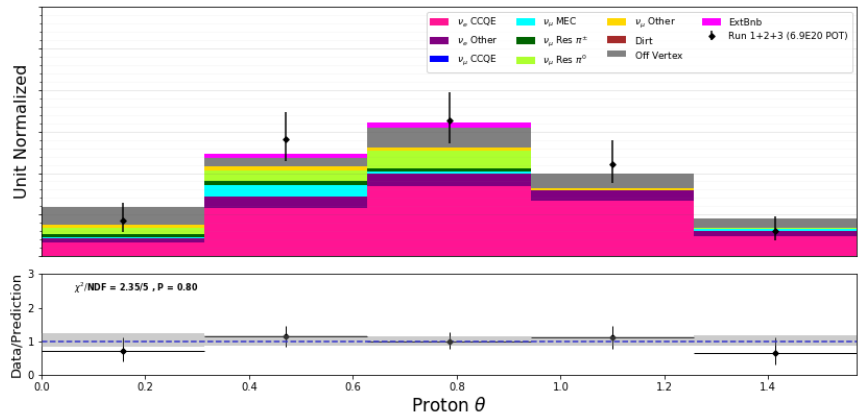

Figure 9-39: Blind plot: Proton $\theta$

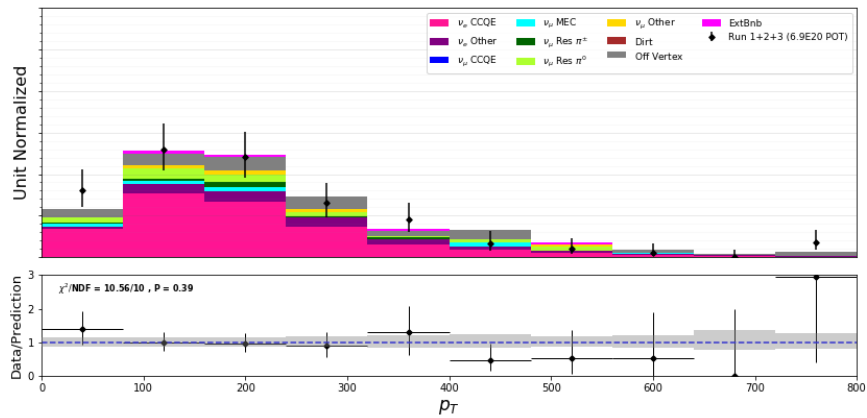

Figure 9-40: Blind plot: Final state transverse momentum of the event

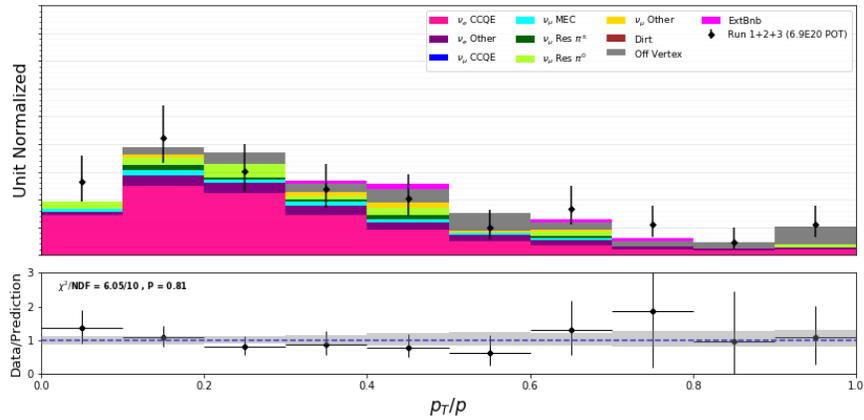

Figure 9-41: Blind plot: Final state transverse momentum of the event divided by total momentum

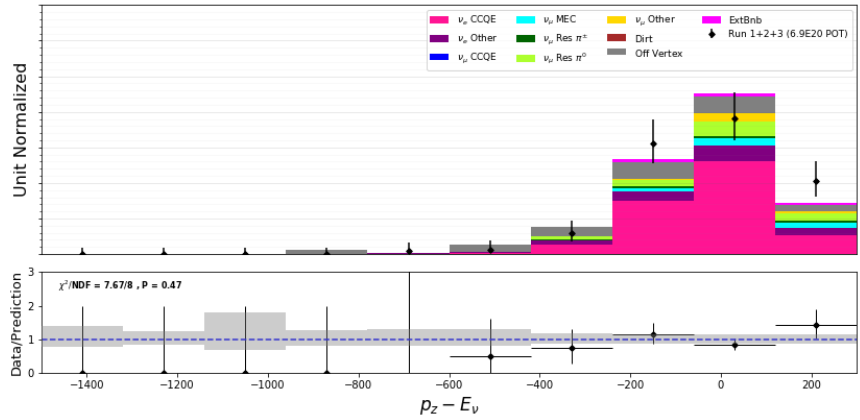

Figure 9-42: Blind plot: $p_z - E_\nu$

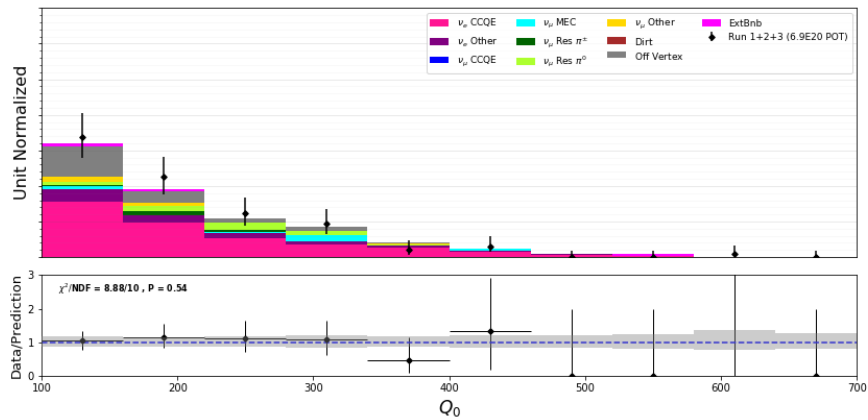

Figure 9-43: Blind plot: $Q_0$

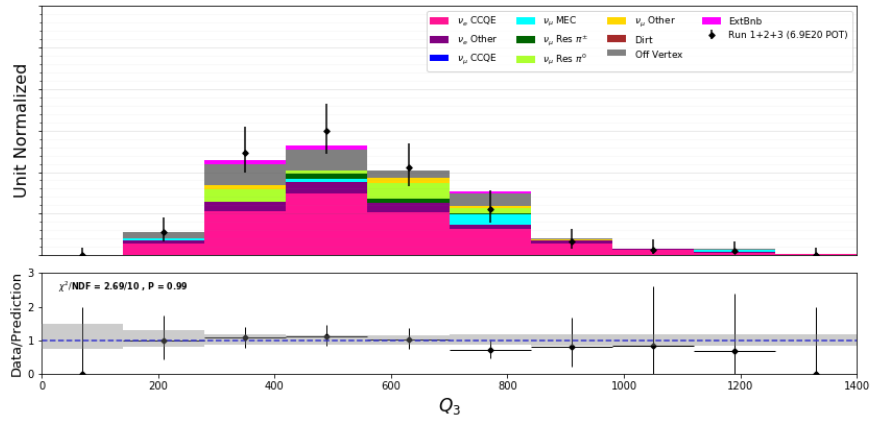

Figure 9-44: Blind plot: $Q_3$

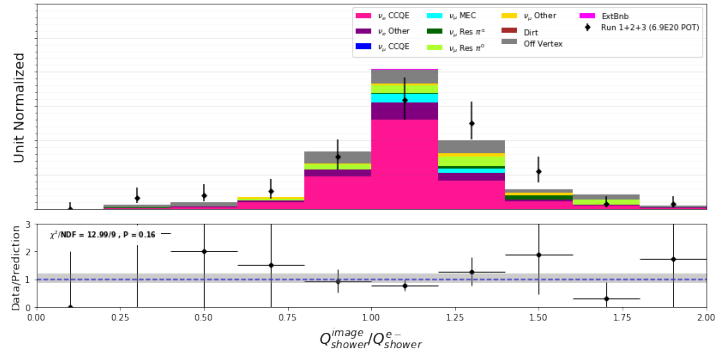

Figure 9-45: Blind plot: Shower charge in image divided by shower charge in electron cluster

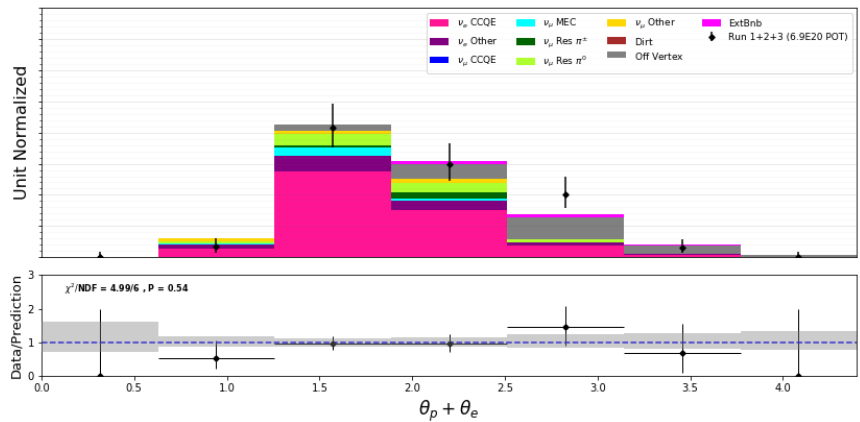

Figure 9-46: Blind plot: $\theta_p + \theta_e$

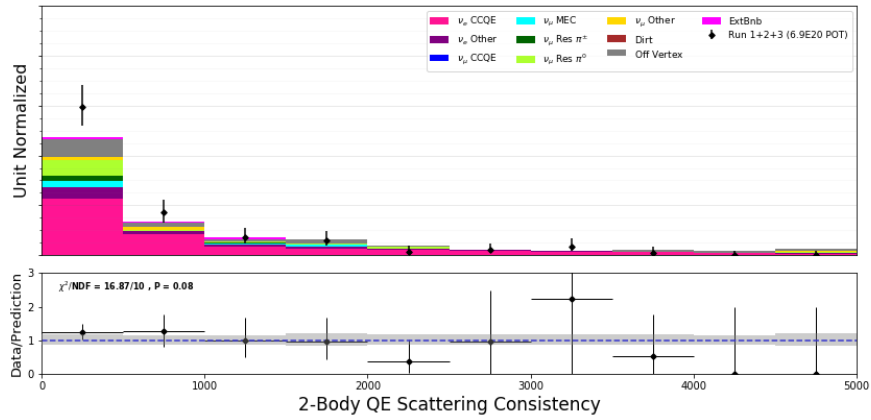

Figure 9-47: Blind plot: Consistency between three $E_\nu$ reconstruction formulae

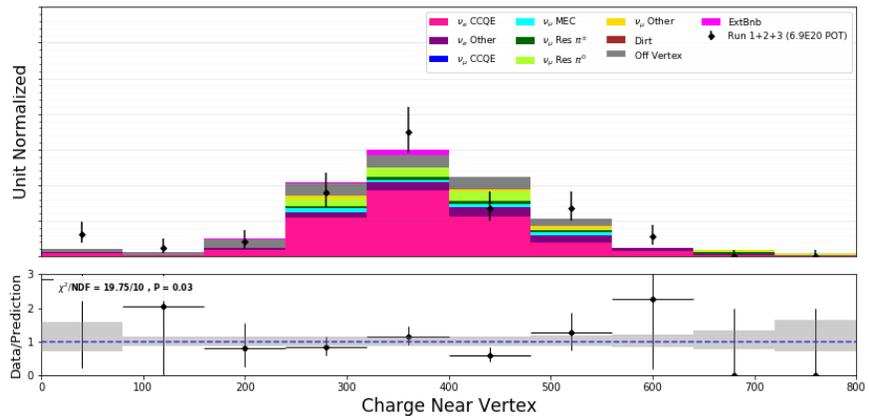

Figure 9-48: Blind plot: Charge near vertex

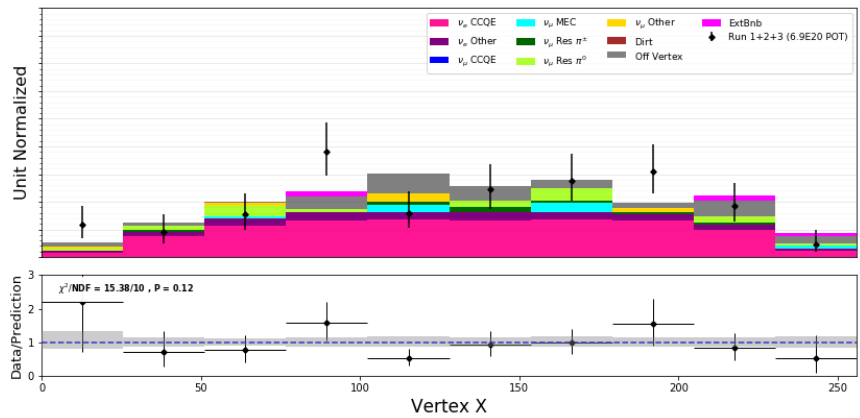

Figure 9-49: Blind plot: Vertex X

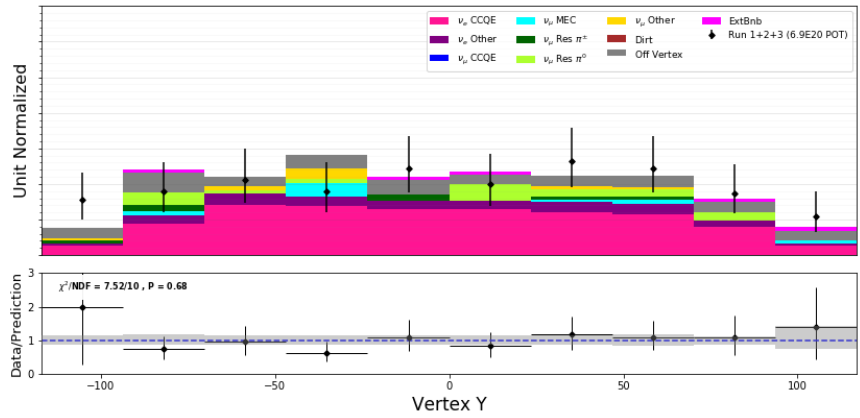

Figure 9-50: Blind plot: Vertex Y

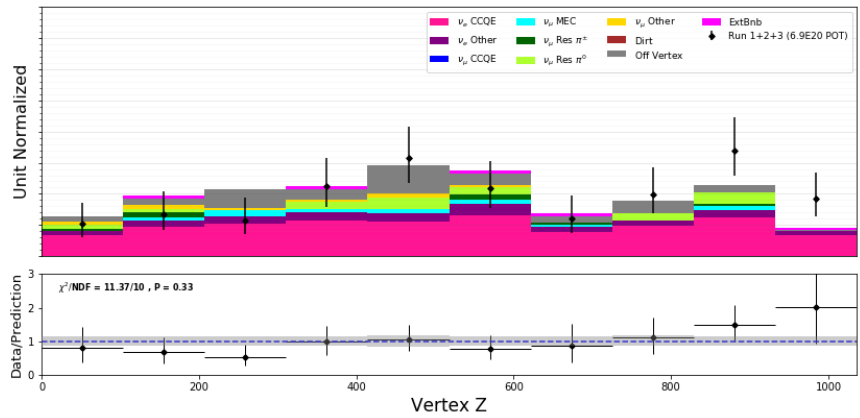

Figure 9-51: Blind plot: Vertex Z

## 9.4 Prediction: The Energy Spectrum and Sensitivity Estimate

Given the excellent results from the above cross checks, we can predict the results of MicroBooNE once the data is unblinded. That result will come from BNB Runs 1-3 and corresponds to $6.9 \times 10^{20}$ POT.

The anticipated spectrum for the standard model (colored histogram) and the for MiniBooNELow Energy Excess prediction for Runs 1-3 is shown in Fig. 9-52.

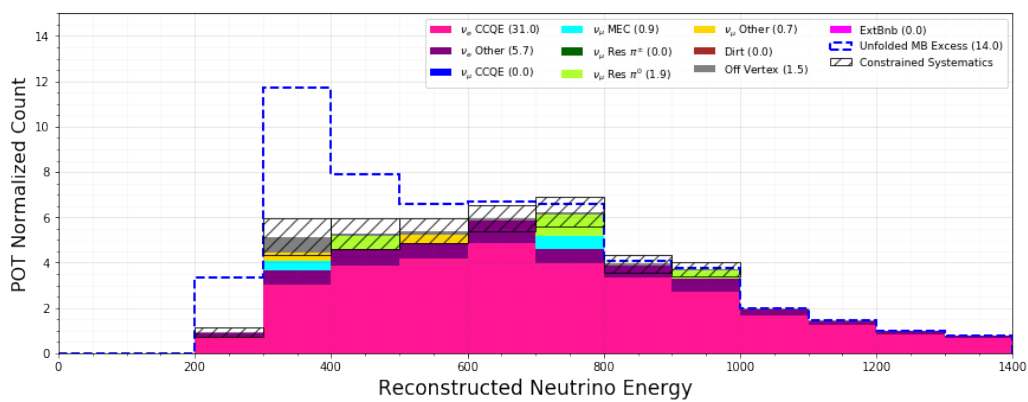

Figure 9-52: Anticipated reconstructed neutrino energy spectrum for BNB Runs 1-3 resulting from the application of the entire reconstruction and selection. The standard model spectrum is illustrated as a stacked histogram. The unfolded MiniBooNE excess is shown as a dashed blue line.

Based on these predicted spectra, we compute the probability that the standard model can fluctuate up to a level that is described by the predicted LEE excess. Thus, if we measure a result in agreement with the standard model, this is the probability that we will exclude the MiniBooNE result. Accounting for all uncertainties we compute this sensitivity assuming data from Runs 1-3 ($6.9 \times 10^{20}$ POT), which we expect will be made in the near future.

The sensitivity to exclude a MiniBooNE-like low energy excess is found using frequentist studies performed within the "SBNfit" framework. [54] This is code that will be utilized beyond MicroBooNE, by the entire short baseline program at Fermilab. For each sensitivity study, $10^5$ pseudo-experiments with statistical and systematic fluctuatiosn permitted are thrown with two different hypotheses. $H_0$ is the null universe which contains only standard model physics and no excess, only intrinsic $\nu_e$ and $\nu_\mu$ backgrounds. $H_1$ corresponds to a universe in which there is additionally a MiniBooNE like $\nu_e$ excess. In each pseudo-experiment, compute how well the fluctuation agrees with either hypothesis via a Combined Neyman-Pearson $\chi^2$ ($\chi^2_{CNP}$) (Eq. 9.1) This metric is specifically chosen because of it's good behavior for low statistics bins. The whole ensemble of pseudo experiments is then used to build up a distribution of $\Delta\chi^2$ between $H_0$ and $H_1$. These distributions permit one to ask "If $H_0$ is correct, how frequently would we obtain a $\Delta\chi^2$ which is greater than the median $\Delta\chi^2$ obtained if $H_1$ is correct?" With a known number of pseudo-experiments, this frequency allows you to compute the probability $\alpha$. The median sensitivity is then quoted as n-$\sigma$ corresponding to that $\alpha$ value for a normal distribution.

The results of these pseudo-exents for the analysis presented in this thesis are illustrated in Fig. 9-53. The red distributions illustrate the $\Delta\chi^2$ given $H_0$ while the blue are given $H_1$. The black lines illustrate the quantiles of the $H_1$ distribution. The value of $\alpha$ corresponds to the fraction of the $H_0$ distribution which falls to the right of the median of the $H_1$. This study indicates a median of $3.2\sigma$ with Runs 1-3. In other words, given an unfolded prediction derived from the latest MiniBooNE observation (2018 results) if MicroBooNE observes an excess consistent with this signal then we can exclude it being from a background fluctuation at $3.2\sigma$.

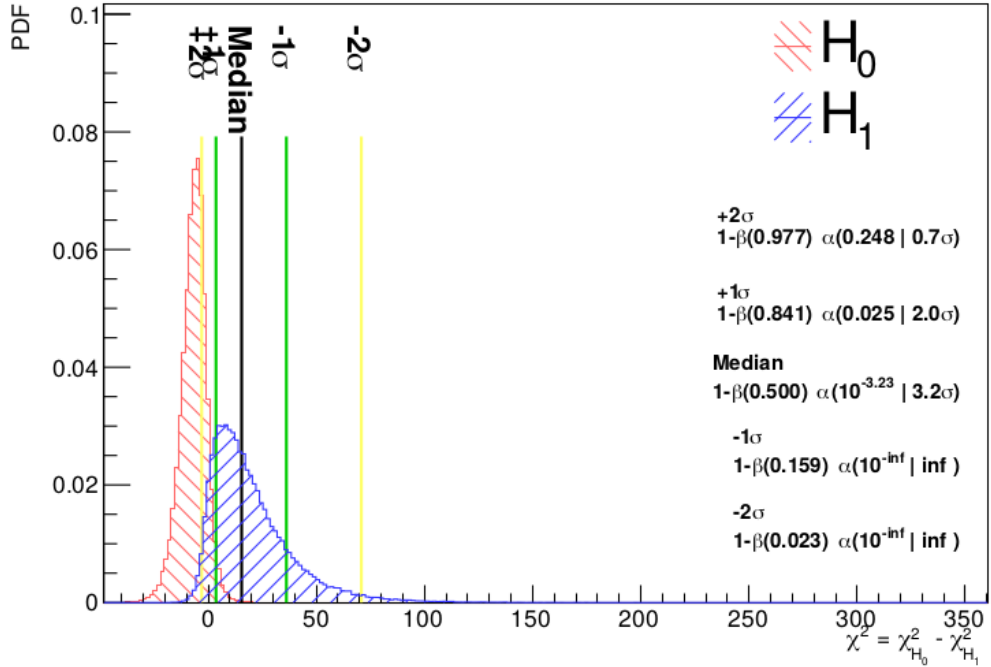

Figure 9-53: Results of frequentist study for sensitivity to exclude a MiniBooNE $\nu_e$ like excess with BNB Runs 1-3

This leads to the final result of this thesis: The sensitivity of this analysis to exclude the MiniBooNE Low Energy Excess signal is $3.2\sigma$.

# Chapter 10

# Conclusions

The primary result of this thesis is an analysis that has $> 3\sigma$ sensitivity to the MiniBooNE Low Energy Excess for a data set corresponding to $7 \times 10^{20}$ POT. We have presented a result that indicates that a $3.2\sigma$ fluctuation of the Standard Model backgrounds is required for MicroBooNE to observe a MiniBooNE-like excess. The credibility of this prediction has been established by extensive comparisons between data and simulation. Based on this work, the selection cuts for this analysis are now frozen.

Before the analysis is unblinded and this prediction is tested, some work remains from the collaboration. Some systematic uncertainties are still to be delivered that may slightly reduce this expectation, although this is not expected to be a large effect. On the other hand, some additional constraints beyond the $1\mu 1p$ sample described here are under consideration that may slightly increase the final sensitivity. Although some work remains, the conclusion of this thesis is that this analysis can deliver an "observation" of the MiniBooNE signal, if it is within the data set at the level predicted by the MiniBooNE data.

The data may reveal any one of the following scenarios:

- The result may be in agreement with the Standard Model prediction in this thesis. In this case, the MicroBooNE result will cause the global fits to sterile neutrino oscillation models to be so poor as to be ruled out.

- The result may be in agreement with MiniBooNE. This will be particularly interesting since the MiniBooNE result does not perfectly agree with a sterile neutrino scenario. This will indicate that there is some source of additional $\nu_e$ in the beam, but that not all of it can be explained with oscillations.

- The result may lie between the Standard Model and the MiniBooNE prediction. The MicroBooNE result will then motivate further studies by the future experiments in the Fermilab Short Baseline Program.

All of these scenarios are exciting, and the author looks forward to unblinding of the MicroBooNE data.

While the work presented in this thesis represents a complete analysis which can achieve significant sensitivity, there nonetheless remains further work that the author would like to see incorporated in future versions of the analysis. One of the most exciting involves more significant use of deep learning to supplement or supplant steps in the analysis that are currently done algorithmically. Estimation of shower energy using deep learning, for instance, has already begun to be explored and shows significant potential to outperform the clustering algorithm used here. Migration of certain networks such as SparseSSNet and MPID to 3 dimensions rather than three 1 dimensional scores would also likely yield improvement. Revisiting the vertexing algorithm would also be a fruitful investment of time as it is currently one of the least efficient steps. Implementing systematic constraints in addition to the $1\mu 1p$ currently used may also be helpful. Two of the most systematically sensitive backgrounds include MEC and resonant $\pi^0$ final states. Selections to isolate such events have been explored and it may improve our sensitivity if these can be refined to the point where they can be incorporated as additional systematic constraint samples.

# Appendix A

# List of Contributions

This thesis has described an analysis based on Deep Learning combined with other standard algroithmic techniques that will allow the MicroBooNE experiment to search for the MiniBooNE Low Energy Excess. As a member of a large collaboration, many people contributed to aspects of this work. The aspects that are directly attributable to the author are:

- Development of the pre-selection requirements, including those on light collection which became the "common optical filter" for the collaboration.

- Authorship of the original shower reconstruction code which was then futher honed by other collaborators.

- Introduction of BDTs into the analysis, early development of the $1\mu1p$ BDT, and full responsibility for the development of the $1e1p$ BDT.

- Development of QE consistency, boosting technique, and kinematic consistency based selection techniques.

- Establishment of cuts ot isolate the Michel electron, High Energy Box, and the Low BDT Samples.

- Development of the concept and the cuts for the "Blindness-safe plots."

- Study of Data-to-Simulation comparisons for open data sets.

The author also contributed to the commissioning of the MicroBooNE light collection system and the subsequent gain calibration that is provided in Appendix B. (As a graduate student, the author also was involved in developing lightguides, a detector for potential future use, and this is described in Appendix C, as well as authored a phenomeology paper on dark photon signatures in DUNE [56]).

# Appendix B

# Optical System Calibration

A pulse on a PMT readout waveform fundamentally links to a certain amount of optical activity seen by the PMT, i.e. there is a link between integrated charge (measured in ADC · time ticks) and incident number of photons. The *gain* captures the ratio of charge per photoelectron (PE) produced at the PMT cathode. In order to maintain the stability of optical triggering and optical selection cuts, it is necessary to have a time dependent calibration to ensure that reconstruction level cuts, which are fundamentally placed on observed charge, correspond to the same amount of real optical activity.

This calibration is achieved by leveraging a high rate of single photoelectron (SPE) noise within MicroBooNE. A rate of $\approx 200$ kHz is observed by all PMTs within the detector. While the source remains of unknown origin, it is uniformly distributed over the detector volume and does not vary with time. This background noise provides a unique opportunity for calibration because it is not necessary to halt normal detector operations to collect the calibration data. This provides ample statistics taken at constant time intervals. Specifically, the data comes from the Unbiased BNB stream to ensure that the SPE rate is not biased by requiring something brighter in the event.

Extracting the SPE pulses from the unbiased readout waveforms employs a version of a constant fraction discriminator. This algorithm was chosen because it performs well on pulses which are variable in size but very similar in shape. The triggering threshold is set to 10 ADC, which corresponds to approximately 0.5 PE at normal operating voltage. This is also about 20 times the RMS of the waveform baseline.

To select well defined pulses which are of the correct magnitude to be SPE pulses, the following criteria are required.

1. Amplitude < 50 ADC

2. Integrated area < 500 ADC·ticks

3. Baseline RMS before pulse < 2 ADC

4. Baseline RMS after pulse < 2 ADC

5. Difference in baseline mean before / after is at most 5%

After this selection criteria we are left primarily with SPE pulses with some contribution from nPE pulses. The number of PE to reach the first dynode of a PMT for a given number N of emittted PE is poisson distributed. The dynode amplification will introduce gaussian smearing. The observed final spectrum will be the product of these two, the so called Multi Photoelectron (MPE) function.

$$S(x) = \sum_n A_n \frac{N^n e^{-N}}{n!} \frac{1}{\sqrt{2\pi n\sigma^2}} exp\Big[\frac{(x-n\mu)^2}{2n\sigma^2}\Big] \tag{B.1}$$

This function is then fitted to the observed distribution of selected pulses to extract $\mu$, the value for the mean SPE. Fits are done both to the pulse amplitude and pulse area histograms. This provides an effective gain factor linking ADC and PE. An example fit is show in Fig. B-1.

Data is measured in this fashion in 1-week chunks. This time span is sufficient to accumulate enough statistics but short enough that gain fluctuations are typically negligible.

Gain measurements are saved to a database which defines the appropriate gain values and interval of validity for that calibration point. This database is then used to correct optical data before flashes are reconstructed for analysis.

Validation of the gain calibration has been performed by comparing observed (with and without calibration) and predicted light for a sample of anode or cathode crossing cosmic tracks. This is shown summed over all PMTs in Fig. B-2 and for one example PMT in Fig. B-3.

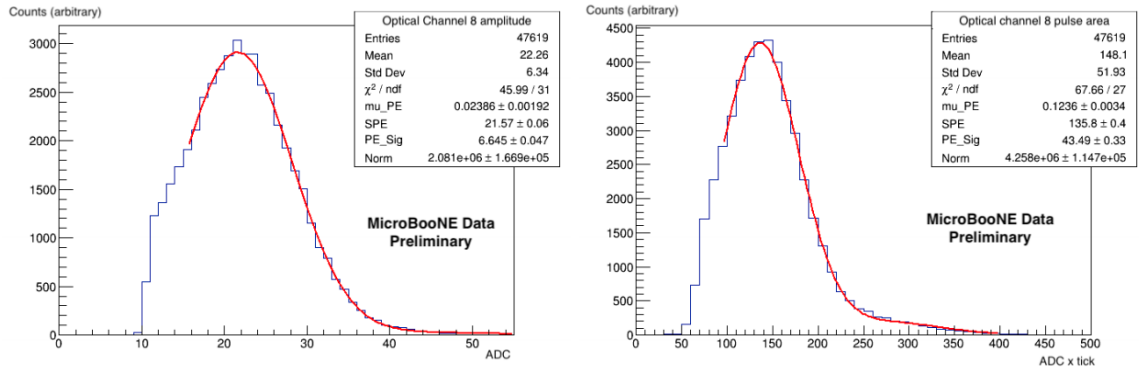

Figure B-1: An example of the Multi Photoelectron fit applied to a spectrum of selected pulse amplitudes (left) and to areas (right).

The application of this calibration has proved to be vital to ensuring high quality agreement between data and simulation as well as ensuring that temporal variability is not introduced into analysis.

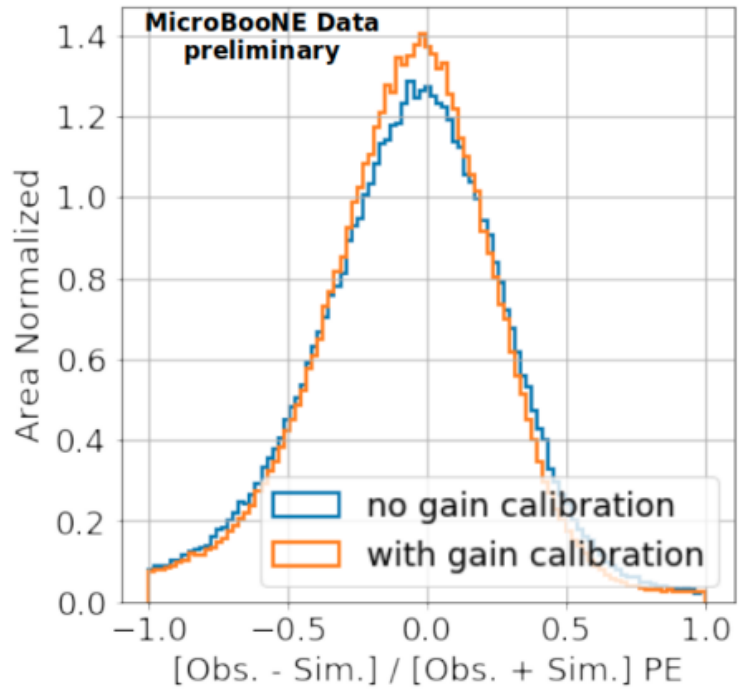

Figure B-2: Observed vs predicted light before and after performing the PMT gain calibration. This is summed over all PMTs. The distribution is visibly narrowed.

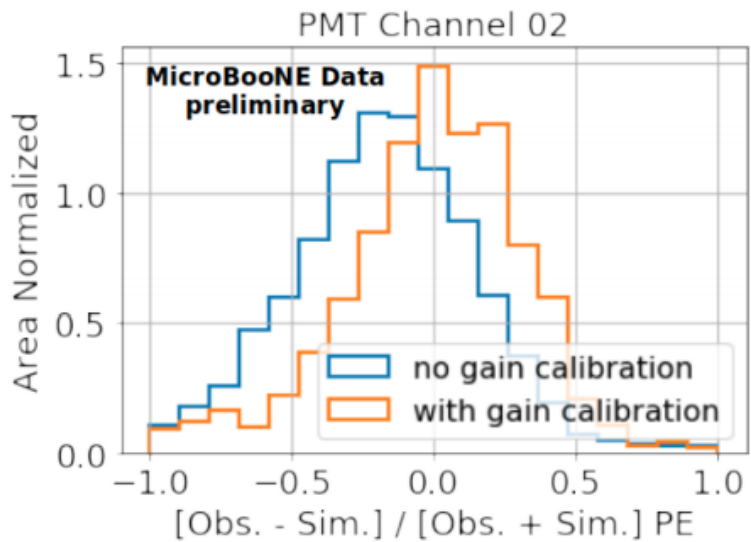

Figure B-3: Observed vs predicted light before and after performing the PMT gain calibration for a specific example PMT. The distribution is visibly narrowed and shifted closer to 0.

# Appendix C

# Development of Light Guides for Future LArTPCs

The development of new TPB coated light guides is a project for future large scale LArTPCs to which the author contributed substantially. Here we reproduce a preprint documenting some of the latest developments. The original can be found at

https://arxiv.org/pdf/1604.03103.pdf

# A Factor of Four Increase in Attenuation Length of Dipped Lightguides for Liquid Argon TPCs Through Improved Coating

Z. Moss[1], J. Moon[1], L. Bugel[1], J.M. Conrad[1], K. Sachdev[2], M. Toups[2], T. Wongjirad[1]

[1]*Department of Physics, Massachusetts Institute of Technology*
[2]*Neutrino Department, Fermi National Accelerator Laboratory*

ABSTRACT: This paper describes new techniques for producing lightguides for detection of scintillation light in liquid argon time projection chambers. These can be used in future neutrino experiments such as SBND and DUNE. These new results build on a dipped-coating technique that was previously reported and is reviewed here. The improvements to the approach indicate a factor of four improvement in attenuation length of the lightguides compared to past studies. The measured attenuation lengths, which are $>2$ m, are consistent with the bulk attenuation length of the material. Schematics for a mechanical dipping system are provided in this paper. This system is shown to result in coatings with $<10\%$ variations.

## Contents

## 1. Introduction

Flat-panel lightguides for light detection are proposed for a number of future Liquid Argon Time Projection Chambers (LArTPCs) including SBND [1] and DUNE [2]. Scintillation light produced in liquid argon (LAr) has a wavelength of 128 nm, too short to be detected by a vast majority of current photodetectors. Therefore, the light must be converted into the visible to be observed. The lightguide technology takes advantage of this requirement by embedding a wavelength shifter, tetraphenyl-butadiene (TPB), into the coating of a bar that will capture and guide the light to the end. The bars can be assembled into a flat panel that requires substantially less space than a more traditional design based on photomultiplier tubes (PMTs), such as has been used in ICARUS [3] and MicroBooNE [4]. The end of the bar can be instrumented with silicon photomultipliers (SiPMs) that have high quantum efficiency ($\sim 40\%$) for the visible light and very low dark rate at the cryogenic temperature of LAr, 87 K.

This paper focuses on a technology for producing lightguides constructed out of clear, polished acrylic bars and covered with an acrylic-embedded TPB coating. TPB efficiently absorbs the 128 nm scintillation light and re-emits at approximately 425 nm [5]. This wavelength corresponds nicely with the peak efficiency of the SiPMs. Development of these lightguides has progressed over several years, and Refs. [6, 7, 8] describe the past steps. In this paper, we present improvements in technique that provide a substantial step forward in absolute brightness and attenuation length of this family of lightguides. In Sec. 2, we provide the relevant historical information needed for the discussion of the improved lightguides. In Sec. 3, we report the improved method for producing the lightguides. In Sec. 4, we present results on tests of the new lightguides in air. In Sec 5, we use the model of the lightguides found in [8] to predict the attenuation length of the new lightguides in liquid argon. In Sec 6, we discuss ongoing tests by SBND and DUNE on these new lightguides in liquid argon and provide a prediction concerning the attenuation length that will result from these tests.

## 2. Relevant Background Information

Our most recent paper describing the technique for producing lightguides and benchmarking their performance was published in 2015 [8]. We will therefore refer to these as the 2015 lightguides. In this paper, we will compare our improved lightguides to the 2015 lightguides using many of the procedures and techniques developed in Ref. [8]. Therefore, a brief review of the relevant information from that paper is required.

The lightguides are made of UTRAN UV-transmitting acrylic [9], with index of refraction of 1.49. This is cut into bars of appropriate length and diamond polished on the sides and ends. In our studies we report on bars that are 0.25"×1.00"×20.0". The bulk attenuation length of a single bar has been reported as 160, 260 and 260 cm for 385, 420 and 470 nm light, respectively [10]. The error on the measured bulk attenuation was not reported.

The bars are carefully annealed. For the 2015 bars (and the new bars described in this paper), we use the annealing procedure described in Ref. [8]. The annealing apparatus consists of an insulated tube that houses the acrylic bars and whose inner volume is warmed by a heat-gun inserted into one end of the tube. The temperature of the air near the output of the heat gun is 230°F, while at the opposite end of the tube, the air temperature is measured to be 180°F. Therefore, the air temperature throughout the tube follows a gradient from 230°F to 180°F with most of the tube well below 230°F.

After the annealing procedure, the bars are thoroughly cleaned with ethanol. Next, a vertical cylinder with oval cross section large enough to contain all but the upper few centimeters of the bar is filled with the liquid coating. This cylinder is referred to as the "candlestick" (see Fig 1, left). The bar is then inserted into the candlestick and allowed to soak in the liquid. Finally, the bar is drawn out vertically and allowed to air-dry.

In Ref. [8] we had found that the lightguide performance did not correlate with the order in which up to 5 lightguides were dipped in the same batch of solution. We now find, though, that particulates may contaminate liquid that has been used many times. Therefore, we recommend pouring the liquid out after five dips and either filtering or making a new liquid batch. Also, to minimize particulate in the candlestick, the candlestick is now washed with ethanol immediately before the dipping process as well as after the dipping process, as was done for the 2015 lightguides.

The recipe that was used for the dipping liquid for the 2015 lightguides combined 1 g acrylic pellets, 0.5 g TPB, 50 mL toluene and 10 mL of ethanol. The acrylic pellets assure that the coating's index of refraction matches well to that of the bar. The TPB becomes embedded in the acrylic matrix and shifts the 128 nm light. The toluene is the primary solvent. The added ethanol produces a more uniform coating when the bar dries, which is essential to a long attenuation length. The amount of ethanol required in the mix is determined experimentally by testing the attenuation length of the lightguides.

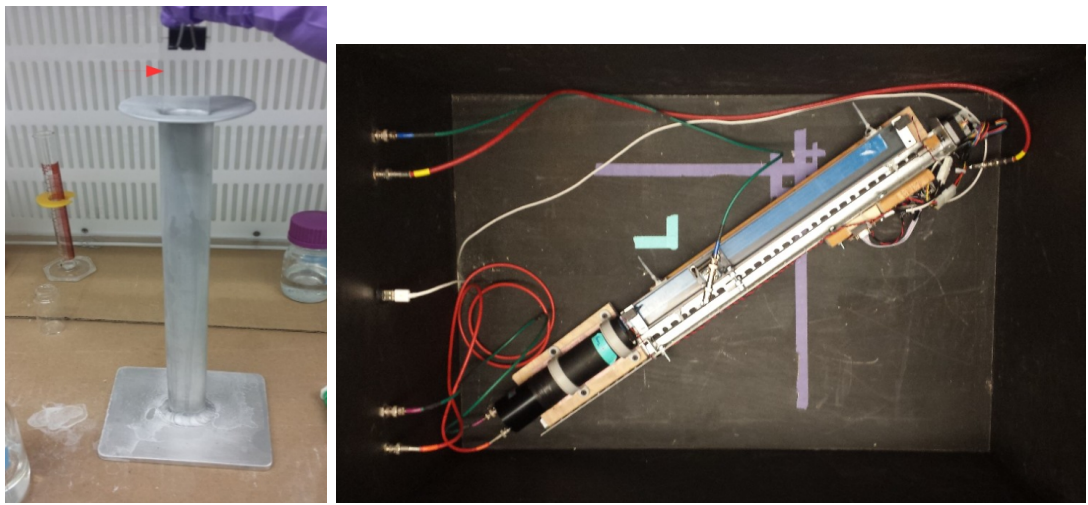

**Figure 1.** *Left: Acrylic guide being dipped in the candlestick filled with TPB coating; Right: Attenuation test setup. Images from Ref. [8].*

The attenuation length of the lightguides was tested using an attenuation length tester described in detail in Ref. [8] and shown in Fig 1, right. This tester uses a 286 nm LED. The LED is moved along the bar by a stepper motor in well-calibrated steps. The LED is pulsed $10^4$ times, and the wavelength-shifted light exiting the end of the bar is measured by a PMT. The waveforms of these PMT pulses are recorded, and their integrals are histogrammed. The brightness value corresponding to a particular distance along the bar is the mean of this distribution of integrals (charges, in ADC×ns). The corresponding standard error on the mean is small due to the large number of measurements and is negligible in comparison to systematic errors on the charge. The set of mean charges recorded at each distance is then fit to extract bar performance parameters. We also developed a model to describe the attenuation length measurement [8, 11].

The attenuation length measurement in air is sensitive to the thickness of the coating on the bar because the 286 nm light penetrates into the bar and excites all of the TPB in its path. Thus, with a thicker coating, more TPB will be excited and the light output of the bar will increase. This is unlike the expected behavior of the bar when exposed to 128 nm LAr scintillation light, which does not penetrate beyond near the surface of the coating. As a result, because the dipping process tends to yield a thicker coat at the base of the bar than at the top of the bar, the coating thickness affects the measured attenuation length. If the attenuation length is measured with top closest to the light detector and base farthest from the light detector ("forward"), the result will be artificially long due to the increase in coating thickness. On the other hand, if one reverses the bar, so that the measurement proceeds with base nearest to the photodetector and top farthest ("backward"), one obtains an attenuation length that is artificially short. Neither will represent the actual attenuation length as obtained when 128 nm light hits the bar. As described in the Sec 4.1, one can fit the

– 3 –

combined forward and backward measurement to simultaneously extract parameters that describe the change in coating thickness along the bar as well as the actual attenuation length.

The attenuation length of the propagating visible light in the lightguide depends on two properties: (1) the bulk attenuation in the lightguide and (2) surface losses. The first property is not related to whether the lightguide is immersed in air or liquid argon, since it is simply a property of the acrylic. The second property leads to a change in the measured attenuation length in air versus liquid argon because the difference in indices of refraction of air and liquid argon result in a change in the angle of total internal reflection. In Ref. [8], we developed a simulation which uses the forward and backward measurements as inputs, and then propagates the visible light through the lightguide with a parameter to describe the fractional light loss per reflection off of the coated surface. This paper showed excellent agreement between the model and measurements of the attenuation length of the lightguides in liquid argon. The 2015 lightguides had attenuation lengths in the 50–60 cm range in liquid argon [8].

The bars discussed in this paper are produced of the same UTRAN acrylic, cut and annealed as described above. The dipping process is the same as for the 2015 lightguides, with a small change to the candlestick design to reduce humidity, described below. However, the coating recipe has changed substantially, as discussed below. The same attenuation tester is used, with the modification that the Alazar Digitizer was replaced with a CAEN DT5740 digitizer, reducing the time required to test each 20" bar in one direction by a factor of six, to 20 minutes. We compared attenuation length measurements taken with the two digitizers and found that they are consistent to better than $1\sigma$ of the measurement errors reported in Ref. [8].

## 3. Improvements to the Previous Techniques

Our goal was to improve brightness and attenuation length. The brightness is affected by the ratio of TPB to acrylic in the coating. If the 128 nm light hits acrylic, it will be immediately absorbed. Therefore, an important goal was to increase the ratio of TPB to acrylic compared to the 2015 lightguide recipe. However, one cannot add so much TPB that the coating loses clarity and becomes white, or light will be lost when guided down the bar.

The attenuation length is affected by the uniformity of the coating. Thick coatings tend to produce a nonuniform, wavy surface when they dry, which can cause light to be lost as it travels along the bar. Thus, a goal was to make a coating that is thinner than the coating used for the 2015 lightguides, which motivated increasing the amount of toluene compared to the TPB and acrylic. We have also found that ethanol is important to producing a smooth surface. Bars produced without ethanol have visibly rough surfaces. In our new recipe we have honed the relative amount of ethanol to toluene through experiment, but have not introduced a major change from the ratio in the recipe used to produce the 2015 lightguides.

### 3.1 New Coating Recipe and Technique

We have adjusted the coating recipe to the following mixture:

- 50 mL toluene,

- 12 ml ethanol,

- 0.1 g acrylic, and

- 0.1 g TPB.

Lightguides with this coating will be called the '2016' lightguides in order to distinguish them from our previous '2015' lightguides in the discussion below.

We had found that the 2015 lightguides had consistently good response if soaked in the coating solution for 5 minutes or more. We find that for the 2016 lightguides, the bars should soak in the candlestick for 10 minutes in order to consistently produce bars with an even coating.

### 3.2 Humidity Control in Laboratory

Trial 2016 bars produced with this improved formula seemed more sensitive to the relative humidity in the lab than the 2015 bars. Specifically, the coatings of these trial bars produced during higher relative humidity summer conditions turned visibly cloudy and gave a shorter measured attenuation length than another batch of identically prepared 2016 bars produced when the relative humidity in the lab was lower. The temperature in the lab is steady at 70°F. The "hand-dipped" lightguide results presented here were gathered from bars made in the lab when the relative humidity was $< 20\%$. In the mechanically-dipped case, the apparatus made use of dry gas, resulting in negligible relative humidity.

### 3.3 Selecting the "Draw Time"

We have observed that bars removed from the candlestick quickly have a longer attenuation length (fit variable $\lambda$ described below) than those drawn out of the candlestick more slowly. Visual inspection also indicated more apparent variation in coating across the bar for those drawn more slowly. To test this hypothesis, we removed three 20 inch bars from the candlestick in total periods of 24 s, 15 s, and 4 s. We found attenuation lengths of 85 cm, 187 cm and 288 cm, respectively. Based on this study, we concluded that it is best to remove the bars from the candlestick relatively quickly. In the results presented below, all 20 inch bars are drawn from the candlestick in 10 s or less.

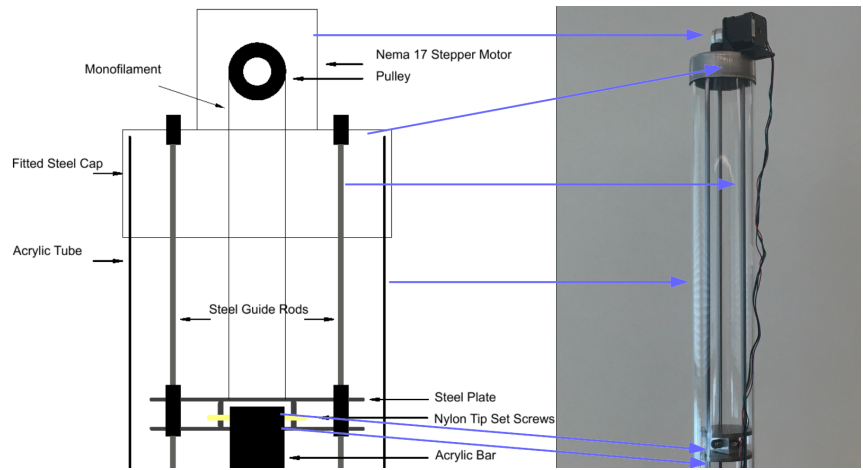

**Figure 2.** *Schematic of the drying and dipping tube*

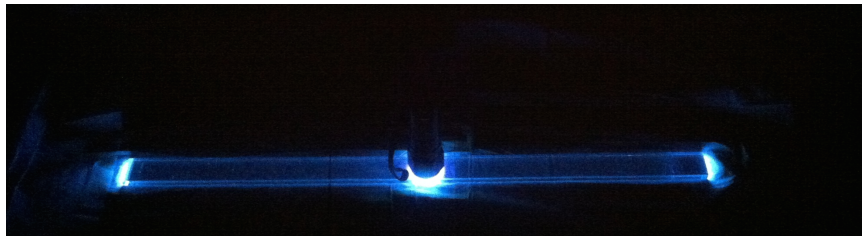

**Figure 3.** *Photograph of a 20" lightguide. The center is illuminated by a UV flashlight. Visible light is guided to the ends with low losses along the guide.*

### 3.4 Mechanized Dipping System

To further refine our production procedure, we constructed a container that automatically dips and dries the bars in near zero relative humidity conditions. See Fig. 2 for a schematic sketch and photograph of the dry tube setup.

The apparatus body is composed of a hollow acrylic tube oriented vertically. The candlestick is inserted at the base of the tube. A metal cap with a Nema 17 stepper motor attached is connected via a pulley and mono-filament system to a platform which moves up and down along three steel guide rods. The acrylic bar to be dipped is attached to this movable platform using set screws. This allows us to insert a new bar and then carefully control the ascent/descent rate of the bar.

The draw speed for the mechanically dipped bars was set to 9 s per 20" bar. This draw speed was limited by the motor. As will be discussed below, in the next generation of dipping machines, a faster motor might be desirable.

To provide a low humidity environment we introduce pressurized dry argon gas into base of the acrylic tube via a small valve. The relative humidity in the drying region is monitored by an Adafruit HTU21D-F capacitive relative humidity sensor, which provides real-time monitoring. Our monitor recorded zero relative humidity at all times during the dipping procedure. However, the technical specifications for the Adafruit monitor give an uncertainty in the relative humidity of $\pm 2\%$ for an optimized range of 5% - 95% relative humidity. Without more detailed information on the near zero performance of the HTU21D-F model, we can only claim that all procedures were done below 5% relative humidity.

### 4. Results in Air

We report measurements from the two techniques: hand dipping and mechanical dipping. All bars were 20" in length. All hand-dipped bars were produced from a single batch of coating recipe. All mechanically-dipped bars were produced from a different single batch of coating recipe. Results are listed in Tab. 1 and shown in Fig. 4. For each of the categories of bars, we report the average, with the standard deviation of the measurement in parenthesis.

Before discussing the results in detail, we note that these 2016 lightguides are visibly of high quality and appear substantially better than the 2015 lightguides. Fig. 3 illustrates this point. A UV-light is placed at the center of a lightguide. The visible light that is captured in the bar is guided to the ends with very little loss along the bar.

| Technique | λ, Attenuation length (cm) | C, Increase in thickness (%/cm) | N, Normalization (arbitrary units) |
|---|---|---|---|
| Mechanically-Dipped | 229 | 0.80 | 17.6 |
| | 286 | 0.84 | 18.9 |
| | 179 | 0.77 | 20.3 |
| | 201 | 0.69 | 20.9 |
| | 213 | 0.64 | 20.7 |
| | 217 | 0.84 | 20.3 |
| mean (std. dev) | 220 (36) | 0.76 (0.08) | 19.8 (1.3) |
| Hand-Dipped | 259 | 0.57 | 23.0 |
| | 339 | 0.50 | 21.0 |
| | 257 | 0.45 | 23.6 |
| | 306 | 0.87 | 21.0 |
| | 205 | 0.49 | 23.2 |
| | 232 | 0.28 | 25.1 |
| | 251 | 0.10 | 26.8 |
| mean (std. dev) | 264 (45) | 0.45 (0.25) | 23.7 (2.7) |

**Table 1.** Extracted attenuation lengths, percentage increase in brightness due to coating thickness from top to bottom of the bar, and normalization for two production techniques: hand-dip (top) and mechanized-dip (middle, bottom). The normalization is in (ADC counts * ns) divided by 1000 – an arbitrary unit, but one that allows for cross comparison.

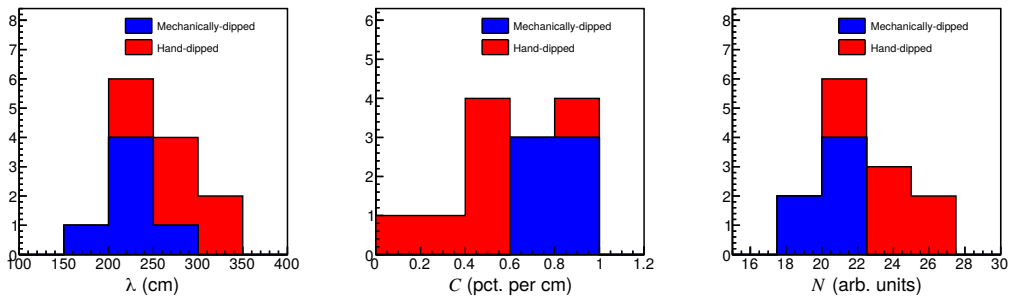

**Figure 4.** *Histograms of data presented in Tab. 1. Red: hand-dipped lightguides, Blue: mechanically-dipped lightguides. Left: λ, attenuation length; Middle: C, percentage change of brightness per centimeter; Right: N, normalization in arbitary units.*

### 4.1 Model Characterizing Guide Performance in Air

In the analysis of the response of the 2016 lightguides to 286 nm LED light measured in air, we characterize lightguide performance using a model that accounts for two effects: 1) the attenuation length of the bar due to imperfections on the surface or in the bulk and 2) the varying thickness of the coating which we assume is linear from one end of the bar to the other. To constrain the latter, the light output of the bars is measured twice in our setup in two different orientations, forward and backward, as described in Sec. 2.

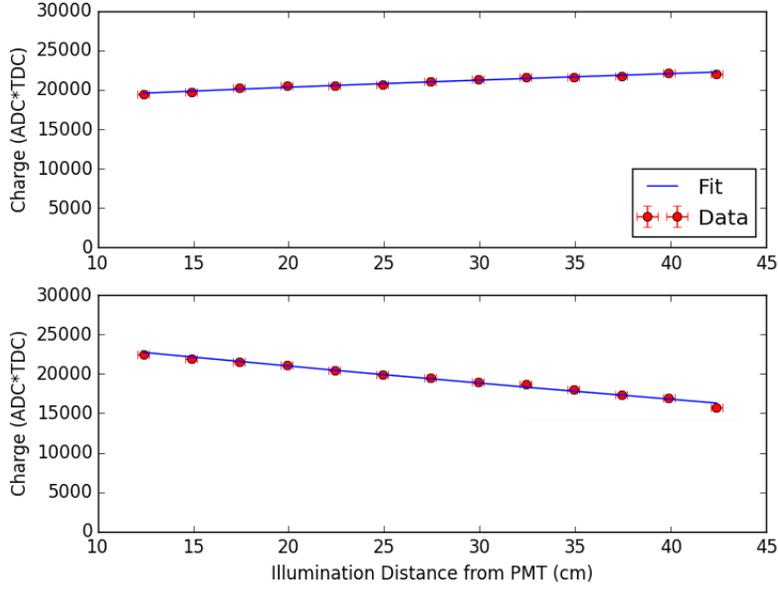

**Figure 5.** *An example of the four parameter fit to one mechanically-dipped bar. Top: forward measurement; Bottom: backward measurement.*

We fit the forward and backward measurements simultaneously to extract four parameters. In the forward direction, the light response is fit to:

$$F(x) = N exp(-\lambda * x) * (1 + C * (x - m)), \quad (4.1)$$

where $x$ is the distance from the PMT to the point measured, $m$ is the distance from the end of the bar to the meniscus where the coating begins. There are three free parameters in this equation:

- $N$ is an overall normalization of the brightness, measured in (ADC counts)*ns. Because this unit is not cross calibrated to a known brightness, we call this an arbitrary unit below.

- $\lambda$ is the attenuation length, measured in cm.

- $C$ is the change in thickness of the coating along the bar, which is expected to increase linearly from top to bottom, and is measured in percentage change of brightness per cm.

The backward data are fit with the equation:

$$B(x) = (\varepsilon * N) exp(-\lambda * x) * (1 + C * (L - m - x)). \quad (4.2)$$

where $L$ is the total length of the bar, from top to bottom. $N$, $\lambda$ and $C$ are the three parameters simultaneously fit to the forward data. The backward fit has an additional, fourth parameter:

- $\varepsilon$, measured as a percentage, allows for differences in absolute light collected by the PMT because in the backward direction the end of the bar is coated, while the in the forward direction it is uncoated.

Of these four fit parameters, three are important to this study. The attenuation length, $\lambda$, will be related to the smoothness of the coating and the bulk attenuation length. This is the primary result reported about the lightguides below. Parameters $N$ and $C$ provide measurements of the consistency of the coating between and along acrylic bars. The results on these parameters will be shown to be more stable with a mechanical dipping system. We find that the $\varepsilon$ efficiency correction is close to unity ($> 96\%$ in all cases) and not relevant to our conclusions. When the fit is performed, the data points are required to be within $> 10$ cm and $< 42$ cm to avoid the edges of the bar in the forward and backward direction.

Fig. 5 shows an example fit to one mechanically dipped bar. Note that in the forward bar arrangement, the overall slope is positive. This is because the change in coating thickness per unit length is dominating the fit. This is typical of all of the 2016 lightguides.

### 4.2 Results on Mechanically-dipped Lightguides

For mechanically dipped bars, we find an average (standard deviation) of the attenuation length of 220 (36) cm. The 16% spread of values encompasses the reported 260 cm bulk attenuation length at $1.1\sigma$. Thus we conclude that the measured $\lambda$ values are consistent with the bulk attenuation length. Through repeated measurements on the same bar, we found that $N$, the normalization, can vary up to 2.5%. Systematic effects that may cause this include variations of the light source and variations in how the bar is seated against the PMT. The PMT was allowed to "warm up" for 30 minutes before each measurement, however variation of the dark rate after this period may also contribute to this systematic error. The average (standard deviation) for $C$ was 0.76 (0.08) and for $N$ was 19.8 (1.3). Thus the spreads of both coating-specific results are $\sim 10\%$ across the lightguide samples.

### 4.3 Comparison of Mechanically-dipped Results to Hand-dipped Results

Tab. 1 and Fig. 4 also present the results of a set of hand-dipped lightguides made with the improved 2016 coating. We find that the hand-dipped results are in agreement with the mechanically-dipped results within the standard-deviation for all three parameters of interest. Interestingly, the hand-dipped results have systematically better parameter values. However, compared to the mechanically-dipped bars, the spread in the measured values of the hand-dipped lightguides is much larger.

Consider, first, the two parameters associated with the coating, $C$ and $N$. For $C$ we obtain an average (standard deviation) of 0.45 (0.25) for the all hand-dipped bars. The spread is very large – 56% compared to 10% for mechanically dipped bars. The normalization fit yields 23.7 (2.7). This standard deviation, which is about twice as large, may be due to less reproducible behavior when hand-dipping. Overall, the mechanical-dipping produces substantially tighter distributions of both coating-related parameters. The mechanically-dipped and hand-dipped results agree within the spread, but it is apparent in Fig. 4, that the hand-dipped results for $C$ are lower, and for $N$ are systematically higher, than for the mechanically-dipped case. This indicates that the hand-dipped bars have marginally better coating performance than the mechanically-dipped bars. This may result from differences in mixing the coating recipe, as the two samples were produced from two different batches. Better control of the coating production process may address this spread.

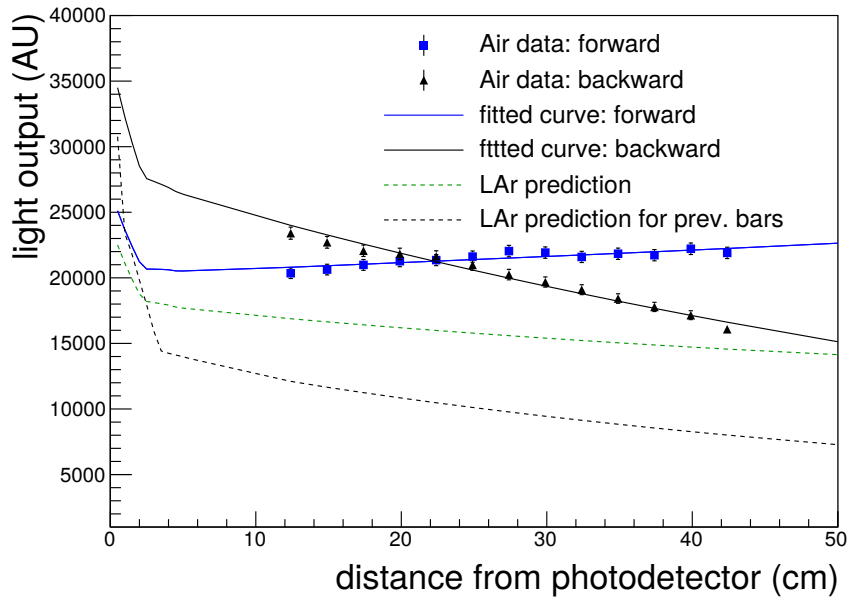

**Figure 6**. *Predicted behavior of one of the 2016 lightguides in liquid argon. Using the simulation from Ref. [8], we input the forward (blue solid) and backward air data (black solid) to characterize the bar and make prediction of its behavior in liquid argon (green dashed) by tracking the simulated bounces along the guide. For comparison, the liquid argon prediction for a previous bar studied in Ref. [8] is shown as well (black dashed).*

On the other hand, the hand-dipped and mechanically-dipped attenuation length measurements agree reasonably well, as seen in Fig. 4. The average (standard deviation) of the attenuation length of the hand-dipped bars, 264 (45), agrees within less than $1\sigma$ with that of the mechanically dipped bars. This is also in excellent agreement with the bulk attenuation length. As with the mechanically-dipped samples, the bulk attenuation appears to dominate. With this said, there may be a small systematic shift to longer attenuation length in the hand-dipped case. We had noted earlier that the hand-dipped bars were removed with a draw of $< 10$ s, but that the exact time was not carefully measured as the bars were constructed. Faster draw-time could lead to a slightly higher attenuation length. We recommend that in a future mechanical drawing system, the draw time, which was 9 s, be reduced to 5 s.

## 5. Predicted Performance In Liquid Argon

In order to verify our results, 2016 lightguides produced using the techniques of this paper are now under study in liquid argon by two subgroups, one from DUNE [12] and one from SBND [13]. The attenuation lengths of the lightguides will be measured using cosmic rays and sources and will improve upon the previous method by taking more granular data, thereby providing more detailed information on the attenuation length of the bars in liquid argon. Results will be available within 6 months.

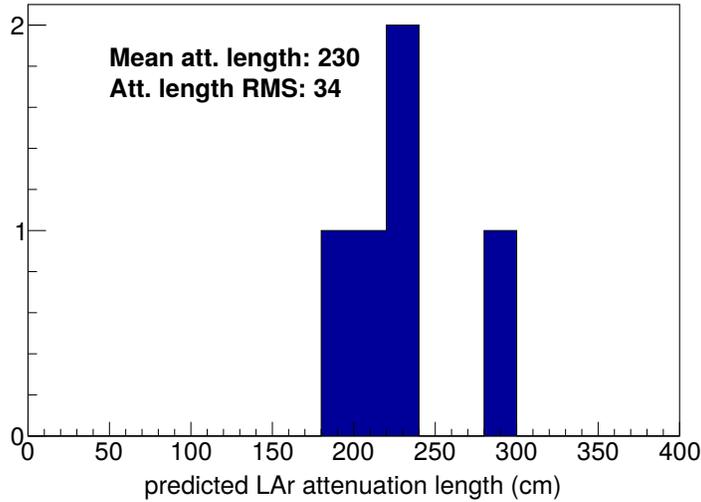

**Figure 7.** *Predicted attenuation lengths of the 2016 lightguides in liquid argon. The simulation from Ref. [8] is used to model the bars' behavior in liquid argon using parameters found from a simultaneous fit of the forward and backward air data.*

In this section, we use the model for performance in liquid argon published in Ref. [8] to make predictions for these results. We input the forward and backward air data to characterize the bar and make prediction of its behavior in liquid argon by tracking the simulated bounces along the guide. Fig. 6 shows the result of the model for the mechanically dipped lightguides in the green-dashed line. Near the end of the lightguide, direct light hitting the sensors causes a deviation from an exponential. But beyond about 10 cm, an exponential can be used to quantify the predicted attenuation length of the 2016 lightguides. As shown in Fig. 7 we find the bars in liquid argon are expected to have an attenuation length greater than 200 cm and consistent with the results reported in Tab. 1. The fact that the air and liquid argon results agree so well is consistent with the conclusions of the previous section that the bulk attenuation length dominates over losses from smoothness of the coating in these 2016 lightguides.

For comparison Fig. 6 also shows the model's prediction for the liquid argon behavior of one 2015 lightguide, as the black-dashed curve. Note that this prediction was shown to agree well with the liquid argon measurements in Ref. [8]. One can see that there is marked improvement from the 2016 lightguides, with the green-dashed curve being significantly flatter than the black-dashed curve. (Note that the normalizations of the green and black dashed curves are arbitrary.) The attenuation lengths of the 2015 lightguides in liquid argon were about 50-60 cm. Thus, the 2016 lightguides have a factor of four longer attenuation length than this previous iteration.

## 6. Future Step: Improved Acrylic

DUNE plans to use lightguides that are 220 cm long and SBND proposes ∼100 cm lightguides. These are on the scale of the bulk attenuation length of the acrylic which is now apparently dominating the light loss along the 2016 lightguides. In this case, if the results of future tests in liquid argon bear out our prediction, then the next step to improve these dipped lightguides will be to obtain acrylic with longer bulk attenuation length than the UTRAN currently used. Ref. [14] provides a study of bulk attenuation length of acrylic on the market, and identifies three that have bulk attenuation > 4 m. If this improved acrylic can be used successfully, then this will lead to the next dramatic improvement in light output of dip-coated lightguides.

## Acknowledgements

The MIT collaborators are funded by NSF grant PHY-1505855. TW gratefully acknowledges the support provided by the Pappalardo Fellowship Program at MIT.